# Theory of Network Wave

Bo LI*, Mao YANG, and Zhongjiang YAN

School of Electronics and Information, Northwestern Polytechnical University,

Xi'an, CHINA, 710072

{libo.npu*, yangmao, zhjyan}@nwpu.edu.cn

***Abstract*: Aiming at the disorder problem (i.e. uncertainty problem) of the utilization of network resources commonly existing in multi-hop transmission networks, the paper proposes the idea and the corresponding supporting theory, i.e. theory of network wave, by constructing volatility information transmission mechanism between the sending nodes and their corresponding receiving nodes of a pair of paths (composed of two primary paths), so as to improve the orderliness of the utilization of network resources. It is proved that the maximum asymptotic throughput of a primary path depends on its intrinsic period, which in itself is equal to the intrinsic interference intensity of a primary path. Based on the proposed theory of network wave, an algorithm for the transmission of information blocks based on the intrinsic period of a primary path is proposed, which can maximize the asymptotic throughput of a primary path. In the cases of traversals with equal opportunities, an algorithm for the cooperative volatility transmission of information blocks in a pair of paths based on the set of maximum supporting elements is proposed. It is proved that the algorithm can maximize the asymptotic joint throughput of a pair of paths. As for the cases of traversals with unequal opportunities, an algorithm for the cooperative volatility transmission of information blocks in a pair of paths based on the set of maximum supporting elements is also proposed. Moreover, the upper bound of the asymptotic joint throughput of a pair of paths achieved by using this algorithm is obtained. Finally, the concept of the Distribution Spectrum of Interference-Spacing of nodes of a primary path is defined and based on it a simple and general algorithm flow for solving the intrinsic period of a primary path is proposed. The research results of the** **paper lay an ideological and theoretical foundation for further exploring more general methods that can improve the orderly utilization of network resources.**

# I. Introduction

As we all know, in today's networks with a certain scale (including wired networks and wireless networks), the multi-hop transmission mode based on the idea of "store and forward" is mainly used to transmit information from one end of the network to the other end (in fact, the idea of "store and forward" constitutes the core of packet switching technology [1]). When implementing multi-hop transmissions based on the idea of "store and forward", the common practice is that an information forwarding node (commonly known as "a router") schedules its access to network resources according to the distribution of its local resources and the dynamic requirements of network traffic. The advantage of this method for accessing network recourses is that it can help to reduce the design and implementation complexity of network protocols as much as possible. However, the problem it brings about is also prominent, that is, because each node only determines the usage of network resources according to its own needs, which can be compared to the way that each node uses shared network resources from the starting point of "myopia" and "selfishness" (Note: as a matter of fact, the multiple access protocol of nodes in a multi-hop network operates based on this basic idea). This will inevitably lead to the disorderly usage of network resources (Note: the "disorder" here actually refers to "uncertainty", which can be measured by the concept of entropy proposed in information theory [2]). And the disorderly usage of network resources leads to two problems: first, the availability of network resources used to transmit network traffic presents great uncertainty, which makes it difficult to ensure the Quality of Services (QoS) guarantee; second, due to the possible disorderly competition for network resources among different nodes, collisions of resource occupations occur from time to time, which further reduces the utilization of network resources.

Based on the starting point of "myopia" and "selfishness", with the continuous expansion of the network scale (i.e. the continuous increase of the number of nodes accessing into the network), the uncertainty of the availability of network resources will also increase. In other words, with the continuous expansion of the network scale, the disorderly usage of network resources will be further amplified. Consider the example shown in **Figure I.1**, in the subfigure (a), node 1 sends three packets, and to some extent the time intervals between the three packets present some uncertainty (i.e. disorder). It is particularly noteworthy that due to the uncertainty of the intervals of these transmissions, the remaining idle access resources also present some uncertainty. And then, another node, i.e. node 2, can only, by following the starting point of "myopia" and "selfishness", access into the remaining resources which have already shown a certain degree of uncertainty (Note: in order to avoid mutual interference, the accesses and transmissions of node 1 and node 2 must be staggered in time), and sends its two packets. As can be seen from the subfigure (b), the access resources occupied by nodes 1 and 2 show greater uncertainty.

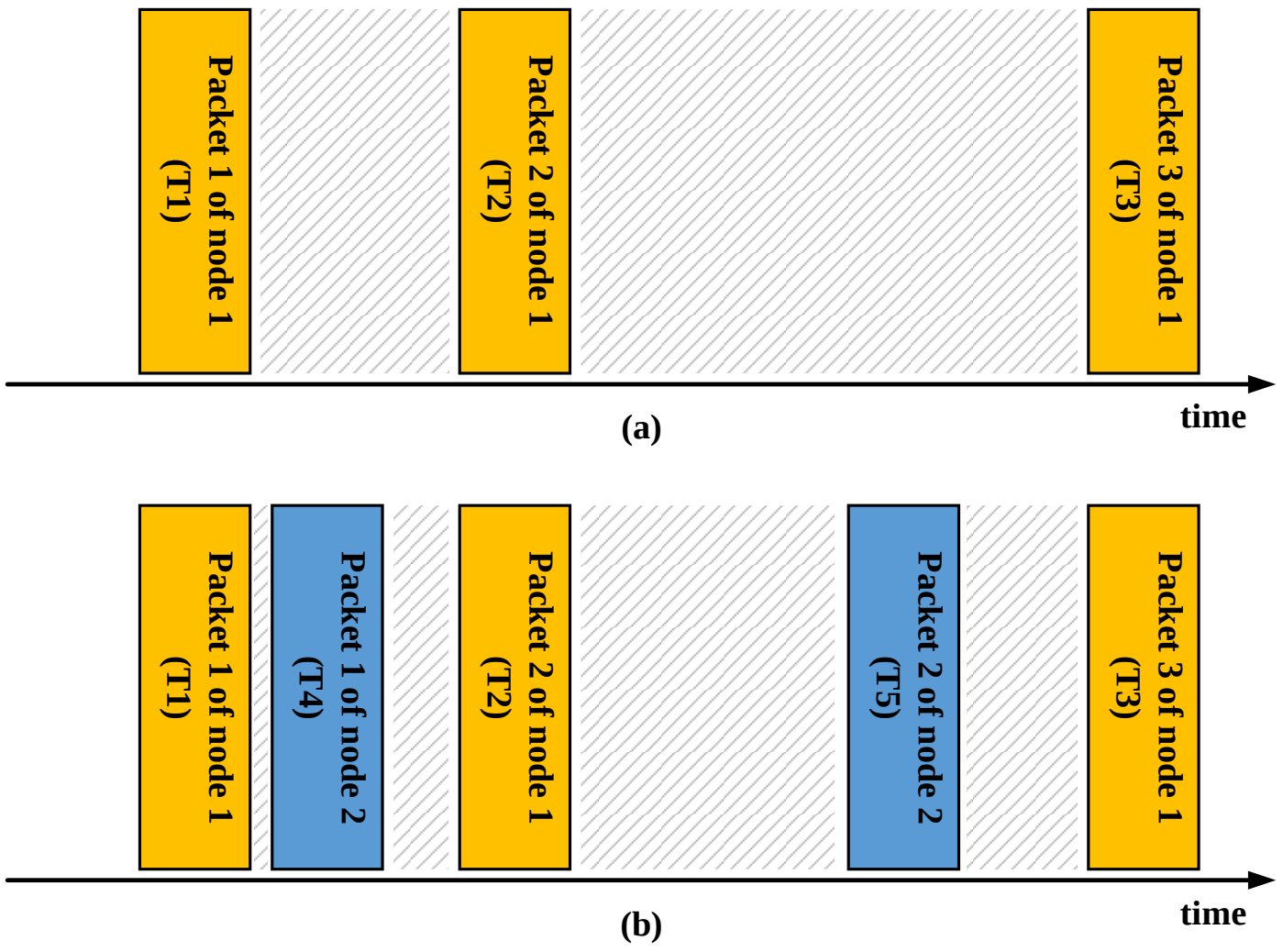

**Figure I.1** The disorder in the usage of network resources

With the help of the concepts of entropy and conditional entropy defined in information theory [2], it is easy to explain why the disorderly usage of network resources is gradually amplified. Suppose that the three transmission time instances, which can be regarded as three random variables, corresponding to the three packets sent by node 1 in **Figure I.1** are $T_1, T_2, T_3$, respectively. And the two transmission time instances corresponding to the two packets sent by node 2 are $T_4, T_5$ (which are also regarded as random variables), respectively. The following relationship can be obtained easily according to the information theory:

$$\begin{aligned} H(T_1, T_2, T_3, T_4, T_5) &= H(T_1, T_2, T_3) + H(T_4, T_5 | T_1, T_2, T_3) \\ &\geq H(T_1, T_2, T_3) \end{aligned} \tag{I.1}$$

Among them, $H(T_1, T_2, T_3)$ is the joint entropy of the transmission time instances corresponding to the three packets sent by node 1, i.e. the uncertainty presented by node 1 with its transmissions. And, $H(T_4, T_5 | T_1, T_2, T_3)$ denotes the conditional entropy corresponding to the other two packets sent by node 2 on the premise that the access time instances of node 1 are known (that is, the uncertainty of the accesses of node 2 with the transmissions of node 1 being given).

$H(T_1, T_2, T_3, T_4, T_5)$ is the joint entropy of the random variables corresponding to the five transmission time instances of both node 1 and node 2, which measures the uncertainty of the usage of the network resources with the transmissions of the two nodes. It can be seen from **Equation (I.1)**, with the further accesses of node 2, the occupation of network resources shows greater uncertainty (i.e. $H(T_1, T_2, T_3, T_4, T_5) \geq H(T_1, T_2, T_3)$). What is worse, in addition to the gradual "amplification" of the above disorder, this kind of disorder can be rapidly spreaded throughout the network with the interactions between adjacent nodes (just like the spread of viruses!). In short, the access mode based on the starting point of "myopia" and "selfishness" makes the disorderly usage of network resources to be continuously amplified and spreaded out in the whole network.

In the paper, the idea of introducing volatility into the network access mechanism is proposed, which reduces the disorder of the usage of network resources and accordingly improves the orderliness of the usage of network resources. Therefore, network throughput and performance for QoS guarantee can be improved. The contributions of the paper can be summarized as follows:

1) The idea of volatility is introduced into the access mechanism of multi-hop networks.
2) For a single multi-hop transmission path (i.e. a primary path), the corresponding basic theoretical framework of network wave is constructed. It is proved that the maximum asymptotic throughput of a primary path depends on its intrinsic period, and the intrinsic period of a primary path is equal to its intrinsic interference intensity.
3) An algorithm for the transmission of information blocks based on the intrinsic period of a primary path is proposed, which can maximize the asymptotic throughput of a primary path.
4) For a pair of paths composed of two primary paths, the corresponding basic theoretical framework of network wave is constructed, and the reachability of the joint period of a pair of paths is proved.
5) As for the cases of traversals with equal opportunities, an algorithm for the cooperative volatility transmission of information blocks in a pair of paths based on the set of maximum supporting elements is proposed. It is proved that the algorithm can maximize the asymptotic joint throughput of a pair of paths.
6) As for the more general cases of traversals with unequal opportunities, an algorithm for the cooperative volatility transmission of information blocks in a pair of paths based on the set of maximum supporting elements is also proposed. Moreover, the upper bound of the asymptotic joint throughput of a pair of paths achieved by using the proposed algorithm is obtained.

7) The concept of the Distribution Spectrum of Interference-Spacing of nodes of a primary path is defined, and based on it a simple and general algorithm flow for solving the intrinsic period of a primary path is proposed.

The paper is organized into 12 parts: in Section I, the disorder of the usage of network resources in the current access mechanisms of the multi-hop networks is pointed out. In Section II, the basic idea of the paper is proposed, that is, volatility is introduced into the access mechanism of a multi-hop network. In Section III, the concepts of a primary path and a pair of paths are defined. In Section IV, the relationships of concurrency and interference between nodes are defined. In Section V and Section VI, the concepts of the intrinsic concurrency intensity and the intrinsic interference intensity are defined, and the related basic properties are analyzed and proved. In Section VII and Section VIII, the concepts of the subset of equally spaced nodes, the reachable period of a primary path and the reachable joint period of a pair of paths are introduced, and the related basic properties are analyzed and proved. In Section IX and Section X, the average (joint) throughput and the asymptotic (joint) throughput of both a primary path and a pair of paths are defined, respectively, and the relevant basic properties are proved. Moreover, algorithms for the volatility transmission of information blocks in a primary path and in a pair of paths are proposed, respectively. In Section XI, the concept of the Distribution Spectrum of Interference-Spacing of nodes of a primary path is defined, and a simple and general algorithm for solving the intrinsic period of a primary path is proposed. Finally, conclusions are given in Section XII.

## II. Basic idea

In order to reduce the disorderly usage of network resources, it is necessary to increase the certainty for behaviors of nodes accessing into network resources. In the example shown in **Figure II.1**, the access behaviors of both node 1 and node 2 show apparent certainty (i.e. notable periodicity). If most of the nodes access into the network by following such kind of notable periodicity, network resources available to be used will inevitably show good certainty (i.e. orderliness).

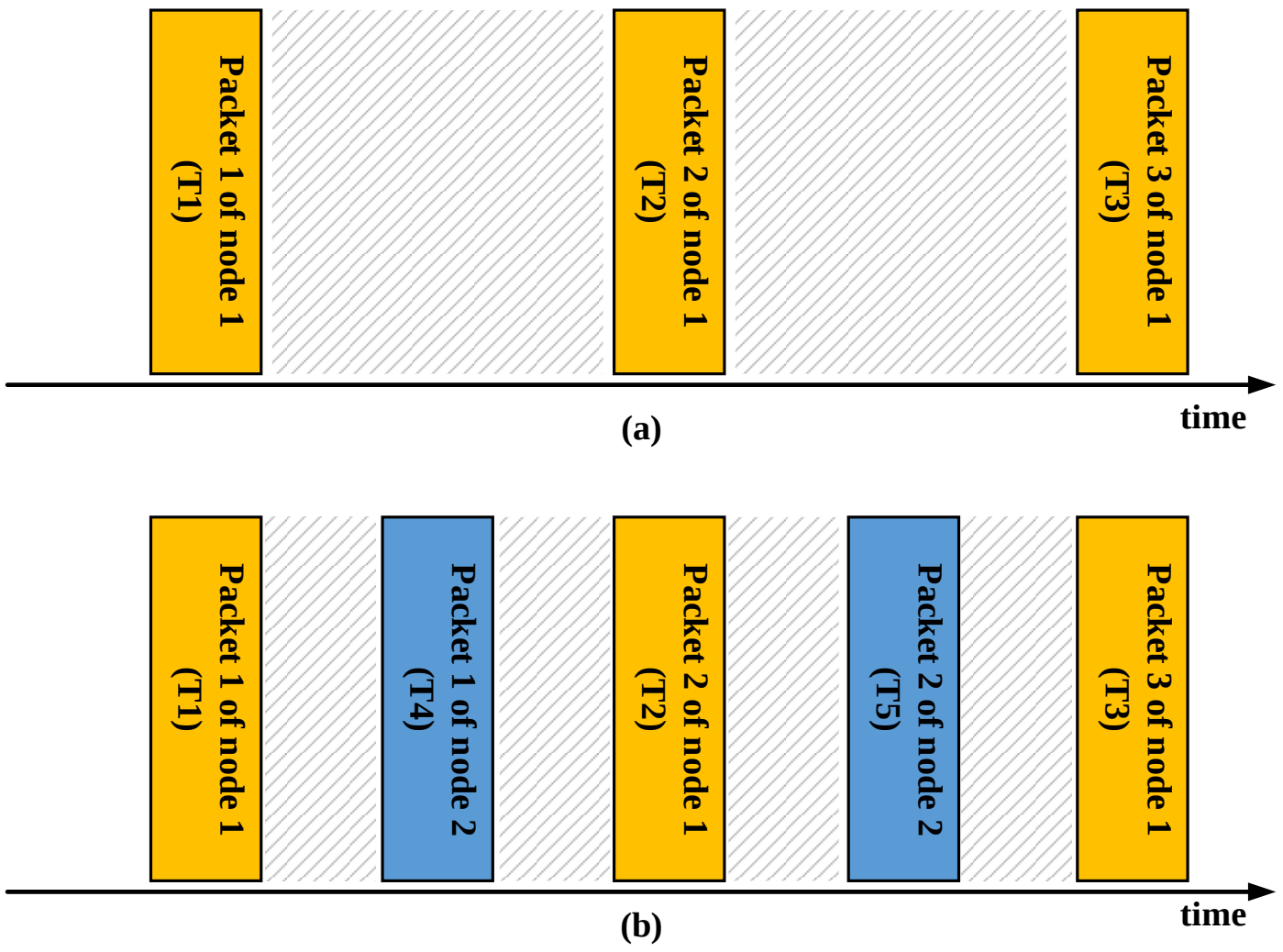

**Figure II.1** Orderliness of the usage of network resources

So, how to achieve the above "attractive" orderliness in networks? There are many possible ways, but the authors' suggestion is to follow nature! As we all know, wave phenomena occur almost everywhere in nature. We can enumerate many different types of waves, such as water wave, sound wave, light wave, seismic wave, gravitational wave, and even the "material wave" studied in quantum mechanics, and so on. Just as the famous physicist Feynman pointed out in his book "the Feynman structures on Physics (Vol. I)" [3], "…except that wave oscillations appear not only as time-oscillations at one place, but propagate in space as well". In other words, if one looks along the propagation path of a wave, she will find that the wave is moving forward. Moreover, if a point at a specific position on the propagation path of the wave being observed is fixed, it will be found that the motion of that point shows a strong regularity, that is, the so-called "vibration phenomenon". As for information network, which is invented by human beings, its core mission is to transmit information from one end of the network to the other. Therefore, is it possible that we imitate the wave mechanism in nature and build a similar “wave” in the network (let us call it as “network wave”), and, based on that, realize the orderly and efficient transmission of information blocks? Based on the research of the paper, the answer we give is yes. Since the physical world itself operates in this way, why can’t this mechanism give us some important enlightenments?!

In short, the core idea of building network wave can be summarized as follows: make each forwarding node on a multi-hop path regularly switch back and forth between “forwarding state” and “silence state”. And the state switching behaviors between adjacent nodes on the path are well coordinated in time (that is, in the same time beat, we make some nodes with a certain spatial distance on the path to be in the “forwarding state” and the other nodes to be in the “silence state”). In this way, the network wave can be established, and the information can be transmitted from one end to the other along the multi-hop path. Obviously, with the establishment of network wave, the access behaviors of nodes on the path do have good certainty (refer to the subsequent chapters of this paper for the detailed theory of network wave and the corresponding proposed algorithms).

## III. The concepts of a primary path and a pair of paths

In this section, the concepts of both a primary path and a pair of paths are given. The universal set of nodes for a primary path is defined. The set and subsets of sending nodes contained in a primary path are defined. The universal set of nodes for a pair of paths is defined. And the set and subsets of sending nodes contained in a pair of paths are defined, etc.

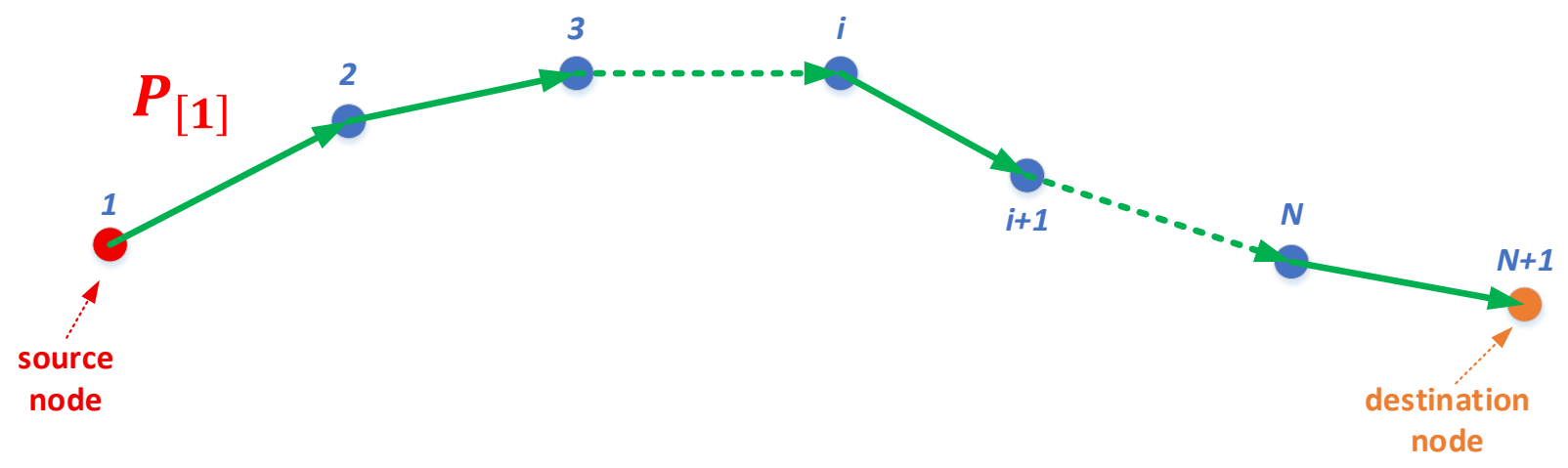

**Figure III.1** A primary path

**Definition III.1: A primary path**

As shown in **Figure III.1**, consider that in a multi-hop path consisting of $N_{[i]}+1$ ($N_{[i]} \geq 1; i \in \mathbb{Z}^{+}$) nodes, information blocks are relayed from the node 1 (i.e. the source node) through the node 2, the node 3,..., the node $i$,..., and the node $N_{[i]}$ to the final node $N_{[i]}+1$ (i.e. the destination node). A multi-hop path, which transmits information blocks from the source node to the destination node, consisting of $N_{[i]}$ sending nodes (whose sequence numbers are set in turn by using incrementing positive integers, i.e., from 1 to $N_{[i]}$) and one destination node (i.e., node $N_{[i]}+1$) arranged in sequence is called as a primary path, which is denoted as $P_{[i]}$ ($i$ ($\in \mathbb{Z}^{+}$) is the index number of a primary path).

**Note:**

1) It is assumed that all the former $N_{[i]}$ nodes have the function of sending/forwarding information blocks, and the last node $N_{[i]}+1$ only has the function of receiving information blocks (i.e., the node will not be in the transmission state);
2) The node in a primary path with a larger sequence number is said to be the “downstream node” of the node with a smaller sequence number, and, conversely, the node with a smaller sequence number is the “upstream node” of the node with a larger sequence number;
3) It is assumed that in a primary path the transmission of an information block from one node to

its immediate downstream node takes only the duration of one time beat, which is denoted as $\Delta t$ (unit: second);

4) Unless otherwise specified, all definitions given hereinafter are time-invariant.

**Definition III.2：The universal set of nodes for a primary path**

The set composed of all nodes in the primary path $P_{[i]}$ is defined as the universal set of nodes of $P_{[i]}$, denoted by $S^U_{P_{[i]}}$.

**Definition III.3: A specific node on a primary path, its predecessor node and its successor node**

Let $n_{i,j}$ $(i \in \mathbb{Z}^+; 1 \leq j \leq |S^U_{P_{[i]}}|)$ denote the node with the sequence number $j$ in the primary path $P_{[i]}$. Furthermore, node $n_{i,j-1}$ $(i \in \mathbb{Z}^+; 1 < j \leq |S^U_{P_{[i]}}|)$ is defined as the predecessor node of node $n_{i,j}$ in $P_{[i]}$, denoted by $n^-_{i,j}$. And node $n_{i,j+1}$ $(i \in \mathbb{Z}^+; 1 \leq j < |S^U_{P_{[i]}}|)$ is defined as the successor node of node $n_{i,j}$ in $P_{[i]}$, denoted by $n^+_{i,j}$.

**Note:**

1) It is stipulated that any sending node $n_{i,j}$ in a primary path $P_{[i]}$ transmits (or forwards) information only to its successor node $n^+_{i,j}$.

**Definition III.4: The spacing between two nodes on a primary path**

Define the spacing between the node $n_{i,k}$ $(\in S^U_{P_{[i]}})$ with sequence number $k$ and the node $n_{i,l}$ $(\in S^U_{P_{[i]}})$ with sequence number $l$ on the primary path $P_{[i]}$ as: $d^{k,l}_{P_{[i]}} \triangleq |k - l|$.

**Note:**

1) In this paper, only when two nodes are located on the same primary path can their spacing be well-defined.

**Definition III.5: The set of sending nodes of a primary path**

The set of nodes, consisting of all the sending nodes in a primary path $P_{[i]}$, is defined as the set of sending nodes of $P_{[i]}$, which is denoted as $S_{P_{[i]}}$ $(\subseteq S^U_{P_{[i]}})$.

**Definition III.6: A subset of sending nodes of a primary path**

A subset of sending nodes of a primary path $P_{[i]}$ is defined as a subset of $S_{P_{[i]}}$, which is denoted as $S'_{P_{[i]}}$ $(\subseteq S_{P_{[i]}})$.

**Definition III.7: A pair of paths**

Consider a holistic primary path system composed of two primary paths, i.e., $P_{[1]}$ and $P_{[2]}$. Such a system is defined as a pair of paths and denoted by $P_{[1,2]}$ (as shown in **Figure III.2**).

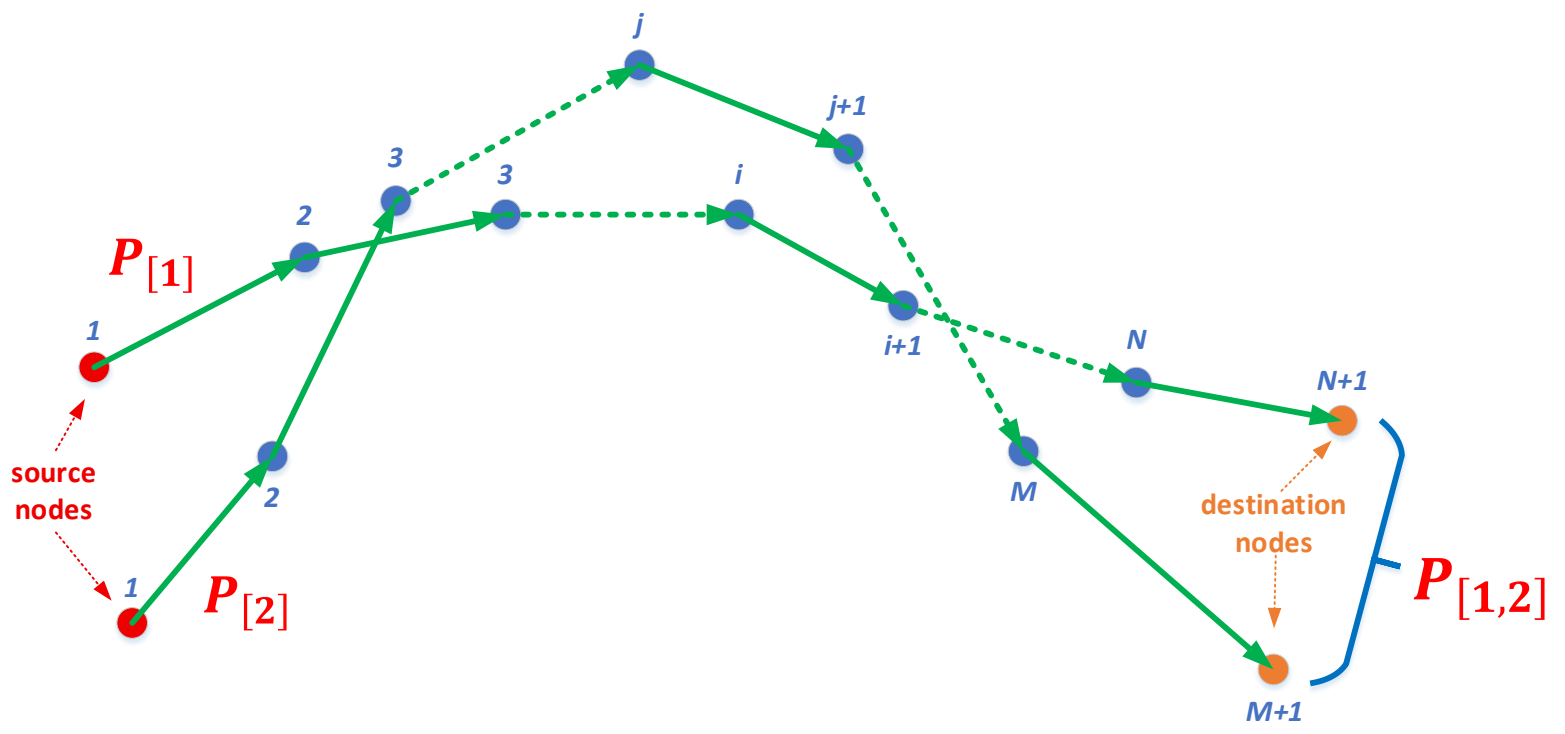

**Figure III.2** A pair of paths

**Note:**

1) Without loss of generality, unless otherwise specified, whenever a pair of paths is considered in this paper, it is always regarded to be composed of primary path $P_{[1]}$ and primary path $P_{[2]}$;
2) In this paper, any two nodes belonging respectively to primary path $P_{[1]}$ and primary path $P_{[2]}$ are regarded as distinct nodes.

**Definition III.8: The universal set of nodes for a pair of paths**

A set composed of all the nodes contained in the pair of paths $P_{[1,2]}$ is defined as the universal set of nodes of $P_{[1,2]}$, denoted as $S^U_{P_{[1,2]}}$. Obviously, according to **Definition III.2**, it follows that $S^U_{P_{[1]}} \subseteq S^U_{P_{[1,2]}}$, $S^U_{P_{[2]}} \subseteq S^U_{P_{[1,2]}}$ and $\left|S^U_{P_{[1,2]}}\right| = \left|S^U_{P_{[1]}}\right| + \left|S^U_{P_{[2]}}\right|$.

**Note:**

1) Unless otherwise specified, the set $S^{U}_{P_{[1,2]}}$ shall be regarded as the universal set of nodes for the research problem in this paper;
2) We say that two nodes $n_{i,j}$ $(\in S^{U}_{P_{[1,2]}})$ and $n_{k,l}$ $(\in S^{U}_{P_{[1,2]}})$ within the pair of paths $P_{[1,2]}$ are distinct if relation $(i \neq k) \vee (j \neq l)$ is satisfied, which is denoted as $n_{i,j} \neq n_{k,l}$.

**Definition III.9: The set of sending nodes of a pair of paths**

The set of nodes, consisting of all the sending nodes in a pair of paths $P_{[1,2]}$, is defined as the set of sending nodes of $P_{[1,2]}$, which is denoted as $S_{P_{[1,2]}}$ $(\subseteq S^{U}_{P_{[1,2]}})$.

**Definition III.10: A subset of sending nodes of a pair of paths**

A subset of sending nodes of a pair of paths $P_{[1,2]}$ is defined as a subset of $S_{P_{[1,2]}}$, which is denoted as $S'_{P_{[1,2]}}$ $(\subseteq S_{P_{[1,2]}})$.

## IV. Definitions of the relationships of concurrency and interference between nodes

In this section, the relationships of concurrency and interference between nodes in a pair of paths are defined; The universal set of concurrency nodes and the universal set of interfering nodes for a sending node are defined; Transmission effectiveness for a sending node is defined. And effectiveness of concurrent transmission for a non-empty subset of sending nodes is also defined; The interference decreasing rule among nodes on a primary path (**Rule IV.1**) is given; The inherent relationships between the universal set of interfering nodes and the universal set of concurrency nodes of a sending node are proved (**Theorem IV.1** and **Theorem IV.2**).

**Definition IV.1: The transmission coverage radius of a sending node**

As for a pair of paths $P_{[1,2]}$, let the transmission coverage radius corresponding to a sending node $n_{i,j}$ $(\in S_{P_{[1,2]}})$ be denoted as $r_{n_{i,j}}^{TC}$ $(r_{n_{i,j}}^{TC} \in \mathbb{R}, r_{n_{i,j}}^{TC} > 0)$ (unit: meter). The transmission coverage area of the sending node $n_{i,j}$ is a circular spatial region with the node as the center and a radius of $r_{n_{i,j}}^{TC}$. It is assumed that the spatial position of the successor node $n_{i,j}^{+}$ $(\in S_{P_{[1,2]}}^{U})$ of any sending node $n_{i,j}$ on the pair of paths $P_{[1,2]}$ lies within the transmission coverage radius $r_{n_{i,j}}^{TC}$ of the node $n_{i,j}$.

**Definition IV.2: A sending node forms a concurrency relationship with itself**

In a pair of paths $P_{[1,2]}$, as for any sending node $n_{i,j}$ $(\in S_{P_{[1,2]}})$, the following definitions are given:

1) Node $n_{i,j}$ forms a concurrency relationship with itself, which is denoted as $n_{i,j} || n_{i,j}$;
2) Node $n_{i,j}$ dose not form an interference relationship with itself.

**Definition IV.3: Concurrency relationship and interference relationship between two distinct sending nodes**

In a pair of paths $P_{[1,2]}$, as for any two distinct sending nodes $n_{i_1,j_1}, n_{i_2,j_2}$ $(n_{i_1,j_1} \neq n_{i_2,j_2})$, the following definitions are made:

1) These two nodes are either in a concurrency relationship (denoted as $n_{i_1,j_1} || n_{i_2,j_2}$) or in an interference relationship (denoted as $n_{i_1,j_1} >< n_{i_2,j_2}$);
2) Concurrency relationships are mutual. That is, if $n_{i_1,j_1} || n_{i_2,j_2}$ holds, then $n_{i_2,j_2} || n_{i_1,j_1}$ also

holds;

3) Interference relationships are also mutual. That is, if $n_{i_1,j_1} >< n_{i_2,j_2}$ holds, then $n_{i_2,j_2} >< n_{i_1,j_1}$ also holds.

**Definition IV.4: The universal set of concurrency nodes of a sending node**

The set composed of all sending nodes within the pair of paths $P_{[1,2]}$ that have concurrency relationships with a sending node $n_{i,j}$ ($\in S_{P_{[1,2]}}$) is defined as the universal set of concurrency nodes for node $n_{i,j}$ on the pair of paths $P_{[1,2]}$, which is denoted as $S_{n_{i,j}}^{CU}$ ($\subseteq S_{P_{[1,2]}}$).

**Note:**

1) Based on **Definition IV.2**, it holds that $n_{i,j} \in S_{n_{i,j}}^{CU}$ for any sending node $n_{i,j}$ ($\in S_{P_{[1,2]}}$).

**Definition IV.5: The universal set of interfering nodes for a sending node**

The set composed of all sending nodes within the pair of paths $P_{[1,2]}$ that have interference relationships with a sending node $n_{i,j}$ ($\in S_{P_{[1,2]}}$) is defined as the universal set of interfering nodes for node $n_{i,j}$ on the pair of paths $P_{[1,2]}$, which is denoted as $S_{n_{i,j}}^{IU}$ ($\subseteq S_{P_{[1,2]}}$).

**Note:**

1) Based on **Definition IV.2** and **Definition IV.3**, it holds that $n_{i,j} \notin S_{n_{i,j}}^{IU}$ for any sending node $n_{i,j}$ ($\in S_{P_{[1,2]}}$).

**Definition IV.6: Effective transmission of a sending node over a specific time interval**

As for a pair of paths $P_{[1,2]}$, given a time interval $[t_b, t_e]$ and a sending node $n_{i,j}$ ($\in S_{P_{[1,2]}}$), the transmission from the node $n_{i,j}$ (the sending node) to its successor node $n_{i,j}^{+}$ ($\in S_{P_{[1,2]}}^{U}$) (the receiving node) is regarded as effective over the time interval $[t_b, t_e]$ if for any time instant $t \in [t_b, t_e]$ either of the following two conditions is satisfied:

1) $S_{n_{i,j}}^{IU} = \emptyset$;
2) $S_{n_{i,j}}^{IU} \neq \emptyset$, and over the time interval $[t_b, t_e]$, all nodes within $S_{n_{i,j}}^{IU}$ are in their silent states.

**Note:**

1) In this paper, an effective transmission is regarded as a correct (i.e., successful) transmission;
2) This definition is reduced to describing the effectiveness of a transmission of a sending node at a specific time instant $t = t_b$ for the case of $t_b = t_e$;
3) If $S^{IU}_{n_{i,j}} = \emptyset$, the transmission from node $n_{i,j}$ to node $n^{+}_{i,j}$ is effective at any time instant;
4) From a physical perspective, if all nodes in $S^{IU}_{n_{i,j}}$ are in their silent states over a time interval $[t_b, t_e]$, it can be equivalently considered that $S^{IU}_{n_{i,j}} = \emptyset$ holds over this time interval;
5) Based on the above definition, we correspondingly formulate the definition of an ineffective transmission of a sending node over a specified time interval, which is stated as follows: As for a pair of paths $P_{[1,2]}$, given a time interval $[t_b, t_e]$ and a sending node $n_{i,j}$ ($\in S_{P_{[1,2]}}$), if $S^{IU}_{n_{i,j}} \neq \emptyset$ holds, and there exists at least one node $n'$ ($\in S^{IU}_{n_{i,j}}$) in $S^{IU}_{n_{i,j}}$ that is in a non-silent state (i.e., the sending state) at a certain time instant $t_n$ ($\in [t_b, t_e]$), then the transmission from node $n_{i,j}$ to its successor node $n^{+}_{i,j}$ is defined as ineffective over this time interval.

**Definition IV.7: Effective concurrent transmission of a non-empty subset of sending nodes over a specific time interval**

For a given time interval $[t_b, t_e]$ and a given non-empty subset of sending nodes $S'_{P_{[1,2]}}$ in the pair of paths $P_{[1,2]}$, the transmission, namely concurrent transmission, of the subset $S'_{P_{[1,2]}}$ is defined as effective if the transmissions of all sending nodes contained in $S'_{P_{[1,2]}}$ are effective within the time interval of $[t_b, t_e]$.

**Note:**

1) This definition is reduced to **Definition IV.6** when only one element is contained in $S'_{P_{[1,2]}}$.

**Rule IV.1: The interference decreasing rule among nodes on a primary path.**

**For any two distinct sending nodes $n_{i,j}$ and $n_{i,k}$ with concurrency relationship on a primary path $P_{[i]}$, namely $n_{i,j} || n_{i,k}$ $(n_{i,j}, n_{i,k} \in S_{P_{[i]}}, j < k)$, both of the following two statements hold:**

**1) If the successor node $n^{+}_{i,k}$ of node $n_{i,k}$ exists, namely $n^{+}_{i,k} \in S_{P_{[i]}}$, there exists a concurrency relationship between node $n_{i,j}$ and node $n^{+}_{i,k}$. That is, $n_{i,j} || n_{i,k} \Rightarrow$**

$\boldsymbol{n_{i,j}||n_{i,k}^{+}}$;

2) **If the predecessor node** $\boldsymbol{n_{i,j}^{-}}$ **of node** $\boldsymbol{n_{i,j}}$ **exists, namely** $\boldsymbol{n_{i,j}^{-} \in S_{P_{[i]}}}$**, there exists a concurrency relationship between node** $\boldsymbol{n_{i,k}}$ **and node** $\boldsymbol{n_{i,j}^{-}}$**. That is,** $\boldsymbol{n_{i,j}||n_{i,k} \Rightarrow n_{i,j}^{-}||n_{i,k}}$**.**

**Note:**

1) The above rule characterizes a common fact prevalent in most multi-hop information transmission paths: as the spacing (Refer to **Definition III.4**) between two sending nodes on a path increases, the potential mutual interference between them decreases accordingly;
2) In the following context, the above rule is deemed to hold unless otherwise specified.

**Lemma IV.1: Necessary and sufficient conditions for the existence of concurrency relationship between sending node(s).**

**Given a pair of paths** $\boldsymbol{P_{[1,2]}}$ **and any two sending node(s)** $\boldsymbol{n_{i,j}, n_{k,l}\ (\in S_{P_{[1,2]}})}$ **thereon, the necessary and sufficient condition that node** $\boldsymbol{n_{i,j}}$ **and node** $\boldsymbol{n_{k,l}}$ **hold a concurrency relationship is:** $\boldsymbol{n_{i,j} \notin S_{n_{k,l}}^{IU}}$ **and** $\boldsymbol{n_{k,l} \notin S_{n_{i,j}}^{IU}}$**.**

**Proof:**

First, we prove that if node $n_{i,j}$ and node $n_{k,l}$ have a concurrency relationship, then $n_{i,j} \notin S_{n_{k,l}}^{IU}$ and $n_{k,l} \notin S_{n_{i,j}}^{IU}$ hold. Since $n_{i,j}||n_{k,l}$ holds, it can be directly deduced from the definition of the universal set of interfering nodes for a sending node (**Definition IV.5**) that both $n_{i,j} \notin S_{n_{k,l}}^{IU}$ and $n_{k,l} \notin S_{n_{i,j}}^{IU}$ are satisfied.

Next, we prove that if $n_{i,j} \notin S_{n_{k,l}}^{IU}$ and $n_{k,l} \notin S_{n_{i,j}}^{IU}$ hold, then node $n_{i,j}$ and node $n_{k,l}$ have a concurrency relationship. Two cases are discussed separately.

**Case 1:** $n_{i,j} = n_{k,l}$

It can be concluded based on the definition of the universal set of interfering nodes of a sending node (**Definition IV.5**) and the definition stating that a sending node forms a concurrency relationship with itself (**Definition IV.2**) that both $n_{i,j} \notin S_{n_{i,j}}^{IU}$ and $n_{i,j}||n_{i,j}$ hold simultaneously.

**Case 2:** $n_{i,j} \neq n_{k,l}$

Proof by contradiction is adopted. If node $n_{i,j}$ and node $n_{k,l}$ do not possess a concurrency relationship, this means that node $n_{i,j}$ and node $n_{k,l}$ are in an interference relationship. According to the definition of the universal set of interfering nodes of a sending node (**Definition IV.5**), $n_{i,j} \in S_{n_{k,l}}^{IU}$ and $n_{k,l} \in S_{n_{i,j}}^{IU}$ hold. This contradicts the given premise. ■

**Theorem IV.1: The universal set of interfering nodes and the universal set of concurrency nodes of a sending node are disjoint.**

**Given a pair of paths $P_{[1,2]}$ and any sending node $n_{i,j}\ (\in S_{P_{[1,2]}})$ thereon. The universal set of interfering nodes and the universal set of concurrency nodes of $n_{i,j}$ are $S_{n_{i,j}}^{IU}$ and $S_{n_{i,j}}^{CU}$ respectively, then $S_{n_{i,j}}^{IU} \cap S_{n_{i,j}}^{CU} = \emptyset$.**

**Proof:**

Proof by contradiction is adopted. Assume $S_{n_{i,j}}^{IU} \cap S_{n_{i,j}}^{CU} \neq \emptyset$, then there must exist at least one element $n\ (\in (S_{n_{i,j}}^{IU} \cap S_{n_{i,j}}^{CU}))$. According to **Definition IV.5** and **Definition IV.4**, $S_{n_{i,j}}^{IU} \subseteq S_{P_{[1,2]}}$ and $S_{n_{i,j}}^{CU} \subseteq S_{P_{[1,2]}}$. Therefore, it can be deduced that $n \in S_{P_{[1,2]}}$. In the following, further proof is given by dividing into two cases.

**Case 1:** $(n \in S_{P_{[1,2]}}) \wedge (n = n_{i,j})$

From $n \in (S_{n_{i,j}}^{IU} \cap S_{n_{i,j}}^{CU})$ and $n = n_{i,j}$, $n_{i,j} \in S_{n_{i,j}}^{IU}$ can be obtained, which contradicts **Definition IV.5**.

**Case 2:** $(n \in S_{P_{[1,2]}}) \wedge (n \neq n_{i,j})$

From $n \in (S_{n_{i,j}}^{IU} \cap S_{n_{i,j}}^{CU})$ and $n \neq n_{i,j}$, it follows that $n >< n_{i,j}\ (n \neq n_{i,j})$ and $n || n_{i,j}\ (n \neq n_{i,j})$, which contradicts **Definition IV.3**. ■

**Theorem IV.2: The union of the universal set of interfering nodes and the universal set of concurrency nodes of a sending node is equal to the universal set of sending nodes of a pair of paths.**

**Given a pair of paths $P_{[1,2]}$ and any sending node $n_{i,j}\ (\in S_{P_{[1,2]}})$ thereon. The universal set of interfering nodes and the universal set of concurrency nodes of $n_{i,j}$ are $S_{n_{i,j}}^{IU}$ and $S_{n_{i,j}}^{CU}$ respectively, then $S_{n_{i,j}}^{IU} \cup S_{n_{i,j}}^{CU} = S_{P_{[1,2]}}$ holds.**

**Proof:**

First, it is proved that $(S_{n_{i,j}}^{IU} \cup S_{n_{i,j}}^{CU}) \subseteq S_{P_{[1,2]}}$ holds.

It follows from **Definition IV.5** and **Definition IV.4** that $S_{n_{i,j}}^{IU} \subseteq S_{P_{[1,2]}}$ and $S_{n_{i,j}}^{CU} \subseteq S_{P_{[1,2]}}$, hence $(S_{n_{i,j}}^{IU} \cup S_{n_{i,j}}^{CU}) \subseteq S_{P_{[1,2]}}$ holds.

Next, it is proved that $S_{P_{[1,2]}} \subseteq (S_{n_{i,j}}^{IU} \cup S_{n_{i,j}}^{CU})$ holds.

For any element $n$ in $S_{P_{[1,2]}}$, consider the following two cases.

**Case 1:** $(n \in S_{P_{[1,2]}}) \wedge (n = n_{i,j})$

It follows from **Definition IV.2** that $n||n_{i,j}$, which implies $n \in S_{n_{i,j}}^{CU}$. Therefore, it can be obtained that $n \in (S_{n_{i,j}}^{IU} \cup S_{n_{i,j}}^{CU})$ holds.

**Case 2:** $(n \in S_{P_{[1,2]}}) \wedge (n \neq n_{i,j})$

It follows from **Definition IV.3** that for two distinct sending nodes $n$ and $n_{i,j}$ in $S_{P_{[1,2]}}$, either $n||n_{i,j}$ holds or $n >< n_{i,j}$ holds. That is, for two distinct sending nodes $n$ and $n_{i,j}$ in $S_{P_{[1,2]}}$, correspondingly, either $n \in S_{n_{i,j}}^{CU}$ holds or $n \in S_{n_{i,j}}^{IU}$ holds. Therefore, it can be concluded that $n \in (S_{n_{i,j}}^{IU} \cup S_{n_{i,j}}^{CU})$ holds.

Based on the discussion of the above two cases, $S_{P_{[1,2]}} \subseteq (S_{n_{i,j}}^{IU} \cup S_{n_{i,j}}^{CU})$ is thus proved.

In summary, it follows that $S_{n_{i,j}}^{IU} \cup S_{n_{i,j}}^{CU} = S_{P_{[1,2]}}$ holds. ■

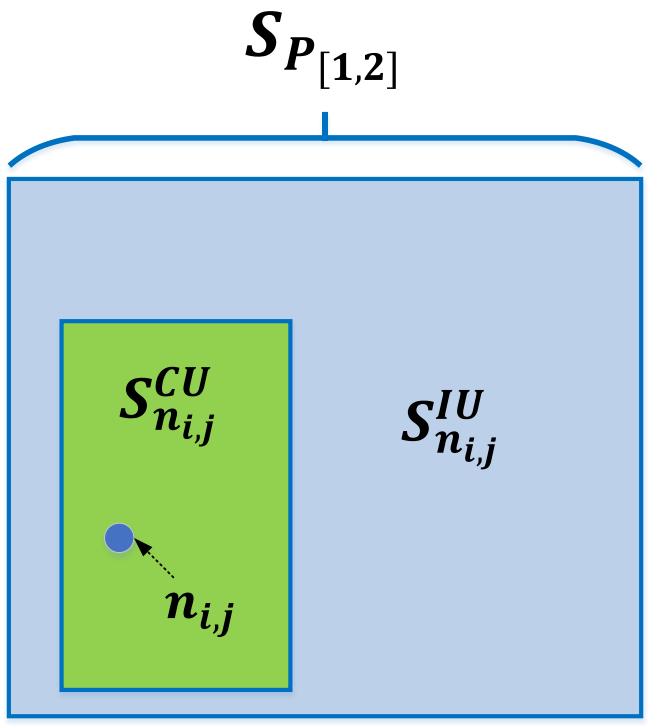

**Figure IV.1** The relationships among sets $S_{n_{i,j}}^{CU}$, $S_{n_{i,j}}^{IU}$ and $S_{P_{[1,2]}}$

$(S_{n_{i,j}}^{IU} \cap S_{n_{i,j}}^{CU} = \emptyset$ and $S_{n_{i,j}}^{IU} \cup S_{n_{i,j}}^{CU} = S_{P_{[1,2]}})$.

**Note:**

1) According to **Theorem IV.1** and **Theorem IV.2**, the inherent relationships among sets $S_{n_{i,j}}^{CU}$, $S_{n_{i,j}}^{IU}$ and $S_{P_{[1,2]}}$ can be illustrated in **Figure IV.1**.

## V. Concepts of the intrinsic concurrency intensity and the intrinsic interference intensity

In this section, the concepts of the intrinsic concurrency intensity for a subset of nodes, the intrinsic interference intensity for a subset of nodes, the intrinsic concurrency connection degree of a node and the intrinsic interference connection degree of a node are proposed. The second-order concurrency subset, the maximum second-order concurrency subset, the second-order interference subset, the maximum second-order interference subset, the subset without interference, the subset with interference, the continuous second-order interference subset, and the subset of nodes with dominant intrinsic interference intensity are defined, respectively. Moreover, the concepts such as the universal set of concurrency nodes, the universal set of interfering nodes, and the complement of a non-empty subset of sending nodes are also defined.

**Definition V.1: A second-order concurrency subset**

If there is a concurrency relationship between any two nodes in a nonempty subset $S'_{P_{[1,2]}}$ of sending nodes in a pair of paths $P_{[1,2]}$, i.e. $n_{i,j}||n_{k,l}$ $(n_{i,j}, n_{k,l} \in S'_{P_{[1,2]}})$, then $S'_{P_{[1,2]}}$ is defined as a second-order concurrency subset of $P_{[1,2]}$. In particular, a subset $S'_{P_{[1,2]}}$ containing only one sending node is also regarded as a second-order concurrency subset.

**Definition V.2: The maximum second-order concurrency subset of a nonempty subset of sending nodes**

In all of the second-order concurrency subsets contained in a nonempty subset $S'_{P_{[1,2]}}$ of sending nodes in a pair of paths $P_{[1,2]}$, the second-order concurrency subset with the largest number of elements is defined as the maximum second-order concurrency subset of $S'_{P_{[1,2]}}$.

**Note:**

1) There may be more than one such the maximum second-order concurrency subset.

**Definition V.3: The intrinsic concurrency intensity of a nonempty subset of sending nodes**

The number of elements contained in a maximum second-order concurrency subset of a nonempty subset $S'_{P_{[1,2]}}$ of sending nodes in a pair of paths $P_{[1,2]}$ is defined as the intrinsic concurrency intensity of the set, which is denoted as $C^*_{S'_{P_{[1,2]}}}$.

**Definition V.4: The intrinsic concurrency intensity of a primary path**

The intrinsic concurrency intensity of the set $S_{P_{[i]}}$ composed of all the sending nodes of a primary path $P_{[i]}$ is defined as the intrinsic concurrency intensity of the primary path $P_{[i]}$, which is denoted as $C^*_{S_{P_{[i]}}}$.

**Definition V.5: The intrinsic concurrency intensity of a pair of paths**

The intrinsic concurrency intensity of the set $S_{P_{[1,2]}}$ composed of all the sending nodes of a pair of paths $P_{[1,2]}$ is defined as the intrinsic concurrency intensity of the pair of paths $P_{[1,2]}$, which is denoted as $C^*_{S_{P_{[1,2]}}}$.

**Definition V.6: The concurrency connection degree of a node**

As for a sending node $n_{i,j}$ ($\in S_{P_{[1,2]}}$), the number of sending nodes (excluding the node $n_{i,j}$ itself) in a nonempty subset $S'_{P_{[1,2]}}$ of sending nodes in a pair of paths $P_{[1,2]}$ having concurrency relationships with it is defined as the concurrency connection degree of the node $n_{i,j}$ in $S'_{P_{[1,2]}}$, which is denoted as $D_{S'_{P_{[1,2]}}||n_{i,j}}$.

**Definition V.7: The intrinsic concurrency connection degree of nodes in a nonempty subset of sending nodes**

The maximum value among the concurrency connection degrees of all the sending nodes $n_{i,j}$ ($\in S'_{P_{[1,2]}}$) in a nonempty subset $S'_{P_{[1,2]}}$ of sending nodes in a pair of path $P_{[1,2]}$ is defined as the intrinsic concurrency connection degree of nodes in $S'_{P_{[1,2]}}$, which is denoted as $D^*_{S'_{P_{[1,2]}}||}$.

**Definition V.8: The intrinsic concurrency connection degree of nodes in a primary path**

The maximum value among the concurrency connection degrees of all the sending nodes $n_{i,j}$ ($\in S_{P_{[i]}}$) in the set $S_{P_{[i]}}$ of nodes in a primary path $P_{[i]}$ is defined as the intrinsic concurrency connection degree of nodes in $P_{[i]}$, which is denoted as $D^*_{S_{P_{[i]}}||}$.

**Definition V.9: The intrinsic concurrency connection degree of nodes in a pair of paths**

The maximum value among the concurrency connection degrees of all the sending nodes $n_{i,j}$ $(\in S_{P_{[1,2]}})$ in the set $S_{P_{[1,2]}}$ of nodes in a pair of paths $P_{[1,2]}$ is defined as the intrinsic concurrency connection degree of nodes in $P_{[1,2]}$, which is denoted as $D^*_{S_{P_{[1,2]}}||}$.

**Definition V.10: The second-order interference subset**

If any two different nodes in a nonempty subset $S'_{P_{[1,2]}}$ of sending nodes in a pair of paths $P_{[1,2]}$ have interference relationship, i.e. $n_{i,j} >< n_{k,l}$ $(n_{i,j}, n_{k,l} \in S'_{P_{[1,2]}}, n_{i,j} \neq n_{k,l})$ holds, then the subset $S'_{P_{[1,2]}}$ is defined as a second-order interference subset of $P_{[1,2]}$.

**Note:**

1) A second-order interference subset contains at least two elements.

**Definition V.11: The subset without interference and the subset with interference**

If a nonempty subset $S'_{P_{[1,2]}}$ of sending nodes in a pair of paths $P_{[1,2]}$ does not contain any second-order interference subset, it is defined as a subset without interference, otherwise it is defined as a subset with interference.

**Definition V.12: The maximum second-order interference subset of a subset with interference**

Among all the second-order interference subsets contained in a subset $S'_{P_{[1,2]}}$ with interference in a pair of paths $P_{[1,2]}$, the subset with the maximum number of elements is defined as a maximum second-order interference subset of $S'_{P_{[1,2]}}$.

**Note:**

1) There may be more than one such maximum second-order interference subset.

**Definition V.13: The intrinsic interference intensity of a nonempty subset of sending nodes**

If a nonempty subset $S'_{P_{[1,2]}}$ of sending nodes in a pair of paths $P_{[1,2]}$ is a subset with interference, the number of elements contained in its maximum second-order interference subset is defined as its intrinsic interference intensity, which is denoted as $I^*_{S'_{P_{[1,2]}}}$. Moreover, if $S'_{P_{[1,2]}}$ is a subset without interference, its intrinsic interference intensity is defined as $I^*_{S'_{P_{[1,2]}}} \triangleq 1$.

**Definition V.14: The intrinsic interference intensity of a primary path**

The intrinsic interference intensity of the set $S_{P_{[i]}}$ composed of all the sending nodes of a primary path $P_{[i]}$ is defined as the intrinsic interference intensity of $P_{[i]}$, which is denoted as $I^*_{S_{P_{[i]}}}$.

**Definition V.15: The intrinsic interference intensity of a pair of paths**

The intrinsic interference intensity of the set $S_{P_{[1,2]}}$ composed of all the sending nodes of a pair of paths $P_{[1,2]}$ is defined as the intrinsic interference intensity of $P_{[1,2]}$, which is denoted as $I^*_{S_{P_{[1,2]}}}$.

**Definition V.16: The interference connection degree of a node**

As for a sending node $n_{i,j}$ ($\in S_{P_{[1,2]}}$), the number of sending nodes in a nonempty subset $S'_{P_{[1,2]}}$ of sending nodes in a pair of paths $P_{[1,2]}$ having interference relationships with it is defined as the interference connection degree of the node $n_{i,j}$ in $S'_{P_{[1,2]}}$, which is denoted as $D_{S'_{P_{[1,2]}}><n_{i,j}}$.

**Definition V.17: The intrinsic interference connection degree of nodes in a nonempty subset of sending nodes**

The maximum value among the interference connection degrees of all the sending nodes $n_{i,j}$ ($\in S'_{P_{[1,2]}}$) in a nonempty subset $S'_{P_{[1,2]}}$ of sending nodes in a pair of paths $P_{[1,2]}$ is defined as the intrinsic interference connection degree of nodes in $S'_{P_{[1,2]}}$, which is denoted as $D^*_{S'_{P_{[1,2]}}><}$.

**Definition V.18: The intrinsic interference connection degree of nodes in a primary path**

The maximum value among the interference connection degrees of all the sending nodes $n_{i,j}$ ($\in S_{P_{[i]}}$) in the set $S_{P_{[i]}}$ of nodes in a primary path $P_{[i]}$ is defined as the intrinsic interference connection degree of nodes in $P_{[i]}$, which is denoted as $D^{*}_{S_{P_{[i]}}}{><}$.

**Definition V.19: The intrinsic interference connection degree of nodes in a pair of paths**

The maximum value among the interference connection degrees of all the sending nodes $n_{i,j}$ ($\in S_{P_{[1,2]}}$) in the set $S_{P_{[1,2]}}$ of nodes in a pair of paths $P_{[1,2]}$ is defined as the intrinsic interference connection degree of nodes in $P_{[1,2]}$, which is denoted as $D^{*}_{S_{P_{[1,2]}}}{><}$.

**Definition V.20: A nonempty subset of sending nodes with dominant intrinsic interference intensity**

In a nonempty subset $S'_{P_{[1,2]}}$ of sending nodes, if its intrinsic interference connection degree $D^{*}_{S'_{P_{[1,2]}}}{><}$ is less than its intrinsic interference intensity $I^{*}_{S'_{P_{[1,2]}}}$, i.e. $D^{*}_{S'_{P_{[1,2]}}}{><} < I^{*}_{S'_{P_{[1,2]}}}$, the subset $S'_{P_{[1,2]}}$ is defined as a subset of sending nodes with dominant intrinsic interference intensity.

**Definition V.21: A primary path with dominant intrinsic interference intensity**

As for a primary path $P_{[i]}$, if its intrinsic interference connection degree $D^{*}_{S_{P_{[i]}}}{><}$ is less than its intrinsic interference intensity $I^{*}_{S_{P_{[i]}}}$, i.e. $D^{*}_{S_{P_{[i]}}}{><} < I^{*}_{S_{P_{[i]}}}$, the primary path $P_{[i]}$ is defined as a primary path with dominant intrinsic interference intensity.

**Definition V.22: A pair of paths with dominant intrinsic interference intensity**

As for a pair of paths $P_{[1,2]}$, if its intrinsic interference connection degree $D^{*}_{S_{P_{[1,2]}}}{><}$ is less than its intrinsic interference intensity $I^{*}_{S_{P_{[1,2]}}}$, i.e. $D^{*}_{S_{P_{[1,2]}}}{><} < I^{*}_{S_{P_{[1,2]}}}$, the pair of paths $P_{[1,2]}$ is defined as a pair of paths with dominant intrinsic interference intensity.

**Definition V.23: The continuous second-order interference subset in a primary path**

Among all the sending nodes in a second-order interference subset $S'_{P_{[i]}}$ of a primary path $P_{[i]}$, let node $n_{i,j}$ ($\in S'_{P_{[i]}}$) have the smallest sequence number (the sequence number of node $n_{i,j}$ is $j$), and node $n_{i,k}$ ($\in S'_{P_{[i]}}, k \neq j$) have the largest sequence number (the sequence number of node $n_{i,k}$ is $k$). If $S'_{P_{[i]}}$ includes all the nodes in $P_{[i]}$ with their sequence numbers between $j$ and $k$, then $S'_{P_{[i]}}$ is defined as a continuous second-order interference subset in $P_{[i]}$.

**Definition V.24: The universal set of concurrency nodes for a subset of sending nodes on a pair of paths**

The union of the universal sets of concurrency nodes of all elements in a non-empty subset of sending nodes $S'\,(\subseteq S_{P_{[1,2]}})$ within the pair of paths $P_{[1,2]}$ is denoted as $S_{S'}^{CU} \triangleq \bigcup_{m\in S'} S_m^{CU}$.

**Definition V.25: The universal set of interfering nodes for a subset of sending nodes on a pair of paths**

The union of the universal sets of interfering nodes of all elements in a non-empty subset of sending nodes $S'\,(\subseteq S_{P_{[1,2]}})$ within the pair of paths $P_{[1,2]}$ is denoted as $S_{S'}^{IU} \triangleq \bigcup_{m\in S'} S_m^{IU}$.

**Definition V.26: The complement of a subset of sending nodes on a pair of paths**

The complement of a non-empty subset of sending nodes $S'\,(\subseteq S_{P_{[1,2]}})$ in the universal set $S_{P_{[1,2]}}$ of the pair of paths $P_{[1,2]}$ is denoted as $S_{P_{[1,2]}}^{\backslash S'} \triangleq S_{P_{[1,2]}} \backslash S'$.

**Definition V.27: Saturated second-order concurrency subset of nodes**

If a second-order concurrency subset of nodes $S'_C$ and its universal set of interfering nodes $S_{S'_C}^{IU}$ in the pair of paths $P_{[1,2]}$ satisfy $S'_C \cup S_{S'_C}^{IU} = S_{P_{[1,2]}}$, then $S'_C$ is referred to as a saturated second-order concurrency subset of nodes in the pair of paths $P_{[1,2]}$.

**Definition V.28: Saturated second-order interference subset of nodes**

If the universal set of concurrency nodes $S_{S'_I}^{CU}$ of a second-order interference subset $S'_I$ in the pair of paths $P_{[1,2]}$ satisfies $S_{S'_I}^{CU} = S_{P_{[1,2]}}$, then $S'_I$ is referred to as a saturated second-order interference subset of nodes in the pair of paths $P_{[1,2]}$.

**Definition V.29: The universal set of (saturated) second-order concurrency subsets of a set of sending nodes**

The set formed by all second-order concurrency subsets of a subset of sending nodes $S'$ in a pair of paths $P_{[1,2]}$ is referred to as the universal set of second-order concurrency subsets of a subset of sending nodes $S'$, denoted by $S_{S'}^{SCU}$.

The set formed by all saturated second-order concurrency subsets of a subset of sending nodes $S'$ in a pair of paths $P_{[1,2]}$ is referred to as the universal set of saturated second-order concurrency subsets of a subset of sending nodes $S'$, denoted by $S_{S'}^{SSCU}$.

**Definition V.30: The universal set of (saturated) second-order interference subsets of a set of sending nodes**

The set formed by all second-order interference subsets of a subset of sending nodes $S'$ in a pair of paths $P_{[1,2]}$ is referred to as the universal set of second-order interference subsets of a subset of sending nodes $S'$, denoted by $S_{S'}^{SIU}$.

The set formed by all saturated second-order interference subsets of a subset of sending nodes $S'$ in a pair of paths $P_{[1,2]}$ is referred to as the universal set of saturated second-order interference subsets of a subset of sending nodes $S'$, denoted by $S_{S'}^{SSIU}$.

**Definition V.31: The number of nodes belonging to a certain primary path within a set of nodes**

The number of nodes in a subset of nodes $S'$ of a pair of paths $P_{[1,2]}$ that belong to a primary path $P_{[i]}$ $(i = 1,2)$ is denoted by $N_{S',i}$.

# VI. Some related properties of the intrinsic concurrency intensity and the intrinsic interference intensity

In this section, the necessary and sufficient conditions for a nonempty subset of nodes to be a second-order concurrency subset are given (**Theorem VI.1**). Both the continuity of a second-order interference subset in a primary path (**Theorem VI.2**) and the continuity of a maximum second-order interference subset (**Corollary VI.2.1**) are proved. The splitting characteristics of a subset of nodes with dominant intrinsic interference intensity are proved (**Theorem VI.3**). The relationship between the intrinsic interference intensity and the intrinsic concurrency intensity of a subset of nodes with dominant intrinsic interference intensity is given (**Corollary VI.3.1**). The concurrent transmission theorem (***Theorem VI.4***) is proved. Moreover, the necessary and sufficient conditions for a second-order concurrency subset of nodes to be a saturated second-order concurrency subset of nodes (**Theorem VI.5**) and for a second-order interference subset of nodes to be a saturated second-order interference subset of nodes (**Theorem VI.6**) are proved.

## *A.* Fundamental properties of second-order concurrency subsets and second-order interference subsets of nodes

**Lemma VI.1.1: The subset without interference forms a second-order concurrency subset.**

**If a nonempty subset $S'_{P_{[1,2]}}$ ($\subseteq S_{P_{[1,2]}}$) of sending nodes is a subset without interference, all the sending nodes in it form a second-order concurrency subset.**

**Proof:**

**Case 1:** Number of sending nodes in $S'_{P_{[1,2]}} \geq 2$

Here, reduction to absurdity is adopted. Assuming that all the sending nodes in $S'_{P_{[1,2]}}$ cannot form a second-order concurrency subset, it is evident that at least two of these sending nodes must have interference relationship with each other. Therefore, these two nodes form a second-order interference subset (**Definition V.10**), which is contrary to the proposition.

**Case 2:** Number of sending nodes in $S'_{P_{[1,2]}} = 1$

According to the definition of a second-order concurrency subset (**Definition V.1**), a subset containing only one sending node forms a second-order concurrency subset. ■

**Lemma VI.1.2: A second-order concurrency subset is a subset without interference.**

**If a nonempty subset $S'_{P_{[1,2]}}$ $(\subseteq S_{P_{[1,2]}})$ of sending nodes is a second-order concurrency subset, it is also a subset without interference.**

**Proof:**

**Case 1:** Number of sending nodes in $S'_{P_{[1,2]}} = 1$

Since $S'_{P_{[1,2]}}$ contains only one element, from the definition of the second-order interference subset (**Definition V.10**), it is known that neither $S'_{P_{[1,2]}}$ nor any of its non-empty subsets can be a second-order interference subset. According to the definition of the subset without interference (**Definition V.11**), the set $S'_{P_{[1,2]}}$ is a subset without interference.

**Case 2:** Number of sending nodes in $S'_{P_{[1,2]}} \geq 2$

Proof by contradiction is adopted. Assume that $S'_{P_{[1,2]}}$ is not a subset without interference, then it contains at least one second-order interference subset. It follows from the definition of the second-order interference subset (**Definition V.10**) that $S'_{P_{[1,2]}}$ contains at least two distinct sending nodes $n\ (\in S'_{P_{[1,2]}})$ and $m\ (\in S'_{P_{[1,2]}})$, and $n >< m$ holds. This contradicts the hypothesis, i.e., $S'_{P_{[1,2]}}$ is a second-order concurrency subset. ■

**Theorem VI.1: The necessary and sufficient condition for a nonempty subset of sending nodes to be a second-order concurrency subset.**

**The necessary and sufficient condition for a nonempty subset $S'_{P_{[1,2]}}$ $(\subseteq S_{P_{[1,2]}})$ of sending nodes to be a second-order concurrency subset is that the subset is a subset without interference.**

**Proof:**

It can be proved based on **Lemma VI.1.1** and **Lemma VI.1.2**.■

**Theorem VI.2: The continuity of a second-order interference subset in a primary path.**

**Given a primary path $P_{[i]}$. It is assumed that two different sending nodes $n_{i,j}$ and**

**$n_{i,k}$ in $P_{[i]}$ do not have concurrency relationship (i.e. $n_{i,j} >< n_{i,k}$ ($n_{i,j}, n_{i,k} \in S_{P_{[i]}}, j < k$)). Then the node set $S'_{P_{[i]}} \triangleq \left\{n_{i,l} | n_{i,l} \in S_{P_{[i]}}, j \le l \le k\right\}$ composed of all the sending nodes (including $j$ and $k$) with sequence numbers between $j$ and $k$ in $P_{[i]}$ is a continuous second-order interference subset.**

**Proof:**

Since the node subset $S'_{P_{[i]}}$ contains all nodes whose sequence numbers are between $j$ and $k$ (including $j$ and $k$), it is only necessary to prove that $S'_{P_{[i]}}$ is a second-order interference subset in the following (**Definition V.23**).

Reduction to absurdity is adopted. Assuming that node subset $S'_{P_{[i]}} = \left\{n_{i,l} | n_{i,l} \in S_{P_{[i]}}, j \le l \le k\right\}$ is not a second-order interference subset, therefore, there are at least two nodes $n_{i,l}$ and $n_{i,m}$ ($j \le l < m \le k$) that can be concurrent, i.e. $n_{i,l} || n_{i,m}$ ($j \le l < m \le k$). According to **Rule IV.1**, we know that $n_{i,l} || n_{i,k}$ holds through several iterations. Since $n_{i,l} || n_{i,k}$, according to **Rule IV.1**, $n_{i,j} || n_{i,k}$ can be obtained through several iterations, which is contradictory to the premise of $n_{i,j} >< n_{i,k}$ in the proposition. Therefore, $S'_{P_{[i]}}$ must be a second-order interference subset. Furthermore, combined with the definition of the continuous second-order interference subset in a primary path (**Definition V.23**), it can be seen that $S'_{P_{[i]}}$ is a continuous second-order interference subset in $P_{[i]}$. ■

**Corollary VI.2.1: The continuity of a maximum second-order interference subset in a primary path.**

**Any maximum second-order interference subset in a primary path $P_{[i]}$ containing at least one second-order interference subset is a continuous second-order interference subset.**

**Proof:**

Reduction to absurdity is adopted. Consider a subset $S'_{P_{[i]}}$ ($\subseteq S_{P_{[i]}}$) is a maximum second-order interference subset in a primary path $P_{[i]}$, and the number of elements of $S'_{P_{[i]}}$ is $I^*_{S'_{P_{[i]}}}$. Without losing generality, assuming that node $n_{i,j}$ has the smallest sequence number (i.e. $j$) in $S'_{P_{[i]}}$ and node $n_{i,k}$ has the largest sequence number (i.e. $k$) in $S'_{P_{[i]}}$. If $S'_{P_{[i]}}$ is not continuous, there must exist at least one node $n_{i,l}$ ($\in S_{P_{[i]}}$) in $P_{[i]}$ whose sequence number $l$ is between $j$ and $k$, and the node $n_{i,l}$ does not belong to $S'_{P_{[i]}}$. Therefore, it can be deduced from **Theorem VI.2** that all nodes with sequence numbers between $j$ and $k$ (including $j$ and $k$) can form a continuous second-order interference subset $S''$, which contains node $n_{i,l}$ ($\notin S'_{P_{[i]}}$) and all elements of $S'_{P_{[i]}}$. It is evident that the number of elements in the second-order interference subset $S''$ must be greater than $I^*_{S'_{P_{[i]}}}$, which contradicts the premise that $S'_{P_{[i]}}$ is a maximum second-order interference subset. ■

**Theorem VI.3: The splitting characteristics of a subset of sending nodes with dominant intrinsic interference intensity.**

**If the intrinsic interference intensity of a subset $S'_{P_{[1,2]}}$ of sending nodes with dominant intrinsic interference intensity is $I^*_{S'_{P_{[1,2]}}}$, all the nodes contained in it can be divided into $I^*_{S'_{P_{[1,2]}}}$ mutually disjoint second-order concurrency subsets of nodes.**

**Proof:**

Reduction to absurdity is adopted. Because the intrinsic interference intensity of $S'_{P_{[1,2]}}$ is $I^*_{S'_{P_{[1,2]}}}$, at least $I^*_{S'_{P_{[1,2]}}}$ sending nodes that can jointly form a second-order interference subset can be found in $S'_{P_{[1,2]}}$. Starting from these $I^*_{S'_{P_{[1,2]}}}$ sending nodes, we can construct $I^*_{S'_{P_{[1,2]}}}$ subsets of sending nodes which contain only one sending node (from one of these $I^*_{S'_{P_{[1,2]}}}$ sending nodes) each and are mutually disjoint to each other. On the premise of keeping these subsets of sending nodes as second-order concurrency subsets, by adding the remaining sending nodes in $S'_{P_{[1,2]}}$ into these subsets one by one, $I^*_{S'_{P_{[1,2]}}}$ mutually disjoint second-order concurrency subsets of nodes can be finally formed. Assuming that at least one sending node $n$ in $S'_{P_{[1,2]}}$ cannot be added into any of these $I^*_{S'_{P_{[1,2]}}}$ second-order concurrency subsets, which indicates that in each of these $I^*_{S'_{P_{[1,2]}}}$ second-order concurrency subsets at least one sending node, which has interference relationship with the node $n$, can be found. Therefore, it can be seen that the interference connection degree of node $n$ in $S'_{P_{[1,2]}}$ must not be less than $I^*_{S'_{P_{[1,2]}}}$, which is contrary to the assumption that the set $S'_{P_{[1,2]}}$ is dominated by intrinsic interference intensity. Therefore, such a node $n$ does not exist at all. That is to say, all nodes contained in $S'_{P_{[1,2]}}$ can be divided into $I^*_{S'_{P_{[1,2]}}}$ mutually disjoint second-order concurrency subsets of nodes. ■

**Corollary VI.3.1: The relationship between the intrinsic interference intensity and the intrinsic concurrency intensity of a set of sending nodes with dominant intrinsic interference intensity.**

**Let the intrinsic interference intensity and intrinsic concurrency intensity of a set $S'_{P_{[1,2]}}$ of sending nodes with dominant intrinsic interference intensity be $I^*_{S'_{P_{[1,2]}}}$ and $C^*_{S'_{P_{[1,2]}}}$, respectively. If the number of elements in the set $S'_{P_{[1,2]}}$ is $\left|S'_{P_{[1,2]}}\right|$, there is $I^*_{S'_{P_{[1,2]}}} \cdot C^*_{S'_{P_{[1,2]}}} \geq \left|S'_{P_{[1,2]}}\right|$.**

**Proof:**

According to **Theorem VI.3**, $S'_{P_{[1,2]}}$ can be divided into $I^*_{S'_{P_{[1,2]}}}$ nonempty and mutually disjoint second-order concurrency subsets, and the number of elements in each subset will not exceed the intrinsic concurrency intensity $C^*_{S'_{P_{[1,2]}}}$ (according to the definition of intrinsic concurrency intensity of nonempty subsets of nodes (**Definition V.3**)), so $I^*_{S'_{P_{[1,2]}}} \cdot C^*_{S'_{P_{[1,2]}}} \geq \left|S'_{P_{[1,2]}}\right|$ is valid. ■

**Note:**

1) **Theorem VI.3** and **Corollary VI.3.1** do not depend on whether **Rule IV.1** is satisfied or not.

**Corollary VI.3.2: Any two distinct maximum second-order interference subsets contained in a set of sending nodes with dominant intrinsic interference intensity do not intersect to each other.**

**If there are multiple distinct maximum second-order interference subsets in a set $S'_{P_{[1,2]}}$ of sending nodes with dominant intrinsic interference intensity, these maximum second-order interference subsets do not intersect to each other.**

**Proof:**

Assuming that the intrinsic interference intensity of $S'_{P_{[1,2]}}$ is $I^*_{S'_{P_{[1,2]}}}$. Let node subsets $S'_A$ ($\subseteq S'_{P_{[1,2]}}$) and $S'_B$ ($\subseteq S'_{P_{[1,2]}}$) are maximum second-order interference subsets of $S'_{P_{[1,2]}}$, and $S'_A \neq S'_B$. Therefore, at least there exist one node $j$ ($\in S'_{P_{[1,2]}}$) which satisfies $((j \notin S'_A) \wedge (j \in S'_B)) \vee ((j \notin S'_B) \wedge (j \in S'_A))$.

Reduction to absurdity is adopted. Assuming that $S'_A \cap S'_B \neq \emptyset$, there is at least one node $k$ $((\in S'_A) \wedge (\in S'_B))$. Since $S'_A \neq S'_B$, without loss of generality, consider the case where at least one node $j$ satisfies $(j \notin S'_A) \wedge (j \in S'_B)$. According to the definition of the maximum second-order interference subset (**Definition V.12**), there is an interference relationship between $k$ ($\in S'_A$) and the other $I^*_{S'_{P_{[1,2]}}} - 1$ nodes in $S'_A$, and the node $k$ ($\in S'_B$) and the node $j$ ($\notin S'_A, \in S'_B$) also interfere with each other. Therefore, it can be deduced that the interference connection degree of node $k$ in $S'_{P_{[1,2]}}$ is not less than $I^*_{S'_{P_{[1,2]}}}$, which contradicts with the premise that $S'_{P_{[1,2]}}$ is dominated by its intrinsic interference intensity, so $S'_A \cap S'_B = \emptyset$.■

### *B*. Concurrent transmission theorem and its related properties

**Lemma VI.4.1: A second-order concurrency subset is disjoint from its universal set of interfering nodes.**

**Let $S'_C$ be an arbitrary second-order concurrency subset on a pair of paths $P_{[1,2]}$, and its universal set of interfering nodes is $S_{S'_C}^{IU}$. Then $S'_C \cap S_{S'_C}^{IU} = \emptyset$ holds.**

**Proof:**

Proof by contradiction is adopted. Assume that there exists at least one sending node $n$ that belongs to both set $S'_C$ and set $S_{S'_C}^{IU}$. Accordingly, there exists a sending node $n' (\in S'_C)$ such that $n \in S_{n'}^{IU}$.

**Case 1:** $n \neq n'$

In accordance with **Definition IV.5**, it follows that $n >< n'$ holds, which contradicts the hypothesis (i.e., $n||n' (n, n' \in S'_C)$ holds).

**Case 2:** $n = n'$

It can be further derived from relation $n \in S_{n'}^{IU}$ that $n \in S_{n}^{IU}$, which contradicts **Definition IV.2**. ■

**Lemma VI.4.2: The universal set of interfering nodes of a second-order concurrency subset is a subset of the complement of this second-order concurrency subset.**

**Let $S'_C$ be an arbitrary second-order concurrency subset on the pair of paths $P_{[1,2]}$. And its universal set of interfering nodes and its complement are $S_{S'_C}^{IU}$ and $S_{P_{[1,2]}}^{\backslash S'_C}$, respectively. Then $S_{S'_C}^{IU} \subseteq S_{P_{[1,2]}}^{\backslash S'_C}$ holds.**

**Proof:**

**Case 1:** $S_{S'_C}^{IU} = \emptyset$

Since the empty set is a subset of any arbitrary set, it follows that $S_{S'_C}^{IU} \subseteq S_{P_{[1,2]}}^{\backslash S'_C}$ holds.

**Case 2:** $S_{S'_C}^{IU} \neq \emptyset$

From **Lemma VI.4.1**, it follows that $S'_C \cap S_{S'_C}^{IU} = \emptyset$. Therefore, for any element $n\ (\in S_{S'_C}^{IU})$ in $S_{S'_C}^{IU}$, both $n \in S_{P_{[1,2]}}$ and $n \notin S'_C$ hold. Hence, it can be concluded that $n \in S_{P_{[1,2]}}^{\backslash S'_C}$, that is, $S_{S'_C}^{IU} \subseteq S_{P_{[1,2]}}^{\backslash S'_C}$. ■

**Lemma VI.4.3: Effective concurrent transmission cannot be achieved by a subset of nodes with interference.**

**Let $S'_{P_{[1,2]}}$ be any non-empty subset of sending nodes with interference in the pair of paths $P_{[1,2]}$. Not all transmissions of nodes contained in $S'_{P_{[1,2]}}$ are effective within any given time interval $[t_b, t_e]$ (i.e. the concurrent transmission of $S'_{P_{[1,2]}}$ is not effective).**

**Proof:**

According to the definition of subset with interference (**Definition V.11**), at least one second order interference subset of nodes $S'$ is contained in $S'_{P_{[1,2]}}$. It is known from the definition of second order interference subset of nodes (**Definition V.10**) that at least two sending nodes with interference relation are included in $S'\ (\subseteq S'_{P_{[1,2]}})$. Accordingly, two sending nodes with interference relation are contained in $S'_{P_{[1,2]}}$ at least. Without loss of generality, it is assumed that an interference relation exists between node $n_1$ and node $n_2$ within $S'_{P_{[1,2]}}$. As stated in the definition of interference relation between two nodes (**Definition IV.3**) and **Lemma IV.1**, for node $n_1$ and node $n_2$, at least one of these two nodes belongs to the universal set of interfering nodes of the other node. Without loss of generality, let $n_1 \in S_{n_2}^{IU}$. Furthermore, according to the definition of effective transmission of a sending node within a specific time interval $[t_b, t_e]$ (**Definition IV.6**), the transmission of node $n_2$ is confirmed to be ineffective while node $n_1 \in S_{n_2}^{IU}$ is kept in its sending state. Finally, in accordance with the definition of effective concurrent transmission of a non-empty subset of sending nodes within a specific time interval (**Definition IV.7**), the concurrent transmission of $S'_{P_{[1,2]}}$ is proven ineffective over the time interval $[t_b, t_e]$. ■

**Theorem VI.4: Concurrent transmission theorem.**

**Let $S'_C$ be an arbitrary second-order concurrency subset on the pair of paths $P_{[1,2]}$, and its universal set of interfering nodes and complement are $S_{S'_C}^{IU}$ and $S_{P_{[1,2]}}^{\backslash S'_C}$, respectively.**

**Define a set of sending nodes $S$ as $S_{S'_C}^{IU} \subseteq S \subseteq S_{P_{[1,2]}}^{\setminus S'_C}$. Within any given time interval $[t_b, t_e]$, if all nodes contained in $S'_C$ are in the sending state, all transmissions generated by the nodes contained in $S'_C$ during this time interval are effective under either of the following two cases holding:**

**Case 1: $S = \emptyset$;**

**Case 2: $S \neq \emptyset$, and all nodes in $S$ are in silent state.**

**Proof:**

**Case 1:**

First, it follows from the definition of $S$ and **Lemma VI.4.2** that set $S'_C$ and set $S$ are disjoint to each other. Therefore, it is possible in this case to let $S = \emptyset$ and require all nodes in set $S'_C$ to be in the sending state.

Next, proof by contradiction is adopted. Assume that there exists at least one sending node $n\ (\in S'_C)$ whose corresponding transmission is ineffective within $[t_b, t_e]$. According to **Definition IV.6**, it follows that $S_n^{IU} \neq \emptyset$. Since $S_n^{IU} \subseteq S_{S'_C}^{IU} \subseteq S$, it can be concluded that $S \neq \emptyset$, which contradicts the hypothesis under this case (i.e., $S = \emptyset$).

**Case 2:**

First, because set $S'_C$ and set $S$ are disjoint, therefore, it is possible in this case to assume that all nodes in set $S$ remain in the silent state and all nodes in set $S'_C$ remain in the sending state.

Next, proof by contradiction is employed. Assume that there exists at least one sending node $n\ (\in S'_C)$ whose corresponding transmission is ineffective within$[t_b, t_e]$. According to **Definition IV.6**, it follows that $S_n^{IU} \neq \emptyset$, and there exists at least one node $n'$ within $S_n^{IU}$ that is in the non-silent state (i.e., in the sending state) at some time instant $t_n\ (\in [t_b, t_e])$. Since $S_n^{IU} \subseteq S_{S'_C}^{IU} \subseteq S$, we have $n' \in S$. That is to say, at least at some time instant $t_n\ (\in [t_b, t_e])$, there exists a node $n'$ belonging to $S$ in the sending state, which contradicts the hypothesis for this case. ■

**Note:**

1) The physical implication of the concurrent transmission theorem is clear. That is, for a given second-order concurrency set of nodes, as long as the remaining sending nodes on the pair of paths (which shall at least include all nodes in the universal set of interfering nodes of this second-order concurrency set of nodes) are guaranteed to be in their silent states, the concurrent transmission of this second-order concurrency set of nodes is effective. Furthermore, in an extreme case, for a given second-order concurrency set of nodes, effectiveness of the concurrent transmission of the set can be guaranteed provided that all sending nodes in the pair of paths excluding those contained in this set are kept in their silent states. This scenario will be frequently adopted in subsequent discussions;

2) Combined further with **Lemma VI.4.3**, it can be seen that effective concurrent transmission cannot be realized for any subset of nodes with interference. On the contrary, effective concurrent transmission can be achieved for a subset without interference under some specified conditions, which have already been presented in the concurrent transmission theorem;

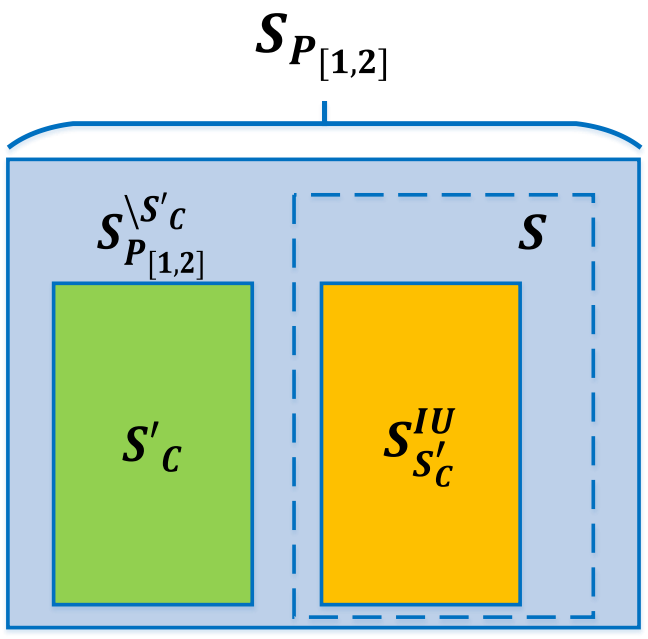

**Figure VI.1** The relationships among sets $S'_C$ (a second-order concurrency subset), $S_{S'_C}^{IU}$, $S_{P_{[1,2]}}^{\backslash S'_C}$, $S$ and $S_{P_{[1,2]}}$ ($S'_C \cap S_{S'_C}^{IU} = \emptyset$, $S_{S'_C}^{IU} \subseteq S_{P_{[1,2]}}^{\backslash S'_C}$ and $S_{S'_C}^{IU} \subseteq S \subseteq S_{P_{[1,2]}}^{\backslash S'_C}$)

3) According to **Lemma VI.4.1**, **Lemma VI.4.2**, and **Theorem VI.4**, the relationships among sets $S'_C$ (a second-order concurrency subset), $S_{S'_C}^{IU}$, $S_{P_{[1,2]}}^{\backslash S'_C}$, $S$ and $S_{P_{[1,2]}}$ are summarized in **Figure VI.1**;

4) This theorem does not depend on whether **Rule IV.1** is satisfied or not;

5) This theorem serves as the theoretical foundation for **Algorithm X.1**, **Algorithm X.2**, **Algorithm X.3** and other related algorithms.

**Theorem VI.5: A second-order concurrency subset of nodes is a saturated second-order**

**concurrency subset of nodes if and only if it cannot be further extended into a larger second-order concurrency subset of nodes.**

**Let $S'_C$ be a second-order concurrency subset of nodes in the pair of paths $P_{[1,2]}$. It is a saturated second-order concurrency subset of nodes if and only if there does not exist a second-order concurrency subset of nodes that contains set $S'_C$ and is not equal to $S'_C$ within the set of sending nodes $S_{P_{[1,2]}}$ of the pair of paths $P_{[1,2]}$.**

**Proof:**

We first prove that if $S'_C$ is a saturated second-order concurrency subset of nodes, then there does not exist a second-order concurrency subset of nodes that contains set $S'_C$ and is not equal to $S'_C$. Proof by contradiction is adopted. Assume that at least one second-order concurrency subset of nodes $S'$ exists which contains set $S'_C$ and is not equal to $S'_C$, i.e., $S'_C \subset S' \, (S' \subseteq S_{P_{[1,2]}})$. According to the definition of a saturated second-order concurrency subset of nodes (**Definition V.27**), at least one element $n$ in set $S^{IU}_{S'_C}$ $(S^{IU}_{S'_C} = S_{P_{[1,2]}} \backslash S'_C)$ also belongs to the set $S'$, namely $n \in S^{IU}_{S'_C}$ and $n \in S'$. Since $n \in S'$, and $S' \, (S'_C \subset S')$ forms a second-order concurrency subset of nodes, it can be obtained that $n$ has a concurrency relationship with any element in $S'_C$, which contradicts $n \in S^{IU}_{S'_C}$.

We then prove that if there does not exist a second-order concurrency subset of nodes that contains set $S'_C$ and is not equal to $S'_C$, then $S'_C$ is a saturated second-order concurrency subset of nodes. Proof by contradiction is adopted. Assume that $S'_C$ is not a saturated second-order concurrency subset of nodes, then $S'_C \cup S^{IU}_{S'_C} \neq S_{P_{[1,2]}}$ holds (according to **Definition V.27**). From **Lemma VI.4.1**, we have $S'_C \cap S^{IU}_{S'_C} = \emptyset$. Therefore, at least one element $n$ in set $S_{P_{[1,2]}}$ satisfies relations: $n \notin S^{IU}_{S'_C}$ and $n \notin S'_C$. It can be obtained that $n$ has a concurrency relationship with any element in $S'_C$. Accordingly, let set $S' = \{n\} \cup S'_C$, then $S'$ is a second-order concurrency subset of nodes (according to **Definition V.1**). That is to say, there exists a second-order concurrency subset of nodes $S'$ that contains set $S'_C$ and is not equal to $S'_C$, which contradicts the given condition. ■

**Corollary VI.5.1: A maximum second-order concurrency subset of nodes is also a saturated second-order concurrency subset of nodes.**

**Let $S'_C$ be a maximum second-order concurrency subset of nodes within the set of sending nodes $S_{P_{[1,2]}}$ in the pair of paths $P_{[1,2]}$, then $S'_C$ is also a saturated second-order concurrency subset of nodes in $P_{[1,2]}$.**

**Proof:**

According to the definition of a maximum second-order concurrency subset of nodes in the pair of paths $P_{[1,2]}$ (**Definition V.2**), there does not exist a second-order concurrency subset of nodes that contains set $S'_C$ and is not equal to $S'_C$ within the set of sending nodes $S_{P_{[1,2]}}$ of $P_{[1,2]}$. From **Theorem VI.5**, it follows that $S'_C$ is a saturated second-order concurrency subset of nodes in $P_{[1,2]}$. ■

**Theorem VI.6: A second-order interference subset of nodes is a saturated second-order interference subset of nodes if and only if it cannot be further extended into a larger second-order interference subset of nodes.**

**Let $S'_I$ be a second-order interference subset of nodes in the pair of paths $P_{[1,2]}$. It is a saturated second-order interference subset of nodes if and only if there does not exist a second-order interference subset of nodes that contains set $S'_I$ and is not equal to $S'_I$ within the set of sending nodes $S_{P_{[1,2]}}$ of the pair of paths $P_{[1,2]}$.**

**Proof:**

We first prove that if $S'_I$ is a saturated second-order interference subset of nodes, then there does not exist a second-order interference subset of nodes that contains set $S'_I$ and is not equal to $S'_I$. Proof by contradiction is adopted. Assume that at least one second-order interference subset of nodes $S'$ exists which contains set $S'_I$ and is not equal to $S'_I$, i.e., $S'_I \subset S'\,(S' \subseteq S_{P_{[1,2]}})$. Therefore, at least one element $n$ in set $S_{P_{[1,2]}}^{\backslash S'_I} \triangleq S_{P_{[1,2]}} \backslash S'_I$ also belongs to $S'$, namely $n \in S_{P_{[1,2]}}^{\backslash S'_I}$ and $n \in S'$. On the one hand, according to the definition of a saturated second-order interference subset of nodes (**Definition V.28**), we have $S_{S'_I}^{CU} = S_{P_{[1,2]}}$. Furthermore, $n \in S_{S'_I}^{CU}$ can be derived from $n \in S_{P_{[1,2]}}^{\backslash S'_I}$. That is to say, $n$ has a concurrency relationship with at least one element in $S'_I$. On the other hand, since $n \in S'$, and $S'\,(S'_I \subset S')$ forms a second-order interference subset of nodes, $n$ has an interference relationship with any element in $S'_I$, which contradicts the previously obtained relationship, i.e. $n \in S_{S'_I}^{CU}$.

We then prove that if there does not exist a second-order interference subset of nodes that contains set $S'_I$ and is not equal to $S'_I$, then $S'_I$ is a saturated second-order interference subset of nodes. Proof by contradiction is adopted. Assume that $S'_I$ is not a saturated second-order interference subset of nodes, then $S^{CU}_{S'_I} \neq S_{P_{[1,2]}}$ holds according to **Definition V.28**. Therefore, at least one element $n$ in set $S_{P_{[1,2]}}$ satisfies relation $n \notin S^{CU}_{S'_I}$. From **Definition V.24** (Universal set of concurrency nodes for a subset of sending nodes in a pair of paths) and **Definition IV.4** (Universal set of concurrency nodes for a single sending node), it follows that $S'_I \subseteq S^{CU}_{S'_I}$, and hence $n \notin S'_I$. Based on $n \notin S'_I$ and $n \notin S^{CU}_{S'_I}$, $n$ has an interference relationship with any element in $S'_I$. Accordingly, let set $S' = \{n\} \cup S'_I$, then $S'$ is a second-order interference subset of nodes (according to **Definition V.10**). That is to say, there exists at least one second-order interference subset of nodes $S'$ that contains set $S'_I$ and is not equal to $S'_I$, which contradicts the given condition. ■

**Corollary VI.6.1: A maximum second-order interference subset of nodes is also a saturated second-order interference subset of nodes.**

**Let $S'_I$ be a maximum second-order interference subset of nodes within the set of sending nodes $S_{P_{[1,2]}}$ of the pair of paths $P_{[1,2]}$, then $S'_I$ is also a saturated second-order interference subset of nodes in $P_{[1,2]}$.**

**Proof:**

According to the definition of a maximum second-order interference subset of nodes in the pair of paths $P_{[1,2]}$ (**Definition V.12**), there does not exist a second-order interference subset of nodes that contains set $S'_I$ and is not equal to $S'_I$ within the set of sending nodes $S_{P_{[1,2]}}$ of the pair of paths $P_{[1,2]}$. From **Theorem VI.6**, it follows that $S'_I$ is a saturated second-order interference subset of nodes on $P_{[1,2]}$. ■

## VII. Concepts of the subset of equally spaced nodes, reachable period, and reachable joint period

In this section, the subset of equally spaced nodes with a given initial phase in a primary path is defined. The concepts of reachable period and intrinsic period of a primary path are given. The concepts of reachable joint period and intrinsic joint period of a pair of paths are proposed. The matrix of joint concurrency relationship between subsets of equally spaced nodes of a pair of paths is defined. The binary matrix and its corresponding set of supporting elements are defined, and the maximum reachable supporting number of a binary matrix is further defined.

**Definition VII.1: The subset of equally spaced nodes with a given initial phase in a primary path**

As for a primary path $P_{[i]}\,(i=1,2)$, given the spacing between adjacent nodes as $T_{S_{P_{[i]}}}\,(T_{S_{P_{[i]}}}\in\mathbb{Z}^+, 1\le T_{S_{P_{[i]}}}\le|S_{P_{[i]}}|, i=1,2)$ and the initial phase as $\theta_{S_{P_{[i]}}}\,(\theta_{S_{P_{[i]}}}\in\mathbb{Z}^+, 1\le \theta_{S_{P_{[i]}}}\le T_{S_{P_{[i]}}}, i=1,2)$, respectively. The subset of nodes $\Phi'_{S_{P_{[i]}}}(\theta_{S_{P_{[i]}}},T_{S_{P_{[i]}}})\triangleq\left\{n_{i,j}|j=\theta_{S_{P_{[i]}}}+k\cdot T_{S_{P_{[i]}}}, 1\le j\le\left|S_{P_{[i]}}\right|, k\in\mathbb{Z}, k\ge 0\right\}$ in $P_{[i]}$ is defined as a subset of equally spaced nodes with its initial phase being $\theta_{S_{P_{[i]}}}$ and the spacing between adjacent nodes being $T_{S_{P_{[i]}}}$. If the set $\Phi'_{S_{P_{[i]}}}(\theta_{S_{P_{[i]}}},T_{S_{P_{[i]}}})$ is also a second-order concurrency subset of nodes, it is defined as a second-order concurrency subset with equally spaced nodes in the primary path $P_{[i]}$.

**Note:**

1) In order to make $\Phi'_{S_{P_{[i]}}}(\theta_{S_{P_{[i]}}},T_{S_{P_{[i]}}})$ contain at least one valid element of node, only the case of $T_{S_{P_{[i]}}}\le|S_{P_{[i]}}|$ is considered in the paper.

**Definition VII.2: The reachable period of a primary path**

For a given value $T_{S_{P_{[i]}}}\,(T_{S_{P_{[i]}}}\in\mathbb{Z}^+, 1\le T_{S_{P_{[i]}}}\le|S_{P_{[i]}}|, i=1,2)$, if all the subsets $\Phi'_{S_{P_{[i]}}}(\theta_{S_{P_{[i]}}},T_{S_{P_{[i]}}})\,(1\le\theta_{S_{P_{[i]}}}\le T_{S_{P_{[i]}}})$ of equally spaced nodes in a primary path $P_{[i]}\,(i=1,2)$ with their initial phases between 1 and $T_{S_{P_{[i]}}}$ (including 1 and $T_{S_{P_{[i]}}}$) are second-order concurrency subsets, then $T_{S_{P_{[i]}}}$ is defined as a reachable period of $P_{[i]}$, otherwise $T_{S_{P_{[i]}}}$ is regarded as an unreachable period of $P_{[i]}$.

**Definition VII.3: The reachable joint period of a pair of paths**

Given the corresponding reachable period $T_{S_{P_{[i]}}}$ $(i = 1,2)$ of the primary path $P_{[i]}$ $(i = 1,2)$, for a certain value $T_{S_{P_{[1,2]}}}$ $(\in \mathbb{Z}^+, \geq 1)$, consider a sequence consisting of $T_{S_{P_{[1,2]}}}$ time beats arranged in chronological order (these time beats are not necessarily contiguous). If all the following conditions can be satisfied for a pair of paths $P_{[1,2]}$, $T_{S_{P_{[1,2]}}}$ is defined as a reachable joint period of $P_{[1,2]}$:

1) **Non-emptiness:** at least one subset $\Phi'_{S_{P_{[i]}}}(\theta_{S_{P_{[i]}}}, T_{S_{P_{[i]}}})$ $(1 \leq \theta_{S_{P_{[i]}}} \leq T_{S_{P_{[i]}}}, i = 1,2)$ of equally spaced nodes is active in each time beat, i.e. all the nodes of the subset are in sending state;
2) **Concurrency:** all the nodes activated simultaneously in each time beat form a second-order concurrency subset;
3) **Uniqueness:** in each time beat, the number of subsets of equally spaced nodes, which belong to the same primary path and are activated at the same time, shall not exceed one;
4) **Uniform ergodicity:** within $T_{S_{P_{[1,2]}}}$ time beats, all subsets of equally spaced nodes $\Phi'_{S_{P_{[i]}}}(\theta_{S_{P_{[i]}}}, T_{S_{P_{[i]}}})$ $(1 \leq \theta_{S_{P_{[i]}}} \leq T_{S_{P_{[i]}}}, i = 1,2)$ should be activated at least once, and all subsets of equally spaced nodes belonging to the same primary path should be activated the same number of times (subsets of equally spaced nodes belonging to different primary paths can be activated different number of times).

**Note:**

1) Within a given time beat, the minimal unit for node activation is defined as a complete subset of equally spaced nodes derived from one primary path, namely $\Phi'_{S_{P_{[i]}}}(\theta_{S_{P_{[i]}}}, T_{S_{P_{[i]}}})$ $(1 \leq$

$\theta_{S_{P_{[i]}}} \leq T_{S_{P_{[i]}}}, i = 1,2)$.

**Definition VII.4: The intrinsic period of a primary path**

The minimum value of all the reachable periods of a primary path $P_{[i]}$ is defined as the intrinsic period of $P_{[i]}$, which is denoted as $T^*_{S_{P_{[i]}}}$.

**Definition VII.5: The intrinsic period of a pair of paths**

The minimum value of all the reachable joint periods of a pair of paths $P_{[1,2]}$ is defined as the intrinsic joint period of $P_{[1,2]}$, which is denoted as $T^*_{S_{P_{[1,2]}}}$.

**Definition VII.6: The joint concurrency coefficient between subsets of equally spaced nodes of two primary paths**

Given periods of the primary paths $P_{[1]}$ and $P_{[2]}$ are $T_{S_{P_{[1]}}}$ and $T_{S_{P_{[2]}}}$, respectively. If a subset $\Phi'_{S_{P_{[1]}}}(\theta_{S_{P_{[1]}}}, T_{S_{P_{[1]}}})$ in $P_{[1]}$ and a subset $\Phi'_{S_{P_{[2]}}}(\theta_{S_{P_{[2]}}}, T_{S_{P_{[2]}}})$ in $P_{[2]}$ can jointly form a second-order concurrency subset, the joint concurrency coefficient of these two subsets of nodes is defined as: $c(\theta_{S_{P_{[1]}}}, \theta_{S_{P_{[2]}}}) \triangleq 1$, otherwise it is defined as $c(\theta_{S_{P_{[1]}}}, \theta_{S_{P_{[2]}}}) \triangleq 0$.

**Note:**

1) With the variations of the periods $T_{S_{P_{[1]}}}$ and $T_{S_{P_{[2]}}}$, the corresponding joint concurrency coefficients may vary accordingly.

**Definition VII.7: The matrix of joint concurrency relationship between subsets of equally spaced nodes of a pair of paths**

Given periods of the primary paths $P_{[1]}$ and $P_{[2]}$ are $T_{S_{P_{[1]}}}$ and $T_{S_{P_{[2]}}}$, respectively. Define a matrix $\boldsymbol{C}$ containing $T_{S_{P_{[1]}}}$ rows and $T_{S_{P_{[2]}}}$ columns of elements as the matrix of joint concurrency relationship between the subsets of equally spaced nodes of $P_{[1,2]}$, i.e.

$$
\boldsymbol{C} \triangleq \begin{pmatrix} c(1,1) & \dots\dots\dots. & c(1,j) & \dots\dots.. & c\left(1,T_{S_{P_{[2]}}}\right) \\ \dots\dots & \dots\dots\dots. & \dots\dots\dots. & \dots\dots.. & \dots\dots\dots. \\ c(i,1) & \dots\dots\dots. & c(i,j) & \dots\dots.. & c\left(i,T_{S_{P_{[2]}}}\right) \\ \dots\dots & \dots\dots\dots. & \dots\dots\dots. & \dots\dots.. & \dots\dots\dots. \\ c\left(T_{S_{P_{[1]}}},1\right) & \dots\dots\dots. & c\left(T_{S_{P_{[1]}}},j\right) & \dots\dots.. & c\left(T_{S_{P_{[1]}}},T_{S_{P_{[2]}}}\right) \end{pmatrix}
$$

(VII.1)

where $\boldsymbol{C}[\theta_{S_{P_{[1]}}},\theta_{S_{P_{[2]}}}] \triangleq c(\theta_{S_{P_{[1]}}},\theta_{S_{P_{[2]}}})\ (1 \leq \theta_{S_{P_{[i]}}} \leq T_{S_{P_{[i]}}}, i = 1,2)$.

**Note:**

1) With the variations of the periods $T_{S_{P_{[1]}}}$ and $T_{S_{P_{[2]}}}$, the matrix $\boldsymbol{C}$ may vary accordingly.

**Definition VII.8: The matrix of intrinsic joint concurrency relationship between subsets of equally spaced nodes of a pair of paths**

Given the intrinsic interference intensities of the primary paths $P_{[1]}$ and $P_{[2]}$ are $I^*_{S_{P_{[1]}}}$ and $I^*_{S_{P_{[2]}}}$, respectively. Construct $I^*_{S_{P_{[1]}}}$ and $I^*_{S_{P_{[2]}}}$ subsets of equally spaced nodes respectively, i.e. $\Phi'_{S_{P_{[1]}}}(\theta_{S_{P_{[1]}}}, I^*_{S_{P_{[1]}}})\ (1 \leq \theta_{S_{P_{[1]}}} \leq I^*_{S_{P_{[1]}}})$ and $\Phi'_{S_{P_{[2]}}}(\theta_{S_{P_{[2]}}}, I^*_{S_{P_{[2]}}})\ (1 \leq \theta_{S_{P_{[2]}}} \leq I^*_{S_{P_{[2]}}})$. Based on these constructed subsets of equally spaced nodes, a matrix $\boldsymbol{C}^*$ consisting of $I^*_{S_{P_{[1]}}}$ rows and $I^*_{S_{P_{[2]}}}$ columns of elements is defined as the matrix of intrinsic joint concurrency relationship between subsets of equally spaced nodes of $P_{[1,2]}$, i.e.

$$
\boldsymbol{C}^* \triangleq \begin{pmatrix} c(1,1) & \dots\dots\dots & c(1,j) & \dots\dots. & c(1,I^*_{S_{P_{[2]}}}) \\ \dots\dots & \dots\dots\dots & \dots\dots\dots & \dots\dots. & \dots\dots\dots \\ c(i,1) & \dots\dots\dots & c(i,j) & \dots\dots. & c(i,I^*_{S_{P_{[2]}}}) \\ \dots\dots & \dots\dots\dots & \dots\dots\dots & \dots\dots. & \dots\dots\dots \\ c(I^*_{S_{P_{[1]}}},1) & \dots\dots\dots & c(I^*_{S_{P_{[1]}}},j) & \dots\dots. & c(I^*_{S_{P_{[1]}}},I^*_{S_{P_{[2]}}}) \end{pmatrix}
$$

(VII.2)

where $\boldsymbol{C}^*[\theta_{S_{P_{[1]}}},\theta_{S_{P_{[2]}}}] \triangleq c(\theta_{S_{P_{[1]}}},\theta_{S_{P_{[2]}}})\ (1 \leq \theta_{S_{P_{[i]}}} \leq I^*_{S_{P_{[i]}}}, i = 1,2)$.

**Definition VII.9: A binary matrix and its corresponding set of supporting elements**

A matrix whose elements are either 0 or 1 is defined as a binary matrix. Given a binary matrix $\boldsymbol{B}$, a set $S_{\boldsymbol{B}_{sp}}$ is defined as the set of supporting elements of matrix $\boldsymbol{B}$, if it satisfies all the following conditions:

1) **Corresponding to non-zero elements:** for binary matrix $\boldsymbol{B}$ whose elements are all 0, define the set $S_{\boldsymbol{B}_{sp}} \triangleq \emptyset$. Moreover, for the binary matrix $\boldsymbol{B}$ with non-zero elements, the set $S_{\boldsymbol{B}_{sp}}$ is composed of several 2-tuples, and any element in $S_{\boldsymbol{B}_{sp}}$, that is, a 2-tuple $(i_k, j_k)$ $(\in S_{\boldsymbol{B}_{sp}}, k \in \mathbb{Z}^+)$, corresponds to the non-zero element $\boldsymbol{B}[i_k, j_k]$ $(= 1)$ in the matrix $\boldsymbol{B}$;
2) **Covering non-zero elements:** for any non-zero element $\boldsymbol{B}[i, j]$ $(= 1)$ in matrix $\boldsymbol{B}$, there is at least one element $(i_k, j_k)$ in the set $S_{\boldsymbol{B}_{sp}}$, which makes $i_k = i$ or $j_k = j$ hold;
3) **Not covering with each other:** for any two elements (if any) in $S_{\boldsymbol{B}_{sp}}$, i.e. $(i_k, j_k), (i_l, j_l)$ $(\in S_{\boldsymbol{B}_{sp}}, k, l \in \mathbb{Z}^+, k \neq l)$, both $i_k \neq i_l$ and $j_k \neq j_l$ hold.

**Note:**

1) Given a binary matrix $\boldsymbol{B}$, there may exist more than one set $S_{\boldsymbol{B}_{sp}}$ satisfying the above conditions. That is, for a binary matrix $\boldsymbol{B}$, it may have multiple sets of supporting elements.

**Definition VII.10: The set of maximum supporting elements of a binary matrix**

Among all possible sets of supporting elements of a binary matrix $\boldsymbol{B}$, the set with the largest number of elements is defined as the set of maximum supporting elements of $\boldsymbol{B}$, which is denoted as $S^*_{\boldsymbol{B}_{sp}}$.

**Note:**

1) Given a binary matrix $\boldsymbol{B}$, there may be more than one such set.

**Definition VII.11: The reachable supporting number and the maximum reachable supporting number of a binary matrix**

The number of elements contained in a set $S_{\boldsymbol{B}_{sp}}$ of supporting elements of a binary matrix $\boldsymbol{B}$ is defined as a reachable supporting number of $\boldsymbol{B}$, which is denoted as $U_{\boldsymbol{B}_{sp}}$. The number of elements contained in a set $S^*_{\boldsymbol{B}_{sp}}$ of maximum supporting elements of $\boldsymbol{B}$ is defined as the maximum reachable supporting number of $\boldsymbol{B}$, which is denoted as $U^*_{\boldsymbol{B}_{sp}}$.

**Definition VII.12: A pair of paths that can form an optimal pairing**

As for a pair of paths $P_{[1,2]}$ whose intrinsic interference intensity is $I^*_{S_{P_{[1,2]}}}$, if its joint period can reach $I^*_{S_{P_{[1,2]}}}$, it is defined as a pair of paths which can form an optimal pairing.

## VIII. Some related basic properties of the subset of equally spaced nodes, reachable period, and reachable joint period

In this section, the reachability of the period of a primary path is proved, and the reachable intrinsic interference intensity theorem for a primary path is obtained (***Theorem VIII.1***). It is further pointed out that subsets of equally spaced nodes corresponding to the intrinsic period of a primary path cannot form a second-order concurrency relationship between each other (**Corollary VIII.1.1**). Moreover, the lower bound (**Theorem VIII.2**) and the upper bound (**Theorem VIII.3**) of the intrinsic interference intensity of a pair of paths are given, respectively. The reachability of the joint period of a pair of paths is proved (**Theorem VIII.4 - Theorem VIII.5**), and the necessary and sufficient condition for the reachable intrinsic interference intensity of a pair of paths is given (***Theorem VIII.5***).

### *A*. Properties of the period and intrinsic interference intensity of a primary path

**Lemma VIII.1.1: The unreachability of the period of a primary path.**

**If the intrinsic interference intensity of a primary path $P_{[i]}$ $(i=1,2)$ with $N_{S_{P_{[i]}}}$ $(i=1,2)$ sending nodes is $I^*_{S_{P_{[i]}}}$, the period of $T_{S_{P_{[i]}}} < I^*_{S_{P_{[i]}}}$ cannot be reached.**

**Proof:**

**Case 1:** $I^*_{S_{P_{[i]}}} > 1$

Let the primary path $P_{[i]}$ be composed of $N_{S_{P_{[i]}}}$ sending nodes, and the sequence numbers of these nodes are 1, 2,..., $N_{S_{P_{[i]}}}$, respectively. Consider the case of $T_{S_{P_{[i]}}} < I^*_{S_{P_{[i]}}}$. First, a matrix $\boldsymbol{M}$ is constructed:

$$\boldsymbol{M} \triangleq \begin{pmatrix} 1 & \cdots\cdots\cdots & 1+k\cdot T_{S_{P_{[i]}}} & \cdots\cdots & 1+\left\lfloor \frac{N_{S_{P_{[i]}}}}{T_{S_{P_{[i]}}}} \right\rfloor \cdot T_{S_{P_{[i]}}} \\ \cdots\cdots & \cdots\cdots\cdots & \cdots\cdots\cdots & \cdots\cdots & \cdots\cdots\cdots \\ j & \cdots\cdots\cdots & j+k\cdot T_{S_{P_{[i]}}} & \cdots\cdots & N_{S_{P_{[i]}}} \\ \cdots\cdots & \cdots\cdots\cdots & \cdots\cdots\cdots & \cdots\cdots & \times\times\times\times \\ T_{S_{P_{[i]}}} & \cdots\cdots\cdots & T_{S_{P_{[i]}}}+k\cdot T_{S_{P_{[i]}}} & \cdots\cdots & \times\times\times\times \end{pmatrix}$$

(VIII.1)

1) The matrix has $T_{S_{P_{[i]}}}$ $(1 \le T_{S_{P_{[i]}}} \le N_{S_{P_{[i]}}})$ rows and $\left\lceil \frac{N_{S_{P_{[i]}}}}{T_{S_{P_{[i]}}}} \right\rceil$ columns;

2) In ascending order, put the sequence number of the node $j$ $(1 \le j \le N_{S_{P_{[i]}}})$ in $P_{[i]}$ into the corresponding unit in row $\left(j - T_{S_{P_{[i]}}} \cdot \left\lfloor \frac{j}{T_{S_{P_{[i]}}}} \right\rfloor\right)$ and column $\left\lceil \frac{j}{T_{S_{P_{[i]}}}} \right\rceil$ of the matrix one by one. That is, set the value of the unit in row $\left(j - T_{S_{P_{[i]}}} \cdot \left\lfloor \frac{j}{T_{S_{P_{[i]}}}} \right\rfloor\right)$ and column $\left\lceil \frac{j}{T_{S_{P_{[i]}}}} \right\rceil$ of $\boldsymbol{M}$ to be $j$;

3) Ignore the settings of matrix units that may not be filled by sequence numbers of nodes.

According to the definition of the subset of equally spaced nodes, it can be seen that row $j$ of matrix $\boldsymbol{M}$ is composed of all the elements of the subset $\Phi'_{S_{P_{[i]}}}(j, T_{S_{P_{[i]}}})$ with initial phase being $j$ and spacing between adjacent nodes being $T_{S_{P_{[i]}}}$.

Now let us prove that if $T_{S_{P_{[i]}}} < I^*_{S_{P_{[i]}}}$, it can be deduced that not all the $T_{S_{P_{[i]}}}$ subsets $\Phi'_{S_{P_{[i]}}}(\theta_{S_{P_{[i]}}}, T_{S_{P_{[i]}}})$ $(1 \le \theta_{S_{P_{[i]}}} \le T_{S_{P_{[i]}}})$ are second-order concurrency subsets, which suggests that $T_{S_{P_{[i]}}} < I^*_{S_{P_{[i]}}}$ is not reachable (**Definition VII.2**). Assuming that $S'_{P_{[i]}}$ is a maximum second-order interference subset of $P_{[i]}$ with the number of its elements being $I^*_{S_{P_{[i]}}}$, and the set is continuous according to **Corollary VI.2.1**. Since the number of rows of matrix $\boldsymbol{M}$ is $T_{S_{P_{[i]}}}$ $(< I^*_{S_{P_{[i]}}})$, the corresponding distributions of the elements of $S'_{P_{[i]}}$ in $\boldsymbol{M}$ will stretch across (i.e. occupy) at least two adjacent columns of $\boldsymbol{M}$, so that there must be at least two adjacent elements on some row $k$ $(1 \le k \le T_{S_{P_{[i]}}})$ of matrix $\boldsymbol{M}$ belonging to the set of $S'_{P_{[i]}}$, which makes it is evident that the two nodes corresponding to these two adjacent elements in $\boldsymbol{M}$ cannot be concurrent. That is to say, the subset $\Phi'_{S_{P_{[i]}}}(k, T_{S_{P_{[i]}}})$ of equally spaced nodes corresponding to the row $k$ of matrix $\boldsymbol{M}$ is not a second-order concurrency subset, which indicates that among $T_{S_{P_{[i]}}}$ $(< I^*_{S_{P_{[i]}}})$ subsets $\Phi'_{S_{P_{[i]}}}(\theta_{S_{P_{[i]}}}, T_{S_{P_{[i]}}})$ $(1 \le \theta_{S_{P_{[i]}}} \le T_{S_{P_{[i]}}})$ of equally spaced nodes, at least one subset is not a second-order concurrency subset. Therefore, according to the definition of the reachability of the period of a primary path (**Definition VII.2**), it can be concluded that the period $T_{S_{P_{[i]}}}$ $(< I^*_{S_{P_{[i]}}})$ is unreachable.

**Case 2:** $I^*_{S_{P_{[i]}}} = 1$

According to the definition of a reachable period (**Definition VII.2**), the reachable period should not be smaller than one. That is to say, as for the case of $I^*_{S_{P_{[i]}}} = 1$, the reachable period $T_{S_{P_{[i]}}}$ should not be smaller than $I^*_{S_{P_{[i]}}}$.■

**Note:**

1) Although the above proof is based on the interference decreasing rule between nodes (**Rule IV.1**), according to the definition of intrinsic interference intensity (**Definition V.14**), it is easy to know that the above conclusion also applies to more general cases where the interference decreasing rule between nodes does not necessarily hold. In short, a very brief proof is carried out by contradiction as follows. If period $T_{S_{P_{[i]}}} < I^*_{S_{P_{[i]}}}$ is reachable, at least two sending nodes belonging to the same maximum second-order interference subset $S'_{P_{[i]}}$ of the primary path are assigned to some subset of equally spaced nodes. Accordingly, this subset of equally spaced nodes cannot be a second-order concurrency subset of nodes.

**Lemma VIII.1.2: The reachability of the period of a primary path.**

**If the intrinsic interference intensity of a primary path $\boldsymbol{P_{[i]}}$ $\boldsymbol{(i = 1, 2)}$ with $\boldsymbol{N_{S_{P_{[i]}}}}$ $\boldsymbol{(i = 1, 2)}$ sending nodes is $\boldsymbol{I^*_{S_{P_{[i]}}}}$, it is reachable for the period of $\boldsymbol{I^*_{S_{P_{[i]}}} \leq T_{S_{P_{[i]}}} \leq N_{S_{P_{[i]}}}}$.**

**Proof:**

Reduction to absurdity is adopted. Suppose that a period $T_{S_{P_{[i]}}}$ $(\geq I^*_{S_{P_{[i]}}})$ is not reachable. First, let us construct the matrix $\boldsymbol{M}$ (that is, the matrix has $T_{S_{P_{[i]}}}$ rows and $\left\lceil \frac{N_{S_{P_{[i]}}}}{T_{S_{P_{[i]}}}} \right\rceil$ columns) according to the same method adopted in **Lemma VIII.1.1**. Because it is assumed that the period $T_{S_{P_{[i]}}}$ $(\geq I^*_{S_{P_{[i]}}})$ is unreachable, according to the definition of the unreachable period of a primary path **(Definition VII.2)**, it can be concluded that at least one subset $\Phi'_{S_{P_{[i]}}}(k, T_{S_{P_{[i]}}})$ corresponding to some row $k$ $(1 \leq k \leq T_{S_{P_{[i]}}})$ of the matrix $\boldsymbol{M}$ is not a second-order concurrency subset, which suggests that at least two nodes in the subset $\Phi'_{S_{P_{[i]}}}(k, T_{S_{P_{[i]}}})$ cannot be concurrent (i.e., an interference relationship exists). According to **Theorem VI.2**, these two nodes and all the other nodes between them form a continuous second-order interference subset (**Definition V.23**), and the number of elements contained in the second-order interference subset will not be less than $T_{S_{P_{[i]}}} + 1$. That is, the number of elements of the formed second-order interference subset is larger than the intrinsic interference intensity $I^*_{S_{P_{[i]}}}$ (since the case of $I^*_{S_{P_{[i]}}} \leq T_{S_{P_{[i]}}} \leq N_{S_{P_{[i]}}}$ is considered here), which contradicts the premise that the intrinsic interference intensity of $P_{[i]}$ is $I^*_{S_{P_{[i]}}}$. ■

**Note:**

1) The condition $T_{S_{P_{[i]}}} \leq N_{S_{P_{[i]}}}$ specified in the premise of this lemma is derived in accordance with the definition of the reachable period of a primary path (**Definition VII.2**). That is to say, pursuant to this definition, $N_{S_{P_{[i]}}}$ serves as the effective upper limit of the reachable period $T_{S_{P_{[i]}}}$ for a primary path;
2) The proof of this lemma is established on the basis of the hypothesis of "the interference decreasing rule between nodes" (**Rule IV.1**);
3) Based on the concurrent transmission theorem proved previously (**Theorem VI.4**), it can be known that within any given time beat, if an arbitrary second-order concurrency subset of nodes is selected from the $T_{S_{P_{[i]}}}$ second-order concurrency subsets of nodes constructed in the procedure of this proof, and this subset is placed in the concurrent transmission state, namely all nodes within the subset are kept in the simultaneous sending states, while all the other nodes contained in the remaining $T_{S_{P_{[i]}}} - 1$ second-order concurrency subsets of nodes are maintained in their silent states, such concurrent transmission can be regarded as effective.

**Theorem VIII.1: Reachable intrinsic interference intensity theorem for a primary path.**

**If the intrinsic period of a primary path $P_{[i]}$ $(i = 1, 2)$ is $T^*_{S_{P_{[i]}}}$, and the intrinsic interference intensity is $I^*_{S_{P_{[i]}}}$, then $T^*_{S_{P_{[i]}}} = I^*_{S_{P_{[i]}}}$ holds.**

**Proof:**

According to **Lemma VIII.1.1**, the period of $T_{S_{P_{[i]}}} < I^*_{S_{P_{[i]}}}$ is not reachable, and according to **Lemma VIII.1.2**, the period of $T_{S_{P_{[i]}}} \geq I^*_{S_{P_{[i]}}}$ is reachable. Therefore, the minimum reachable period, i.e. the intrinsic period of a primary path $P_{[i]}$, is $T^*_{S_{P_{[i]}}} = I^*_{S_{P_{[i]}}}$. ■

**Note:**

1) This important theorem is constructed upon **Lemma VIII.1.2**, which in turn is derived based on the assumption of "the interference decreasing rule between nodes" **(Rule IV.1)**. Accordingly, this theorem is also based on the assumption of "the interference decreasing rule between nodes" (**Rule IV.1**).

**Corollary VIII.1.1: The subsets of equally spaced nodes corresponding to the intrinsic period of a primary path cannot form a second-order concurrency relationship between each other.**

**If the intrinsic period of a primary path $P_{[i]}$ $(i = 1, 2)$ is $T^*_{S_{P_{[i]}}}$ $(> 1)$, among all of its $T^*_{S_{P_{[i]}}}$ subsets of equally spaced nodes with initial phases being $\theta_{S_{P_{[i]}}}$ $(1 \leq \theta_{S_{P_{[i]}}} \leq T^*_{S_{P_{[i]}}})$ and spacing between adjacent nodes being $T^*_{S_{P_{[i]}}}$, any two of them cannot jointly form a second-order concurrency subset.**

**Proof:**

Firstly, a matrix $\boldsymbol{M}$ with the number of rows being $T^*_{S_{P_{[i]}}}$ and the number of columns being $\left\lceil \frac{N_{S_{P_{[i]}}}}{T^*_{S_{P_{[i]}}}} \right\rceil$ is constructed by following the method adopted in **Lemma VIII.1.1**. According to **Theorem VIII.1**, the intrinsic interference intensity of the primary path $P_{[i]}$ is $I^*_{S_{P_{[i]}}} = T^*_{S_{P_{[i]}}}$, and according to **Corollary VI.2.1**, all possible maximum second-order interference subsets of $P_{[i]}$ are continuous. Therefore, if we take any two rows (corresponding to two subsets of equally spaced nodes with different initial phases) from the matrix $\boldsymbol{M}$, it can be found that these two rows must contain two different elements, each of which belongs to a common maximum second-order interference subset, and the corresponding two nodes must interfere with each other. Therefore, the two subsets of equally spaced nodes corresponding to these two rows selected arbitrarily from the matrix $\boldsymbol{M}$ contain nodes that interfere with each other. That is to say, the two arbitrarily selected subsets of equally spaced nodes cannot form a second-order concurrency subset. ■

### *B*. Properties of the joint period and intrinsic interference intensity of a pair of paths

**Theorem VIII.2: The lower bound of the intrinsic interference intensity of a pair of paths.**

**If the intrinsic interference intensity of a pair of paths $P_{[1,2]}$ is $I^*_{S_{P_{[1,2]}}}$, and the intrinsic interference intensities of the primary paths $P_{[1]}$ and $P_{[2]}$ are $I^*_{S_{P_{[1]}}}$ and $I^*_{S_{P_{[2]}}}$, respectively, then there is $I^*_{S_{P_{[1,2]}}} \geq max[I^*_{S_{P_{[1]}}}, I^*_{S_{P_{[2]}}}]$.**

**Proof:**

**Case 1:** $max[I^*_{S_{P_{[1]}}}, I^*_{S_{P_{[2]}}}] = 1$

According to the definition (**Definition V.15**), the intrinsic interference intensity $I^*_{S_{P_{[1,2]}}}$ of a pair of paths $P_{[1,2]}$ must not be less than 1, so $I^*_{S_{P_{[1,2]}}} \geq max[I^*_{S_{P_{[1]}}}, I^*_{S_{P_{[2]}}}]$ holds in this case.

**Case 2:** $max[I^*_{S_{P_{[1]}}}, I^*_{S_{P_{[2]}}}] > 1$

Without losing generality, assume that $I^*_{S_{P_{[1]}}} = max[I^*_{S_{P_{[1]}}}, I^*_{S_{P_{[2]}}}]$, in the following reduction to absurdity is adopted. If $I^*_{S_{P_{[1,2]}}} < I^*_{S_{P_{[1]}}}$ is true, pursuanting to the definition of the intrinsic interference intensity for a pair of paths (**Definition V.15**), it can be concluded that the so-called maximum second-order interference subset $S'_{P_{[1,2]}}$ consisting of $I^*_{S_{P_{[1,2]}}}$ nodes in the pair of paths $P_{[1,2]}$ is not actually the largest second-order interference subset in $P_{[1,2]}$. The reason lies in that there at least exists another second-order interference subset $S'_{P_{[1]}}$ consisting of a greater number of $I^*_{S_{P_{[1]}}}$ nodes in the primary path $P_{[1]}$, which contradicts the premise of the proposition. ■

**Theorem VIII.3: The upper bound of the intrinsic interference intensity of a pair of paths.**

**If the intrinsic interference intensity of a pair of paths $P_{[1,2]}$ is $I^*_{S_{P_{[1,2]}}}$, and the intrinsic interference intensities of the primary paths $P_{[1]}$ and $P_{[2]}$ are $I^*_{S_{P_{[1]}}}$ and $I^*_{S_{P_{[2]}}}$, respectively, then there is $I^*_{S_{P_{[1,2]}}} \leq I^*_{S_{P_{[1]}}} + I^*_{S_{P_{[2]}}}$.**

**Proof:**

Let the nonempty subset $S'_{P_{[1,2]}}$ of nodes be a maximum second-order interference subset of $P_{[1,2]}$. It follows from **Definition V.15** that the number of elements contained in $S'_{P_{[1,2]}}$ is $I^*_{SP_{[1,2]}}$. In the following, the proof proceeds by case analysis.

**Case 1:** All nodes of $S'_{P_{[1,2]}}$ are from a primary path (without losing generality, say from $P_{[1]}$), and $I^*_{SP_{[1,2]}} = 1$:

According to the definition (**Definition V.14**), both $I^*_{SP_{[1]}}$ and $I^*_{SP_{[2]}}$ are not less than 1, so $I^*_{SP_{[1,2]}} < I^*_{SP_{[1]}} + I^*_{SP_{[2]}}$ holds in this case.

**Case 2:** All nodes of $S'_{P_{[1,2]}}$ are from a primary path (without losing generality, say from $P_{[1]}$), and $I^*_{SP_{[1,2]}} > 1$:

According to the definition (**Definition V.15**), the nodes constituting $S'_{P_{[1,2]}}$ form a second-order interference subset of the primary path $P_{[1]}$. Because the intrinsic interference intensity of $P_{[1]}$ is $I^*_{SP_{[1]}}$, there is $I^*_{SP_{[1,2]}} \leq I^*_{SP_{[1]}}$. Furthermore, since $I^*_{SP_{[2]}} \geq 1$, there is $I^*_{SP_{[1,2]}} < I^*_{SP_{[1]}} + I^*_{SP_{[2]}}$.

**Case 3:** One part of the nodes of $S'_{P_{[1,2]}}$ comes from the primary path $P_{[1]}$ (assuming the number of these nodes is $m_1$), and the other part (assuming the number of these nodes is $m_2$) comes from the primary path $P_{[2]}$, and $m_1 > 1,\ m_2 > 1$:

According to the definition of the second-order interference subset (**Definition V.10**), $m_1$ nodes coming from $P_{[1]}$ constitute by themselves a second-order interference subset, and $m_1 \leq I^*_{SP_{[1]}}$; $m_2$ nodes from $P_{[2]}$ also form a second-order interference subset, and $m_2 \leq I^*_{SP_{[2]}}$. Therefore, it can be seen that $I^*_{SP_{[1,2]}} = m_1 + m_2 \leq I^*_{SP_{[1]}} + I^*_{SP_{[2]}}$, i.e. $I^*_{SP_{[1,2]}} \leq I^*_{SP_{[1]}} + I^*_{SP_{[2]}}$ holds.

**Case 4:** One part of the nodes of $S'_{P_{[1,2]}}$ comes from the primary path $P_{[1]}$ (assuming the number of these nodes is $m_1$), and the other part (assuming the number of these nodes is $m_2$) comes from the primary path $P_{[2]}$, and $m_1 > 1,\ m_2 = 1$:

According to the definition of the second-order interference subset (**Definition V.10**), $m_1$ nodes coming from $P_{[1]}$ constitute by themselves a second-order interference subset, and $m_1 \leq$ $I^*_{S_{P_{[1]}}}$; Based on the definition of the intrinsic interference intensity of a nonempty subset of nodes (**Definition V.13**), we have $m_2 \le I^*_{S_{P_{[2]}}}$ ($m_2 = 1$). Therefore, it can be seen that $I^*_{S_{P_{[1,2]}}} = m_1 + m_2 \le I^*_{S_{P_{[1]}}} + I^*_{S_{P_{[2]}}}$, i.e. $I^*_{S_{P_{[1,2]}}} \le I^*_{S_{P_{[1]}}} + I^*_{S_{P_{[2]}}}$ holds.

**Case 5:** One part of the nodes of $S'_{P_{[1,2]}}$ comes from the primary path $P_{[1]}$ (assuming the number of these nodes is $m_1$), and the other part (assuming the number of these nodes is $m_2$) comes from the primary path $P_{[2]}$, and $m_1 = 1,\ m_2 > 1$:

The proof here is the same as that of case 4 (omitted).

**Case 6:** One part of the nodes of $S'_{P_{[1,2]}}$ comes from the primary path $P_{[1]}$ (assuming the number of these nodes is $m_1$), and the other part (assuming the number of these nodes is $m_2$) comes from the primary path $P_{[2]}$, and $m_1 = 1,\ m_2 = 1$:

In this case, $I^*_{S_{P_{[1,2]}}} = m_1 + m_2 = 2$, and considering the fact that both $I^*_{S_{P_{[1]}}}$ and $I^*_{S_{P_{[2]}}}$ are not less than 1, so $I^*_{S_{P_{[1,2]}}} \le I^*_{S_{P_{[1]}}} + I^*_{S_{P_{[2]}}}$ is established. ■

**Theorem VIII.4: The unreachability of the joint period of a pair of paths.**

**If the intrinsic interference intensity of a pair of paths $P_{[1,2]}$ is $I^*_{S_{P_{[1,2]}}}$, the joint period $T_{S_{P_{[1,2]}}} < I^*_{S_{P_{[1,2]}}}$ cannot be reached.**

**Proof:**

**Case 1:** $I^*_{S_{P_{[1,2]}}} = 1$

According to the required ergodicity defined by the reachable joint period of a pair of paths (**Definition VII.3**), it can be seen that in this case, the joint period $T_{S_{P_{[1,2]}}}$ should not be less than 1, i.e. $T_{S_{P_{[1,2]}}} \ge I^*_{S_{P_{[1,2]}}}$ holds.

**Case 2:** $I^*_{S_{P_{[1,2]}}} > 1$

Let the nonempty subset $S'_{P_{[1,2]}}$ of nodes be a maximum second-order interference subset of the pair of paths $P_{[1,2]}$, which is composed of $I^*_{S_{P_{[1,2]}}}$ sending nodes. According to the requirement of concurrency in the definition of the reachable joint period of a pair of paths, these $I^*_{S_{P_{[1,2]}}}$ sending nodes must belong to $I^*_{S_{P_{[1,2]}}}$ different subsets of equally spaced nodes respectively (otherwise, the concurrency will not be met). According to the required concurrency and ergodicity by the definition of the reachable joint period of a pair of paths, it needs at least $I^*_{S_{P_{[1,2]}}}$ time beats to traverse these $I^*_{S_{P_{[1,2]}}}$ different subsets of equally spaced nodes, that is, the joint period $T_{S_{P_{[1,2]}}}$ should not be less than $I^*_{S_{P_{[1,2]}}}$. Therefore, $T_{S_{P_{[1,2]}}} \geq I^*_{S_{P_{[1,2]}}}$ holds. ■

**Lemma VIII.5.1: The reachability of the joint period of a pair of paths.**

**Assuming that the reachable periods of the primary paths $\boldsymbol{P_{[1]}}$ and $\boldsymbol{P_{[2]}}$ in a pair of paths $\boldsymbol{P_{[1,2]}}$ are $\boldsymbol{T_{S_{P_{[1]}}}}$ and $\boldsymbol{T_{S_{P_{[2]}}}}$, respectively. Accordingly, let $\boldsymbol{C}$ be the matrix of joint concurrency relationship between subsets of equally spaced nodes of $\boldsymbol{P_{[1,2]}}$ (Definition VII.7), and its maximum reachable supporting number is $\boldsymbol{U^*_{C_{sp}}}$. The joint period $\boldsymbol{T_{S_{P_{[1,2]}}} = T_{S_{P_{[1]}}} + T_{S_{P_{[2]}}} - U^*_{C_{sp}}}$ is reachable.**

**Proof:**

Let us determine one set $S^*_{\boldsymbol{C}_{sp}}$ as the set of maximum supporting elements of $\boldsymbol{C}$ (**Definition VII.10**), and assign incremental sequence numbers to all of its $U^*_{\boldsymbol{C}_{sp}}$ elements (i.e. element No. 1, element No. 2,..., and element No. $U^*_{\boldsymbol{C}_{sp}}$).

First, let us consider the No. 1 element in $S^*_{\boldsymbol{C}_{sp}}$, i.e. $(i_1, j_1)$ $(\in S^*_{\boldsymbol{C}_{sp}})$, assume that its corresponding non-zero element in $\boldsymbol{C}$ is $\boldsymbol{C}[i_1, j_1]$ $(= 1)$, and arrange the subset $\Phi'_{S_{P_{[1]}}}(i_1, T_{S_{P_{[1]}}})$ of equally spaced nodes in the primary path $P_{[1]}$ and the subset $\Phi'_{S_{P_{[2]}}}(j_1, T_{S_{P_{[2]}}})$ of equally spaced nodes in the primary path $P_{[2]}$ to be activated concurrently within the time beat No. 1 (Since $\boldsymbol{C}[i_1, j_1] = 1$, such concurrency is feasible). Likewise, for the No. 2 element in $S^*_{\boldsymbol{C}_{sp}}$, i.e. $(i_2, j_2)$ $(\in S^*_{\boldsymbol{C}_{sp}})$, assume that its corresponding non-zero element in $\boldsymbol{C}$ is $\boldsymbol{C}[i_2, j_2]$ $(= 1)$, and arrange the subset $\Phi'_{S_{P_{[1]}}}(i_2, T_{S_{P_{[1]}}})$ of $P_{[1]}$ and the subset $\Phi'_{S_{P_{[2]}}}(j_2, T_{S_{P_{[2]}}})$ of $P_{[2]}$ to be activated concurrently within the time beat No. 2. By analogy, we can allocate $U^*_{\boldsymbol{C}_{sp}}$ time beats arranged in chronological order to realize the concurrent activation of the subsets of equally spaced nodes respectively from $P_{[1]}$ and $P_{[2]}$ (i.e. $U^*_{\boldsymbol{C}_{sp}}$ subsets are from $P_{[1]}$, and $U^*_{\boldsymbol{C}_{sp}}$ subsets are from $P_{[2]}$).

Through the above processing for arranging time beats, there are still $T_{S_{P_{[1]}}} - U^*_{\boldsymbol{C}_{sp}}$ and $T_{S_{P_{[2]}}} - U^*_{\boldsymbol{C}_{sp}}$ subsets of equally spaced nodes in $P_{[1]}$ and $P_{[2]}$, respectively, having not been assigned to their corresponding time beats. Therefore, the total number of subsets of equally spaced nodes waiting to be arranged with time beats is $T_{S_{P_{[1]}}} + T_{S_{P_{[2]}}} - 2U^*_{\boldsymbol{C}_{sp}}$. In order to avoid mutual interference, the remaining $T_{S_{P_{[1]}}} + T_{S_{P_{[2]}}} - 2U^*_{\boldsymbol{C}_{sp}}$ subsets of equally spaced nodes are arranged to be activated in the subsequent $T_{S_{P_{[1]}}} + T_{S_{P_{[2]}}} - 2U^*_{\boldsymbol{C}_{sp}}$ time beats arranged in chronological order without overlapping to each other and without any omission (i.e. only one subset of equally spaced nodes is activated in each time beat).

In summary, based on the above processing for arranging time beats, a total of $T_{S_{P_{[1,2]}}} = U^*_{\boldsymbol{C}_{sp}} + (T_{S_{P_{[1]}}} + T_{S_{P_{[2]}}} - 2U^*_{\boldsymbol{C}_{sp}}) = T_{S_{P_{[1]}}} + T_{S_{P_{[2]}}} - U^*_{\boldsymbol{C}_{sp}}$ time beats arranged in chronological order are assigned to realize one round activation (i.e. one round traversal) of all the subsets of equally spaced nodes in $P_{[1,2]}$. According to the definition of the reachable joint period (**Definition VII.3**), in this case, the joint period $T_{S_{P_{[1,2]}}} = T_{S_{P_{[1]}}} + T_{S_{P_{[2]}}} - U^*_{\boldsymbol{C}_{sp}}$ is reachable. ■

**Note:**

1) In the above proof, the set of maximum supporting elements $S^*_{\boldsymbol{C}_{sp}}$ of matrix $\boldsymbol{C}$ is adopted to construct a feasible sequence of time beats for concurrent transmissions, which aims to form as many pairs of second-order concurrency subsets of nodes capable of performing joint concurrent transmissions as possible between primary path $P_{[1]}$ and $P_{[2]}$. Therefore, the reachable joint period can be minimized.

**Lemma VIII.5.2: The unreachability of the joint period of a pair of paths.**

**Assuming that the reachable periods of the primary paths $\boldsymbol{P}_{[1]}$ and $\boldsymbol{P}_{[2]}$ in a pair of**

**paths $P_{[1,2]}$ are $T_{S_{P_{[1]}}}$ and $T_{S_{P_{[2]}}}$, respectively. Accordingly, let $C$ be the matrix of joint concurrency relationship between subsets of equally spaced nodes of $P_{[1,2]}$, and its maximum reachable supporting number is $U^*_{C_{sp}}$. The joint period $T_{S_{P_{[1,2]}}} < T_{S_{P_{[1]}}} + T_{S_{P_{[2]}}} - U^*_{C_{sp}}$ is unreachable.**

**Proof:**

According to the definition of reachable joint period of a pair of paths (**Definition VII.3**), it is known that the time beats contained in a joint period $T_{S_{P_{[1,2]}}}$ must belong to one of the following three categories (empty time beats that activate no subset of equally spaced nodes are not considered), that is, in time beats of category 1, one subset of equally spaced nodes from the primary path $P_{[1]}$ and one subset of equally spaced nodes from the primary path $P_{[2]}$ are activated at the same time, and the two subsets of nodes together constitute a second-order concurrency subset of nodes; In time beats of category 2, only one subset of equally spaced nodes from $P_{[1]}$ is activated; In time beats of category 3, only one subset of equally spaced nodes from $P_{[2]}$ is activated. For the construction of the smallest achievable joint period, pursuant to **Definition VII.3**, an exhaustive traversal activation over the subsets of equally spaced nodes belonging respectively to primary path $P_{[1]}$ and primary path $P_{[2]}$ is necessary and sufficient. Accordingly, executed in accordance with the construction method presented in **Lemma VIII.5.1**, it can be seen that if the number of time beats of category 1 within a joint period is $U$, the numbers of time beats of category 2 and category 3 are $T_{S_{P_{[1]}}} - U$ and $T_{S_{P_{[2]}}} - U$, respectively. Therefore, the total number of time beats within a joint period is $T_{S_{P_{[1,2]}}} = T_{S_{P_{[1]}}} + T_{S_{P_{[2]}}} - U$. It can be observed from this formula that in order to construct the minimum reachable joint period, only $U$ needs to be maximized. According to the definition of the maximum reachable supporting number $U^*_{C_{sp}}$ of a matrix $\boldsymbol{C}$ (**Definition VII.11**), the maximum value of $U$ is $U^*_{C_{sp}}$. That is, the period $T_{S_{P_{[1,2]}}} = T_{S_{P_{[1]}}} + T_{S_{P_{[2]}}} - U^*_{C_{sp}}$ is the minimum reachable joint period. Hence, if a joint period $T_{S_{P_{[1,2]}}} < T_{S_{P_{[1]}}} + T_{S_{P_{[2]}}} - U^*_{C_{sp}}$, it is unreachable. ■

**Lemma VIII.5.3: The reachability of the joint period of a pair of paths.**

**If the intrinsic interference intensity of a pair of paths $P_{[1,2]}$ is $I^*_{S_{P_{[1,2]}}}$, and the reachable periods of the primary paths $P_{[1]}$ and $P_{[2]}$ are $T_{S_{P_{[1]}}}$ and $T_{S_{P_{[2]}}}$, respectively. Let $C$ be the matrix of joint concurrency relationship between subsets of equally spaced nodes of a pair of paths $P_{[1,2]}$. And the maximum reachable supporting number of $C$ is $U^*_{C_{sp}}$. If $U^*_{C_{sp}} = T_{S_{P_{[1]}}} + T_{S_{P_{[2]}}} - I^*_{S_{P_{[1,2]}}}$, then the joint period $T_{S_{P_{[1,2]}}} = I^*_{S_{P_{[1,2]}}}$ is reachable.**

**Proof:**

It can be known from **Lemma VIII.5.1** that the joint period $T_{S_{P_{[1,2]}}} = T_{S_{P_{[1]}}} + T_{S_{P_{[2]}}} - U^*_{C_{sp}}$ is reachable. Combined with the given condition $U^*_{C_{sp}} = T_{S_{P_{[1]}}} + T_{S_{P_{[2]}}} - I^*_{S_{P_{[1,2]}}}$, it is concluded that the joint period $T_{S_{P_{[1,2]}}} = T_{S_{P_{[1]}}} + T_{S_{P_{[2]}}} - (T_{S_{P_{[1]}}} + T_{S_{P_{[2]}}} - I^*_{S_{P_{[1,2]}}}) = I^*_{S_{P_{[1,2]}}}$ is reachable. ■

**Lemma VIII.5.4: The upper bound of the maximum reachable supporting number $U^*_{C_{sp}}$ of the matrix $C$ of joint concurrency relationship.**

**If the intrinsic interference intensity of a pair of paths $P_{[1,2]}$ is $I^*_{S_{P_{[1,2]}}}$, and the reachable periods of the primary paths $P_{[1]}$ and $P_{[2]}$ are $T_{S_{P_{[1]}}}$ and $T_{S_{P_{[2]}}}$, respectively. Let $C$ be the matrix of joint concurrency relationship between subsets of equally spaced nodes of a pair of paths $P_{[1,2]}$. And the maximum reachable supporting number of $C$ is $U^*_{C_{sp}}$. There is $U^*_{C_{sp}} \leq T_{S_{P_{[1]}}} + T_{S_{P_{[2]}}} - I^*_{S_{P_{[1,2]}}}$.**

**Proof:**

The reduction to absurdity is adopted. If $U^*_{C_{sp}} > T_{S_{P_{[1]}}} + T_{S_{P_{[2]}}} - I^*_{S_{P_{[1,2]}}}$ is possible, according to the construction method used in the proof of the above **Lemma VIII.5.1**, the case of a joint period $T_{S_{P_{[1,2]}}} = T_{S_{P_{[1]}}} + T_{S_{P_{[2]}}} - U^*_{C_{sp}} < I^*_{S_{P_{[1,2]}}}$ can be constructed out (i.e. the joint period $T_{S_{P_{[1,2]}}} < I^*_{S_{P_{[1,2]}}}$ is a reachable joint period), which is contrary to the conclusion about the unreachability of the joint period obtained in **Theorem VIII.4**. ■

**Lemma VIII.5.5: The reachability of the joint period of a pair of paths ($I^*_{S_{P_{[1,2]}}} = 1$).**

**If the intrinsic interference intensity of a pair of paths $P_{[1,2]}$ is $I^*_{S_{P_{[1,2]}}} = 1$, the joint period $T_{S_{P_{[1,2]}}} = I^*_{S_{P_{[1,2]}}}$ is reachable.**

**Proof:**

In this case, it follows from **Definition V.13** that the set $S_{P_{[1,2]}}$ composed of all the sending nodes of the pair of paths $P_{[1,2]}$ is a second-order concurrency set of nodes, so the subsets $S_{P_{[1]}}$ and $S_{P_{[2]}}$ composed of all the sending nodes of the primary paths $P_{[1]}$ and $P_{[2]}$ must also be a second-order concurrency subset of nodes, respectively. As for $P_{[1]}$ and $P_{[2]}$, two subsets of equally spaced nodes can be constructed, i.e. the subset $\Phi'_{S_{P_{[1]}}}(1,1)$ (including all the nodes of $S_{P_{[1]}}$) and the subset $\Phi'_{S_{P_{[2]}}}(1,1)$ (including all the nodes of $S_{P_{[2]}}$), and these two subsets of equally spaced nodes can be activated in a same time beat to realize the concurrent activation of all the nodes of $P_{[1,2]}$. According to the definition of the reachable joint period of a pair of paths (**Definition VII.3**), in this case, the joint period $T_{S_{P_{[1,2]}}} = I^*_{S_{P_{[1,2]}}} = 1$ is reachable. ■

**Lemma VIII.5.6: Mutual exclusivity of subsets of equally spaced nodes among primary paths in a pair of paths ($I^*_{S_{P_{[1,2]}}} = I^*_{S_{P_{[1]}}} + I^*_{S_{P_{[2]}}}$).**

**If the intrinsic interference intensity of a pair of paths $P_{[1,2]}$ is $I^*_{S_{P_{[1,2]}}}$, and the intrinsic interference intensities of the primary paths $P_{[1]}$ and $P_{[2]}$ are $I^*_{S_{P_{[1]}}}$ and $I^*_{S_{P_{[2]}}}$, respectively. The matrix of intrinsic joint concurrency relationship between subsets of equally spaced nodes of $P_{[1,2]}$ is $C^*$ (Definition VII.8), and its maximum reachable supporting number is $U^*_{C^*sp}$. If $I^*_{S_{P_{[1,2]}}} = I^*_{S_{P_{[1]}}} + I^*_{S_{P_{[2]}}}$, then $U^*_{C^*sp} = 0$ holds.**

**Proof:**

It follows from **Lemma VIII.5.4** that $U^*_{C^*sp} \leq I^*_{S_{P_{[1]}}} + I^*_{S_{P_{[2]}}} - I^*_{S_{P_{[1,2]}}} = 0$. According to the definition of maximum reachable supporting number (**Definition VII.11**), $U^*_{C^*sp} \geq 0$ can be obtained. Thus, $U^*_{C^*sp} = 0$ holds. ■

**Note:**

1) It can be readily proved that matrix $\boldsymbol{C}^*$ is a matrix with all zero elements under the condition of $U^*_{\boldsymbol{C}^*sp} = 0$ (The proof is omitted). Namely, under the circumstance of $I^*_{S_{P_{[1,2]}}} = I^*_{S_{P_{[1]}}} + I^*_{S_{P_{[2]}}}$, none of the equally spaced second-order concurrency subsets of nodes contained in primary path $P_{[1]}$ (total $I^*_{S_{P_{[1]}}}$ subsets) and those contained in primary path $P_{[2]}$ (total $I^*_{S_{P_{[2]}}}$ subsets) can jointly form a second-order concurrency subset of nodes, which means joint concurrent transmission cannot be achieved between any two equally spaced second-order concurrency subsets derived from the two primary paths within any given time beat.

**Lemma VIII.5.7: The reachability of the joint period of a pair of paths.**

**If the intrinsic interference intensity of a pair of paths $P_{[1,2]}$ is $I^*_{S_{P_{[1,2]}}}$, and the reachable periods of the primary paths $P_{[1]}$ and $P_{[2]}$ are $T_{S_{P_{[1]}}}$ and $T_{S_{P_{[2]}}}$, respectively. Let $\boldsymbol{C}$ be the matrix of joint concurrency relationship between subsets of equally spaced nodes of a pair of paths $P_{[1,2]}$. And the maximum reachable supporting number of $\boldsymbol{C}$ is $U^*_{\boldsymbol{C}_{sp}}$. If the joint period $I^*_{S_{P_{[1,2]}}}$ is reachable, there must exist a combination of values of $T_{S_{P_{[1]}}}$ and $T_{S_{P_{[2]}}}$ that satisfies relation $U^*_{\boldsymbol{C}_{sp}} = T_{S_{P_{[1]}}} + T_{S_{P_{[2]}}} - I^*_{S_{P_{[1,2]}}}$.**

**Proof:**

Proof by contradiction. If no combination of values of $T_{S_{P_{[1]}}}$ and $T_{S_{P_{[2]}}}$ can make $U^*_{\boldsymbol{C}_{sp}} = T_{S_{P_{[1]}}} + T_{S_{P_{[2]}}} - I^*_{S_{P_{[1,2]}}}$ hold, it can be concluded from **Lemma VIII.5.4** that $U^*_{\boldsymbol{C}_{sp}} < T_{S_{P_{[1]}}} + T_{S_{P_{[2]}}} - I^*_{S_{P_{[1,2]}}}$ (i.e. $I^*_{S_{P_{[1,2]}}} < T_{S_{P_{[1]}}} + T_{S_{P_{[2]}}} - U^*_{\boldsymbol{C}_{sp}}$) holds for all possible combinations of $T_{S_{P_{[1]}}}$ and $T_{S_{P_{[2]}}}$. Furthermore, according to **Lemma VIII.5.2**, it can be obtained that any joint period smaller than $T_{S_{P_{[1]}}} + T_{S_{P_{[2]}}} - U^*_{\boldsymbol{C}_{sp}}$ is unreachable for all feasible value combinations of $T_{S_{P_{[1]}}}$ and $T_{S_{P_{[2]}}}$, which contradicts the given condition that joint period $I^*_{S_{P_{[1,2]}}}$ is reachable. ■

**Theorem VIII.5: Necessary and sufficient condition for the reachable intrinsic interference intensity of a pair of paths.**

**If the intrinsic interference intensity of a pair of paths $P_{[1,2]}$ is $I^*_{S_{P_{[1,2]}}}$, and the reachable periods of the primary paths $P_{[1]}$ and $P_{[2]}$ are $T_{S_{P_{[1]}}}$ and $T_{S_{P_{[2]}}}$, respectively. Let $C$ be the matrix of joint concurrency relationship between subsets of equally spaced nodes of a pair of paths $P_{[1,2]}$. And the maximum reachable supporting number of $C$ is $U^*_{C_{sp}}$. The necessary and sufficient condition for the reachability of joint period $I^*_{S_{P_{[1,2]}}}$ is that there exists a value combination of $T_{S_{P_{[1]}}}$ and $T_{S_{P_{[2]}}}$ satisfying relation $U^*_{C_{sp}} = T_{S_{P_{[1]}}} + T_{S_{P_{[2]}}} - I^*_{S_{P_{[1,2]}}}$.**

**Proof:**

It can be readily proved by combining **Lemma VIII.5.3** and **Lemma VIII.5.7**. ■

**Theorem VIII.6: The intrinsic joint period of a pair of paths that can form an optimal pairing.**

**Assume that a pair of paths $P_{[1,2]}$ can form an optimal pairing. If its intrinsic interference intensity is $I^*_{S_{P_{[1,2]}}}$, its intrinsic joint period $T^*_{S_{P_{[1,2]}}} = I^*_{S_{P_{[1,2]}}}$.**

**Proof:**

It can be proved directly according to **Theorem VIII.4** and **Definition VII.12**. ■

## IX. Definitions of throughput of a primary path and throughput of a pair of paths

In this section, the average throughput and the asymptotic throughput of a primary path are defined. Moreover, the average joint throughput and the asymptotic joint throughput of a pair of paths are also defined.

**Definition IX.1: The specific time interval**

Starting from time instant $t$, the time duration of lasting for $b$ $(b \in \mathbb{Z}, b \geq 0)$ time beats is defined as the specific time interval, which is denoted as $D_{t,b}^{int}$, i.e. $D_{t,b}^{int} \triangleq [t, t + b \cdot \Delta t]$ ($\Delta t$ is the duration of one time beat).

**Definition IX.2: The average throughput of a primary path in a specific time interval**

If the destination node of a primary path $P_{[i]}$ $(i = 1,2)$ receives $m_k$ information blocks in the time interval $D_{t,b}^{int}$, the value $R_{S_{P_{[i]}}}^{av} \triangleq \frac{m_k}{b}$ (unit: number of information blocks/time beat) is defined as the average throughput of a primary path $P_{[i]}$ in this time interval.

**Definition IX.3: The average joint throughput of a pair of paths in a specific time interval**

If the destination nodes of a pair of paths $P_{[1,2]}$ receives a total of $m_k$ information blocks in the time interval $D_{t,b}^{int}$, the value $R_{S_{P_{[1,2]}}}^{av} \triangleq \frac{m_k}{b}$ (unit: number of information blocks/time beat) is defined as the average joint throughput of a pair of paths $P_{[1,2]}$ in this time interval.

**Note:**

1) It is evident that $R_{S_{P_{[1,2]}}}^{av} = R_{S_{P_{[1]}}}^{av} + R_{S_{P_{[2]}}}^{av}$.

**Definition IX.4: The asymptotic throughput of a primary path**

Assuming that the destination node of a primary path $P_{[i]}$ $(i = 1,2)$ receives $m_k$ information blocks in the time interval $D_{t,b}^{int}$. The limitation of the average throughput of $P_{[i]}$ in the case of the duration $b$ ($b$ is measured by the number of time beats) of a time interval $D_{t,b}^{int}$ approaching to infinity is defined as the asymptotic throughput of $P_{[i]}$, which is denoted as $R_{S_{P_{[i]}}}^{\infty} \triangleq \lim_{b \to \infty} R_{S_{P_{[i]}}}^{av}$ (unit:

number of information blocks/time beat).

**Definition IX.5: The asymptotic joint throughput of a pair of paths**

Assuming that the destination nodes of a pair of paths $P_{[1,2]}$ receives a total of $m_k$ information blocks in the time interval $D_{t,b}^{int}$. The limitation of the average joint throughput of $P_{[1,2]}$ in the case of the duration $b$ ($b$ is measured by the number of time beats) of a time interval $D_{t,b}^{int}$ approaching to infinity is defined as the asymptotic joint throughput of $P_{[1,2]}$, which is denoted as $R_{S_{P_{[1,2]}}}^{\infty} \triangleq \lim_{b\to\infty} R_{S_{P_{[1,2]}}}^{av}$ (unit: number of information blocks/time beat).

The ratio between the asymptotic throughput $R_{S_{P_{[i]}}}^{\infty}$ of a primary path $P_{[i]}$ in $P_{[1,2]}$ and the asymptotic joint throughput $R_{S_{P_{[1,2]}}}^{\infty}$ of $P_{[1,2]}$ is: $r_{R_{S_{P_{[1,2]}}}^{\infty}}^{i} \triangleq R_{S_{P_{[i]}}}^{\infty} / R_{S_{P_{[1,2]}}}^{\infty}$ $(i = 1,2)$.

**Definition IX.6: Equal opportunity traversals of the primary paths contained in a pair of paths**

In a reachable joint period, if all the subsets of equally spaced nodes belonging to the primary paths $P_{[1]}$ and $P_{[2]}$ contained in a pair of paths $P_{[1,2]}$ are activated $L$ $(\in \mathbb{Z}^{+})$ times, it is said that the traversals of the primary paths contained in a pair of paths (i.e. the "activation" mentioned above) is of equal opportunity. Otherwise, it is said that the traversals of the primary paths contained in a pair of paths is of unequal opportunity.

**Definition IX.7: The $L_1 \times L_2$-th continuation matrix $\boldsymbol{M}^{L_1\times L_2}$ of a matrix $\boldsymbol{M}$**

$$\boldsymbol{M}^{L_1\times L_2} \triangleq \begin{array}{c} \\ 1 \\ 2 \\ \cdots \\ L_1 \end{array} \begin{array}{c} \begin{array}{cccc} 1 & 2 & \cdots\cdots & L_2 \end{array} \\ \begin{pmatrix} \boldsymbol{M}_{n\times o} & \boldsymbol{M}_{n\times o} & \cdots\cdots & \boldsymbol{M}_{n\times o} \\ \boldsymbol{M}_{n\times o} & \boldsymbol{M}_{n\times o} & \cdots\cdots & \cdots \\ \cdots\cdots & \cdots\cdots & \cdots\cdots & \cdots \\ \boldsymbol{M}_{n\times o} & \boldsymbol{M}_{n\times o} & \cdots\cdots & \boldsymbol{M}_{n\times o} \end{pmatrix} \end{array}$$

(IX.1)

Let $\boldsymbol{M}$ be a matrix of $n$ rows and $o$ columns. Repeat it horizontally for $L_2$ times to form a matrix $\boldsymbol{M'}$ of $n$ rows and $o \cdot L_2$ columns, and then repeat the obtained $\boldsymbol{M'}$ vertically for $L_1$ times. The matrix formed in this way (consisting of $n \cdot L_1$ rows and $o \cdot L_2$ columns) is defined as the $L_1 \times L_2$-th continuation matrix of the matrix $\boldsymbol{M}$, which is denoted as $\boldsymbol{M}^{L_1\times L_2}$. If $\boldsymbol{M}$ is a binary matrix, the set of maximum supporting elements of $\boldsymbol{M}^{L_1\times L_2}$ is denoted as $S_{\boldsymbol{M}_{sp}^{L_1\times L_2}}^{*}$, and the corresponding maximum reachable supporting number of $\boldsymbol{M}^{L_1\times L_2}$ is denoted as $U^*_{\boldsymbol{M}_{sp}^{L_1\times L_2}}$.

**Definition IX.8: One time end-to-end delay of an information block on a primary path**

On a primary path $P_{[i]}$ $(i=1,2)$, the total number of time beats $D^{ed}_{S_{P_{[i]}}} \triangleq b_e - b_b$ needed from the starting time instant of time beat $b_b$ an information block entering the source node to the starting time instant of time beat $b_e$ it finally reaching the destination node is defined as the one-time end-to-end delay of the information block on $P_{[i]}$.

## X. Some related basic properties of throughput of a primary path and throughput of a pair of paths

In this subsection, we establish the existence theorem for the upper bound of the asymptotic throughput of a primary path (***Theorem X.1***); propose an algorithm for the volatile transmission of information blocks based on the intrinsic period of a primary path (***Algorithm X.1***); prove the achievability theorem for the upper bound of the asymptotic throughput of a primary path (***Theorem X.2***); prove the theorem on the unique universal optimal partition of the set of sending nodes (***Theorem X.3***); establish the existence theorems for the upper bound of the asymptotic joint throughput of a pair of paths (***Theorem X.4*** and ***Corollary X.4.1***); for equal opportunity traversals, we put forward a cooperative volatile transmission algorithm of information blocks over a pair of paths based on the maximum supporting element set (***Algorithm X.2***), and prove that this algorithm attains the minimal reachable joint period of the pair of paths (***Lemma X.5.3***); We prove the achievability theorem for the upper bound of the asymptotic joint throughput of a pair of paths under equal opportunity traversals (***Theorem X.5***); develop a cooperative volatile transmission algorithm of information blocks over a pair of paths based on the maximum supporting element set (***Algorithm X.3***), and verify that **Algorithm X.3** attains the minimal reachable joint period of a pair of paths (**Theorem X.6**); provide the asymptotic joint throughput of a pair of paths achievable by **Algorithm X.3** (**Corollary X.6.1**) together with its upper bound (**Corollary X.6.2**). Finally, based on a systematic review and summary of the entire paper, we propose a more general algorithmic procedure to maximize the end-to-end asymptotic joint throughput (**Algorithm X.4**). This procedure fully encompasses all factors that may influence the asymptotic joint throughput of a pair of paths, including routing selection, selection of the reachable periods of primary paths, the traversal counts of primary paths within a joint period, and the allocation strategy of time beats.

### *A*. Existence and achievability of the upper bound on the asymptotic throughput of a primary path

Let the duration of one time beat be $\Delta t$ (unit: second). To facilitate the analysis of the properties of the asymptotic (joint) throughput of a primary path and/or a pair of paths, we first define several rules related to information block transmission as follows:

**Rule X.1: Rules for information block transmission over a pair of paths.**

1) **An information block is the fundamental unit processed by each sending node in a pair of paths, and all information blocks carry an equal amount of information;**
2) **For any sending node on a pair of paths, the time required to finish one-time transmission of an information block equals $\Delta t$. In other words, this transmission takes place over the interval $[t, t+\Delta t)$ beginning at time $t$;**
3) **The one-time transmission of an information block from any sending node on a pair of paths is not guaranteed to be effective. Furthermore, each sending node on a pair of paths can conduct at most one effective one-time transmission for a given information block;**
4) **The absolute value of the difference between the start times of any two distinct one-time transmissions from any sending node on a pair of paths shall be no less than $\Delta t$;**
5) **The absolute value of the difference between the start times of any two distinct one-time transmissions corresponding respectively to any two sending nodes with an interference relationship on a pair of paths shall be no less than $\Delta t$;**
6) **Any information block can only originate from the source node of one primary path within a pair of paths, traverse all sending nodes of this primary path hop-by-hop, and finally be transmitted to the destination node of this primary path.**

**Note:**

1) Based on the first of the above rules, we adopt the number of information blocks as one of the primary metrics for measuring the (joint) throughput of a primary path and/or a pair of paths in what follows;
2) According to the second rule above, the transmission rate of information blocks is identical for all sending nodes on a pair of paths;
3) Based on the third of the above rules, the one-time transmission of an information block by a sending node may be ineffective, i.e., an error occurs. If this happens, the common countermeasure is to retransmit the information block. Furthermore, an information block has at most one time interval for effective one-time transmission at any sending node on a pair of paths;

4) Based on the fourth of the above rules, any sending node on a pair of paths can process at most one one-time transmission of an information block at the same time instant;
5) According to the fifth rule above, any two sending nodes with an interference relationship on a pair of paths cannot be in sending state at the same time instant;
6) Based on the sixth of the above rules, an information block can only be transmitted from its source node to its destination node along one primary path within a pair of paths. That is, cross-primary-path transmission of information blocks is not considered here. Furthermore, combined with the third of the above rules, any information block successfully received at the destination node of a certain primary path must have been successfully transmitted once and only once by each sending node of this primary path in sequence.

**Definition X.1: Information block transmission time trace matrix of a primary path**

Let the number of sending nodes of a primary path $P_{[i]}$ $(i = 1,2)$ be $N_{[i]}$. Let $M_{[i]}$ $(\in \mathbb{Z}^+)$ denote the total number of information blocks that have been successfully received by the destination node of primary path $P_{[i]}$ (i.e., node $n_{i,N_{[i]}+1}$) within the time interval $D_{t,b}^{int}$ (**Definition IX.1**). We assign corresponding sequence numbers to these information blocks in ascending order, namely information block 1, …, information block $M_{[i]}$. For these $M_{[i]}$ information blocks, we define the information block transmission time trace matrix $\boldsymbol{T}_{IB,i}(t,b)$ of $N_{[i]} \times M_{[i]}$, which records the start time instants of effective transmission of each information block at each sending node of $P_{[i]}$ within the time interval $D_{t,b}^{int}$. The specific form of $\boldsymbol{T}_{IB,i}(t,b)$ is given as follows:

$$\boldsymbol{T}_{IB,i}(t,b) \triangleq \begin{pmatrix} t_{i_(1,1)} & t_{i_(1,2)} & \cdots\cdots & t_{i_(1,M_{[i]})} \\ \cdots\cdots & \cdots\cdots\cdots & \cdots\cdots & \cdots\cdots\cdots \\ t_{i_(k,1)} & t_{i_(k,2)} & \cdots\cdots & t_{i_(k,M_{[i]})} \\ \cdots\cdots & \cdots\cdots\cdots & \cdots\cdots & \cdots\cdots\cdots \\ t_{i_(N_{[i]},1)} & t_{i_(N_{[i]},2)} & \cdots\cdots & t_{i_(N_{[i]},M_{[i]})} \end{pmatrix} \tag{X.1}$$

where element $t_{i_(k,l)}$ $(\in \mathbb{R}^+, 1 \le k \le N_{[i]}, 1 \le l \le M_{[i]})$ denotes the start time instant of the corresponding time beat at which information block $l$ $(1 \le l \le M_{[i]})$ is effectively transmitted by sending node $n_{i,k}$ $(1 \le k \le N_{[i]})$, with the corresponding time interval for transmitting the information block $l$ $(1 \le l \le M_{[i]})$ being $[t_{i_(k,l)}, t_{i_(k,l)} + \Delta t) \subseteq D_{t,b}^{int}$ $(1 \le k \le N_{[i]}, 1 \le l \le M_{[i]})$.

**Note:**

1) Based on **Rules X.1** Item 3 and Item 6, given the $M_{[i]}\ (\in \mathbb{Z}^+)$ information blocks successfully received by the destination node of the primary path $P_{[i]}$ within the time interval $D_{t,b}^{int}$, the matrix $\boldsymbol{T}_{IB,i}(t,b)$ can be uniquely determined.

**Lemma X.1.1: Suppose the intrinsic interference intensity of a primary path $\boldsymbol{P}_{[i]}\ (\boldsymbol{i}=\mathbf{1},\mathbf{2})$, which contains $\boldsymbol{N}_{[i]}$ sending nodes, is $\boldsymbol{I}^*_{S_{P_{[i]}}}\ (>\mathbf{1})$. Let $\boldsymbol{M}_{[i]}\ (\in \mathbb{Z}^+)$ denote the total number of information blocks successfully received by the destination node of the primary path $\boldsymbol{P}_{[i]}$ within the time interval $\boldsymbol{D}_{t,b}^{int}$, and let the corresponding information block transmission time trace matrix be $\boldsymbol{T}_{IB,i}(\boldsymbol{t},\boldsymbol{b})$. Then among the $\boldsymbol{N}_{[i]}\times\boldsymbol{M}_{[i]}$ matrix elements contained in $\boldsymbol{T}_{IB,i}(\boldsymbol{t},\boldsymbol{b})$, there exist at least $\boldsymbol{M}_{[i]}\cdot\boldsymbol{I}^*_{S_{P_{[i]}}}$ elements such that the corresponding time beats used to transmit information blocks are pairwise non-overlapping.**

**Proof:**

Arbitrarily select $I^*_{S_{P_{[i]}}}$ sending nodes from the set of sending nodes of $P_{[i]}$ that form a maximum second-order interference subset in $P_{[i]}$. Without loss of generality, assume that the selected $I^*_{S_{P_{[i]}}}$ nodes constitute a maximum second-order interference subset $S'_i$ in $P_{[i]}$, namely

$$S'_i \triangleq \left\{n_{i,j_k} \middle| n_{i,j_k} \in S_{P_{[i]}}, 1 \le j_k \le N_{[i]}, 1 \le k \le I^*_{S_{P_{[i]}}}\right\}$$

(X.2)

For the convenience of recording the sequence numbers of nodes in set $S'_i$, we further define set $S_{I,i}$

$$S_{I,i} \triangleq \left\{j_1, \dots, j_{I^*_{S_{P_{[i]}}}}\right\}$$

(X.3)

We first prove the following: in matrix $\boldsymbol{T}_{IB,i}(t,b)$, if $k \ne m\ (k,m \in S_{I,i}, 1 \le l,o \le M_{[i]})$ or $l \ne o\ (k,m \in S_{I,i}, 1 \le l,o \le M_{[i]})$, then $|t_{i_(k,l)} - t_{i_(m,o)}| \ge \Delta t$ holds. The proof is divided into the following three cases:

**Case 1:** $k \neq m\ (k, m \in S_{I,i})$ and $l \neq o\ (1 \leq l, o \leq M_{[i]}, M_{[i]} \geq 2)$

We adopt proof by contradiction. Both node $n_{i,k}$ and node $n_{i,m}$ belong to the maximum second-order interference subset $S'_i$, so an interference relationship exists between them. If $|t_{i_(k,l)} - t_{i_(m,o)}| < \Delta t$, this implies that the time intervals during which node $n_{i,k}$ and node $n_{i,m}$ are in transmission state (transmitting different information blocks) overlap, which contradicts item 5 of **Rule X.1**. Therefore, $|t_{i_(k,l)} - t_{i_(m,o)}| \geq \Delta t$ holds.

**Case 2:** $k \neq m\ (k, m \in S_{I,i})$ and $l = o\ (1 \leq l, o \leq M_{[i]})$

We use proof by contradiction. Similar to **Case 1**, if $|t_{i_(k,l)} - t_{i_(m,o)}| < \Delta t$, it indicates that the time intervals during which node $n_{i,k}$ and node $n_{i,m}$ are in transmission state (transmitting the same information block) overlap, which contradicts item 5 of **Rule X.1**. Therefore, $|t_{i_(k,l)} - t_{i_(m,o)}| \geq \Delta t$ holds.

**Case 3:** $k = m\ (k, m \in S_{I,i})$ and $l \neq o\ (1 \leq l, o \leq M_{[i]}, M_{[i]} \geq 2)$

Based on item 4 of **Rule X.1**, it follows that $|t_{i_(k,l)} - t_{i_(m,o)}| \geq \Delta t$ holds.

Based on the above analysis, we can further draw the following conclusion: from the corresponding elements in matrix $\boldsymbol{T}_{IB,i}(t,b)$ (all $I^*_{SP_{[i]}} \cdot M_{[i]}$ elements in the $I^*_{SP_{[i]}}$ rows of $\boldsymbol{T}_{IB,i}(t,b)$ corresponding to set $S_{I,i}$, i.e., $t_{i_(k,l)}$ $(k \in S_{I,i}, 1 \leq l \leq M_{[i]})$), we can obtain at least $I^*_{SP_{[i]}} \cdot M_{[i]}$ pairwise non-overlapping time intervals for transmitting the corresponding information blocks (i.e., $[t_{i_(k,l)}, t_{i_(k,l)} + \Delta t)$ $(k \in S_{I,i}, 1 \leq l \leq M_{[i]})$). ■

**Lemma X.1.2: Suppose the intrinsic interference intensity of a primary path $\boldsymbol{P}_{[i]}$ $(\boldsymbol{i} = \boldsymbol{1}, \boldsymbol{2})$, which contains $\boldsymbol{N}_{[i]}$ sending nodes, is $\boldsymbol{I}^*_{\boldsymbol{SP}_{[i]}}$ $(= \boldsymbol{1})$. Let $\boldsymbol{M}_{[i]}$ $(\in \mathbb{Z}^+, > \boldsymbol{1})$ denote the total number of information blocks successfully received by the destination node of the primary path $\boldsymbol{P}_{[i]}$ within the time interval $\boldsymbol{D}^{int}_{t,b}$, and let the corresponding information block transmission time trace matrix be $\boldsymbol{T}_{IB,i}(\boldsymbol{t}, \boldsymbol{b})$. Then among the $\boldsymbol{N}_{[i]} \times \boldsymbol{M}_{[i]}$ matrix elements contained in $\boldsymbol{T}_{IB,i}(\boldsymbol{t}, \boldsymbol{b})$, there exist at least $\boldsymbol{M}_{[i]}$ elements such that the corresponding time** **beats used to transmit information blocks are pairwise non-overlapping.**

**Proof:**

In this case, all sending nodes in the primary path $P_{[i]}$ form a second-order concurrency set of nodes. From item 4 of **Rule X.1**, it follows that in matrix $\boldsymbol{T}_{IB,i}(t,b)$, if $l \neq o$ $(1 \leq k \leq N_{[i]}, 1 \leq l, o \leq M_{[i]})$, then $|t_{i_(k,l)} - t_{i_(k,o)}| \geq \Delta t$ holds. Furthermore, we conclude that at least $M_{[i]}$ pairwise non-overlapping time intervals (i.e., $[t_{i_(k,l)}, t_{i_(k,l)} + \Delta t)$ $(1 \leq l \leq M_{[i]})$) for transmitting the corresponding information blocks can be obtained from the relevant elements in matrix $\boldsymbol{T}_{IB,i}(t,b)$ (all $M_{[i]}$ elements in any row $k$ $(1 \leq k \leq N_{[i]})$ of $\boldsymbol{T}_{IB,i}(t,b)$, namely $t_{i_(k,l)}$ $(1 \leq l \leq M_{[i]})$). ■

**Theorem X.1: Existence theorem for the upper bound of the asymptotic throughput of a primary path.**

**If the intrinsic interference intensity of a primary path $\boldsymbol{P}_{[i]}$ $(\boldsymbol{i} = \mathbf{1}, \mathbf{2})$ is $\boldsymbol{I}^*_{S_{P_{[i]}}}$, then the asymptotic throughput of $\boldsymbol{P}_{[i]}$ satisfies the relation**

$$\boldsymbol{R}^{\infty}_{S_{P_{[i]}}} \leq \frac{\mathbf{1}}{\boldsymbol{I}^*_{S_{P_{[i]}}}} \quad (\boldsymbol{i} = \mathbf{1}, \mathbf{2})$$

**(X.4)**

**Proof:**

Let the primary path $P_{[i]}$ consists of $N_{[i]}$ sending nodes, whose sequence numbers are 1, 2, …, and $N_{[i]}$. $M_{[i]}$ $(\in \mathbb{Z}^+)$ denotes the total number of information blocks successfully received by the destination node of the primary path $P_{[i]}$ within time interval $D^{int}_{t,b}$, and let the corresponding information block transmission time trace matrix be $\boldsymbol{T}_{IB,i}(t,b)$.

The proof proceeds by considering two cases: the case of $I^*_{S_{P_{[i]}}} > 1$ and the case of $I^*_{S_{P_{[i]}}} = 1$.

**Case 1:** $I^*_{S_{P_{[i]}}} > 1$

From **Lemma X.1.1**, the total number of time beats $b$ consumed to transmit the $M_{[i]}$ information blocks received by the destination node of the primary path satisfies the relation

$$b \geq I^*_{S_{P_{[i]}}} \cdot M_{[i]}$$

(X.5)

Therefore, the following relation holds

$$\begin{aligned}\frac{M_{[i]}}{b} &\leq \frac{M_{[i]}}{I^*_{S_{P_{[i]}}} \cdot M_{[i]}} \\ &= \frac{1}{I^*_{S_{P_{[i]}}}}\end{aligned} \tag{X.6}$$

As the number $M_{[i]}$ of information blocks received by the destination node increases continuously, the total number of time beats $b$ consumed for transmitting information blocks increases accordingly, and the above inequality always holds. According to the definition of the asymptotic throughput of a primary path (**Definition IX.4**), to obtain the asymptotic throughput of the primary path $P_{[i]}$, we need to take the limit of the total number of time beats $b$ contained in time interval $D^{int}_{t,b}$. From **Equation X.6**, it can be obtained that

$$\begin{aligned}R^{\infty}_{S_{P_{[i]}}} &\triangleq \lim_{b\to\infty}\left(\frac{M_{[i]}}{b}\right) \\ &\leq \frac{1}{I^*_{S_{P_{[i]}}}}\end{aligned} \tag{X.7}$$

Therefore, the proposition holds in this case.

**Case 2:** $I^*_{S_{P_{[i]}}} = 1$

We consider the following two cases:

**Case 2.1:** $M_{[i]} \geq 2$

From **Lemma X.1.2**, the total number of time beats $b$ consumed to transmit the $M_{[i]}$ information blocks received by the destination node of the primary path satisfies the relation

$$b \geq M_{[i]} \tag{X.8}$$

**Case 2.2:** $M_{[i]} = 1$

Obviously, the foregoing conclusion, namely "the total number of time beats $b$ consumed to transmit $M_{[i]}$ $(= 1)$ information block received by the destination node of the primary path is no less than $M_{[i]}$", still holds in this case.

Therefore, combining the above two cases (i.e., **Case 2.1** and **Case 2.2**), the following relation holds

$$\frac{M_{[i]}}{b} \leq \frac{M_{[i]}}{M_{[i]}} = 1$$

(X.9)

As the number $M_{[i]}$ of information blocks received by the destination node increases continuously, the total number of time beats $b$ consumed for transmitting information blocks increases accordingly, and the above inequality always holds. According to the definition of the asymptotic throughput of a primary path (**Definition IX.4**), to obtain the asymptotic throughput of the primary path $P_{[i]}$, we need to take the limit of the total number of time beats $b$ contained in time interval $D_{t,b}^{int}$. From **Equation X.9**, it can be obtained that

$$\begin{aligned} R_{SP_{[i]}}^{\infty} &\triangleq \lim_{b\to\infty}\left(\frac{M_{[i]}}{b}\right) \\ &\leq 1 \\ &= \frac{1}{I_{SP_{[i]}}^{*}} \end{aligned}$$

(X.10)

Hence, the proposition holds for this case. ■

**Note:**

1) The physical implication of this theorem is clear: for a primary path with an intrinsic interference intensity of $I$, its maximum average rate at which the destination node receives information blocks originating from the source node equals to one information block per $I$ time beats.

**Algorithm X.1: The algorithm for the volatile transmission of information blocks based on the intrinsic period of a primary path.**

| | |
|---|---|
| 1: | BEGIN |
| 2: | Set the intrinsic period of a primary path $P_{[i]}$ $(i = 1,2)$ as $T_{SP_{[i]}}^{*}$, the time beat for $P_{[i]}$ |

starting the transmission of the first information block is regarded as the “starting beat”, and the corresponding sequence number of the “starting beat” is set to be 1;

3: WHILE (the sequence number of the beat $\leq$ the upper limit of the sequence number of the time beat) DO

4: IF (the sequence number of the beat = $1 + k \cdot T^*_{S_{P[i]}}$ $(k = 0,1,2,\ldots)$) THEN

5: Activate all the nodes in the equally spaced second-order concurrency subsets of nodes $\Phi'_{S_{P[i]}}(1, T^*_{S_{P[i]}})$. Meanwhile, all nodes in the remaining equally spaced second-order concurrency subsets are kept silent;

6: For node 1 in $\Phi'_{S_{P[i]}}(1, T^*_{S_{P[i]}})$: gets a new information block from the information source (here, it is assumed that there is always an information block waiting to be transmitted at the information source) and sends it to its adjacent downstream node within the current beat. For all the other nodes in $\Phi'_{S_{P[i]}}(1, T^*_{S_{P[i]}})$: send the information block (if any) received from their upstream nodes in the previous beat to their adjacent downstream nodes in the current beat;

7: ELSE IF $(T^*_{S_{P[i]}} > 1)$

//Note: in this case, the sequence number of the beat is $j + k \cdot T^*_{S_{P[i]}}$ $(1 < j \leq T^*_{S_{P[i]}}, k = 0,1,2,\ldots)$

8: Activate all the nodes in the equally spaced second-order concurrency subsets of nodes $\Phi'_{S_{P[i]}}(j, T^*_{S_{P[i]}})$. Meanwhile, all nodes in the remaining equally spaced second-order concurrency subsets are kept silent;

9: For all the nodes in $\Phi'_{S_{P[i]}}(j, T^*_{S_{P[i]}})$: send the information block (if any) received from their upstream nodes in the previous beat to their adjacent downstream nodes in the current beat;

10: END IF

11: Increase the sequence number of the beat by one;

12: END WHILE

13: END

**Note:**

1) The core operation of this algorithm is to cyclically activate $T^*_{S_{P_{[i]}}}$ equally spaced second-order concurrency subsets of nodes with $T^*_{S_{P_{[i]}}}$ as the period, where only one such set is activated in each time beat;
2) During algorithm execution, a sending node is permitted to transmit only one information block within a single time beat in which it is eligible for transmission;
3) It can be known from the concurrent transmission theorem (**Theorem VI.4**) that the concurrent transmission performed in each time beat by this algorithm is effective;
4) By substituting all intrinsic periods $T^*_{S_{P_{[i]}}}$ in the foregoing algorithm description with any reachable period $T_{S_{P_{[i]}}}$, the algorithm can be extended to a more general algorithm, i.e. "the algorithm for the volatile transmission of information blocks based on a reachable period of a primary path".

**Theorem X.2: Reachability theorem for the upper bound of the asymptotic throughput of a primary path.**

**Given that a primary path $P_{[i]}$ $(i = 1, 2)$ has an intrinsic interference intensity of $I^*_{S_{P_{[i]}}}$ and $I^*_{S_{P_{[i]}}} = T^*_{S_{P_{[i]}}}$ holds, the upper asymptotic throughput bound of $\frac{1}{I^*_{S_{P_{[i]}}}}$ can be achieved.**

**Proof:**

**Algorithm X.1** can be adopted. Under the condition that the source node always has pending information blocks for transmission and the system operates in steady state, it can be readily deduced according to the definition of the asymptotic throughput of a primary path (**Definition IX.4**) that the destination node of primary path $P_{[i]}$ achieves the performance of receiving one information block every $T^*_{S_{P_{[i]}}}$ $(= I^*_{S_{P_{[i]}}})$ time beats. In this scenario, the asymptotic throughput of $P_{[i]}$ is $R^{\infty}_{S_{P_{[i]}}} = \frac{1}{T^*_{S_{P_{[i]}}}} = \frac{1}{I^*_{S_{P_{[i]}}}}$. ■

**Theorem X.3: Unique universal optimal partition theorem of a set of sending nodes.**

**If the intrinsic interference intensity of a primary path $P_{[i]}$ $(i = 1, 2)$ is $I^*_{S_{P_{[i]}}}$ (and $I^*_{S_{P_{[i]}}} = T^*_{S_{P_{[i]}}}$ holds). Under this circumstance, the unique universal optimal partitioning scheme to divide the set of sending nodes $S_{P_{[i]}}$ of $P_{[i]}$ into $I^*_{S_{P_{[i]}}}$ second-order concurrency subsets is to partition the set of sending nodes $S_{P_{[i]}}$ into $I^*_{S_{P_{[i]}}}$ equally spaced second-order concurrency subsets, namely the subsets of equally spaced nodes $\Phi'_{S_{P_{[i]}}}(j, I^*_{S_{P_{[i]}}})$ $(1 \le j \le I^*_{S_{P_{[i]}}})$.**

**Proof:**

From **Theorem VIII.1**, if "the interference decreasing rule among nodes on a primary path" (**Rule IV.1**) is satisfied, then $I^*_{S_{P_{[i]}}} = T^*_{S_{P_{[i]}}}$ holds (note that we do not rule out the scenario where $I^*_{S_{P_{[i]}}} = T^*_{S_{P_{[i]}}}$ still holds even if **Rule IV.1** is not satisfied). Since **Rule IV.1** naturally and universally holds on practical effective multi-hop information transmission paths, the case of $I^*_{S_{P_{[i]}}} = T^*_{S_{P_{[i]}}}$ also occurs widely. According to the definition of the intrinsic period of a primary path (**Definition VII.4**), the set of sending nodes $S_{P_{[i]}}$ of $P_{[i]}$ can be partitioned into $I^*_{S_{P_{[i]}}}$ second-order concurrency subsets of equally spaced nodes under circumstances of $I^*_{S_{P_{[i]}}} = T^*_{S_{P_{[i]}}}$. Therefore, such a partition scheme is universal. Furthermore, considering that the intrinsic interference intensity of $P_{[i]}$ is $I^*_{S_{P_{[i]}}}$, such a partitioning scheme also minimizes the number of second-order concurrency subsets when partitioning the set of sending nodes $S_{P_{[i]}}$ of $P_{[i]}$, i.e., the total number of partitioned second-order concurrency subsets equals $I^*_{S_{P_{[i]}}}$. Accordingly, this partitioning scheme constitutes the universal and optimal partition for dividing the set of sending nodes of $P_{[i]}$ into the minimum number of second-order concurrency subsets.

It is hereafter proven that under scenarios corresponding to $I^*_{S_{P_{[i]}}} = T^*_{S_{P_{[i]}}}$, the above partitioning method is the unique universal optimal partition that can split the set of sending nodes $S_{P_{[i]}}$ of $P_{[i]}$ into $I^*_{S_{P_{[i]}}}$ second-order concurrency subsets, which is the minimal achievable quantity. For this purpose, it is only necessary to construct a specific case where the intrinsic interference intensity is $I^*_{S_{P_{[i]}}}$ $(= T^*_{S_{P_{[i]}}})$, and to prove that in that case there is only one way (i.e. the partition $\Phi'_{S_{P_{[i]}}}(j, I^*_{S_{P_{[i]}}})$ $(1 \le j \le I^*_{S_{P_{[i]}}})$ which is proposed in the proposition) of dividing the sending node set $S_{P_{[i]}}$ into $I^*_{S_{P_{[i]}}}$ second-order concurrency subsets. Suppose that $P_{[i]}$ consists of $N_{S_{P_{[i]}}}$ sending nodes, whose sequence numbers are 1, 2,..., $N_{S_{P_{[i]}}}$ in turn. Hereinafter, for convenience, nodes are directly named by their sequence numbers (e.g., the node with sequence number $i$ is referred to as "node $i$"). Furthermore, apart from the condition that the intrinsic interference intensity of $P_{[i]}$ is equal to $I^*_{S_{P_{[i]}}}$ $(= T^*_{S_{P_{[i]}}})$, let us consider the specific case that any two sending nodes $k$ and $l\ (k \ne l, |k-l| < I^*_{S_{P_{[i]}}})$ in $S_{P_{[i]}}$ have interference relationship between them (**Definition IV.3**).

First, in the following equation a matrix $\boldsymbol{M}$ is constructed:

$$\boldsymbol{M} \triangleq \begin{pmatrix} 1 & \cdots\cdots & 1 + kI^*_{S_{P_{[i]}}} & \cdots\cdots & 1 + \left\lfloor \frac{N_{S_{P_{[i]}}}}{I^*_{S_{P_{[i]}}}} \right\rfloor \cdot I^*_{S_{P_{[i]}}} \\ \cdots\cdots & \cdots\cdots & \cdots\cdots & \cdots\cdots & \cdots\cdots \\ j & \cdots\cdots & j + kI^*_{S_{P_{[i]}}} & \cdots\cdots & j + \left\lfloor \frac{N_{S_{P_{[i]}}}}{I^*_{S_{P_{[i]}}}} \right\rfloor \cdot I^*_{S_{P_{[i]}}} \\ \cdots\cdots & \cdots\cdots & \cdots\cdots & \cdots\cdots & \times\times\times\times \\ I^*_{S_{P_{[i]}}} & \cdots\cdots & I^*_{S_{P_{[i]}}} + kI^*_{S_{P_{[i]}}} & \cdots\cdots & \times\times\times\times \end{pmatrix} \tag{X.11}$$

1) The matrix has $I^*_{S_{P_{[i]}}}$ rows and $\left\lceil \frac{N_{S_{P_{[i]}}}}{I^*_{S_{P_{[i]}}}} \right\rceil$ columns;

2) In ascending order, put the sequence number of the node $j\ (1 \le j \le N_{S_{P_{[i]}}})$ in $P_{[i]}$ into the corresponding unit in row $\left(j - I^*_{S_{P_{[i]}}} \cdot \left\lfloor \frac{j}{I^*_{S_{P_{[i]}}}} \right\rfloor\right)$ and column $\left\lceil \frac{j}{I^*_{S_{P_{[i]}}}} \right\rceil$ of the matrix one by one. That is, set the value of the unit in row $\left(j - I^*_{S_{P_{[i]}}} \cdot \left\lfloor \frac{j}{I^*_{S_{P_{[i]}}}} \right\rfloor\right)$ and column $\left\lceil \frac{j}{I^*_{S_{P_{[i]}}}} \right\rceil$ of $\boldsymbol{M}$ to be $j$;

3) Ignore the settings of matrix units that may not be filled by sequence numbers of nodes.

Next, consider the grouping of the nodes with their sequence numbers $n\ (1 \le n \le I^*_{S_{P_{[i]}}})$.

According to the definition of a second-order interference subset (**Definition V.10**), these nodes constitute a second-order interference subset in the specific case considered above. Therefore, in order to ensure that each group of nodes is a second-order concurrency set, these nodes can only be divided into $I^*_{S_{P[i]}}$ groups of nodes (that is, each group contains only one of these $I^*_{S_{P[i]}}$ nodes). Without losing generality and for the convenience of discussion, it is assumed that the node $n$ $(1 \leq n \leq I^*_{S_{P[i]}})$ is put into the group $n$ $(1 \leq n \leq I^*_{S_{P[i]}})$ (that is, the group number is the same as the row number corresponding to the node $n$ $(1 \leq n \leq I^*_{S_{P[i]}})$ in the matrix $\boldsymbol{M}$).

In the following, let us consider the grouping of the remaining sending nodes $n$ $(1+I^*_{S_{P[i]}} \leq n \leq N_{S_{P[i]}})$ in the set $S_{P[i]}$. Here, mathematical induction is used to prove that the node $n = j + mI^*_{S_{P[i]}}$ $(1+I^*_{S_{P[i]}} \leq n \leq N_{S_{P[i]}}, 1 \leq j \leq I^*_{S_{P[i]}}, 1 \leq m \leq \left\lfloor \frac{N_{S_{P[i]}}}{I^*_{S_{P[i]}}} \right\rfloor)$ can and can only be divided into the group $j$ (so as to form a second-order concurrency subset). First, let us consider the grouping of the node $1 + I^*_{S_{P[i]}}$ (this node is located in the first row of the matrix $\boldsymbol{M}$). Considering that we want to find out a way to subdivide the sending node set $S_{P[i]}$ into $I^*_{S_{P[i]}}$ second-order concurrency subsets of nodes, and that $I^*_{S_{P[i]}}$ groups of nodes have already been generated in the above discussion, hence, in order not to generate more groups, we need to find out a grouping method to put the node $1 + I^*_{S_{P[i]}}$ into one of the existing $I^*_{S_{P[i]}}$ groups. Since the node $1 + I^*_{S_{P[i]}}$ has interference relationship with nodes $n'$ $(1<n' \leq I^*_{S_{P[i]}})$ whose sequence numbers are less than $1 + I^*_{S_{P[i]}}$, and only has concurrency relationship with the node 1, the node $1 + I^*_{S_{P[i]}}$ can only be put into the group 1, which is the group of node 1. In this way, we prove that the node $1 + I^*_{S_{P[i]}}$ can only be put into the group 1. Further, assuming that the node $n' = j' + m'I^*_{S_{P[i]}}$ $(1+I^*_{S_{P[i]}} \leq n' \leq n-1, 1 \leq j' \leq I^*_{S_{P[i]}}, 1 \leq m' \leq \left\lfloor \frac{N_{S_{P[i]}}}{I^*_{S_{P[i]}}} \right\rfloor)$ can and can only be put into the group $j'$ is true (to form a second-order concurrency subset), we will prove that the node $n = j + mI^*_{S_{P[i]}}$ $(n \leq N_{S_{P[i]}})$ can and can only be put into the group $j$. With $|n - n'| < I^*_{S_{P[i]}}$ $(n - (I^*_{S_{P[i]}} - 1) \leq n' \leq n - 1)$, there is an interference relationship between node $n$ and node $n'$. These $I^*_{S_{P_{[i]}}} - 1$ nodes (i.e. nodes $n'(n - (I^*_{S_{P_{[i]}}} - 1) \leq n' \leq n - 1)$) have already been divided into $I^*_{S_{P_{[i]}}} - 1$ groups, whose sequence numbers (the group number is the row number corresponding to these nodes in the matrix $\boldsymbol{M}$) are not equal to $j$, respectively. Therefore, in order not to generate more groups of nodes, and to ensure that each group of nodes constituting a second-order concurrency set, the node $n$ can only be put into the group $j$ ($j$ is the row number of the node $n$ in the matrix $\boldsymbol{M}$). Based on the premise that the intrinsic interference intensity of the primary path $P_{[i]}$ is $I^*_{S_{P_{[i]}}}$, it is easy to verify that putting the node $n$ into the group $j$ can indeed form a second-order concurrency subset (the proof is omitted here), which proves that the node $n = j + mI^*_{S_{P_{[i]}}}$ can and can only be put into the group $j$.

It can be concluded from the above proof that, under the aforementioned specific case, the partition of the set of sending nodes formed by the equally spaced second-order concurrency subsets corresponding to the rows of matrix $\boldsymbol{M}$ (namely, the nodes corresponding to elements in each row of $\boldsymbol{M}$ constitute one corresponding equally spaced second-order concurrency subset) is the unique partitioning method that can divide the set of sending nodes $S_{P_{[i]}}$ of $P_{[i]}$ into $I^*_{S_{P_{[i]}}}$ second-order concurrency subsets. Thus far, it has been verified that the node partitioning method stated in the proposition yields the unique universal optimal partition. ■

**Note:**

1) A fundamental inherent law governing multi-hop transmission is uncovered by this theorem: the adoption of network oscillation is not one of optional schemes for efficient multi-hop transmission but the unique universal optimal alternative. This conclusion is consistent with various oscillatory phenomena ubiquitous in nature!

## *B*. Existence and achievability of the upper bound on the asymptotic joint throughput of a pair of paths

**Lemma X.4.1: Suppose the intrinsic interference intensity of a pair of paths $\boldsymbol{P_{[1,2]}}$ is $\boldsymbol{I^*_{S_{P_{[1,2]}}}}$ (> 1), and its set of sending nodes is $\boldsymbol{S_{P_{[1,2]}}}$. Let the total numbers of information blocks successfully received within time interval $\boldsymbol{D^{int}_{t,b}}$ by the respective destination nodes of a primary path $\boldsymbol{P_{[1]}}$ (with $\boldsymbol{N_{[1]}}$ sending nodes) and a primary path $\boldsymbol{P_{[2]}}$ (with $\boldsymbol{N_{[2]}}$ sending nodes) contained in the pair of paths $\boldsymbol{P_{[1,2]}}$ be $\boldsymbol{M_{[1]}}$ $(\in \mathbb{Z}^+)$ and $\boldsymbol{M_{[2]}}$ $(\in \mathbb{Z}^+)$, respectively ( $\boldsymbol{b}$ denotes the total number of time beats spanned by $\boldsymbol{D^{int}_{t,b}}$ ). The corresponding information block transmission time trace matrices of these two primary paths are $\boldsymbol{T_{IB,1}(t,b)}$ and $\boldsymbol{T_{IB,2}(t,b)}$, respectively. Let $\boldsymbol{N_{s,1}}$ and $\boldsymbol{N_{s,2}}$ denote the numbers of nodes in a set of nodes $\boldsymbol{s}$ belonging to the primary path $\boldsymbol{P_{[1]}}$ and the primary path $\boldsymbol{P_{[2]}}$, respectively. $\boldsymbol{S^{SIU}_{S_{P_{[1,2]}}}}$ denotes the universal set of second-order interference subsets of the set of sending nodes $\boldsymbol{S_{P_{[1,2]}}}$. Then we have the relation**

$$\boldsymbol{b \geq \max_{s \in S^{SIU}_{S_{P_{[1,2]}}}} \left\{M_{[1]} \cdot N_{s,1} + M_{[2]} \cdot N_{s,2}\right\}} \tag{X.12}$$

**Proof:**

From the hypothesis (i.e., **Equation X.12**), a set of nodes $s$ $(\in S^{SIU}_{S_{P_{[1,2]}}})$ is a second-order interference subset. According to the definition of a second-order interference subset (**Definition V.10**), the total number of nodes in the set is no less than 2, i.e., $N_{s,1} + N_{s,2} \geq 2$. Therefore, we proceed with the proof by considering two cases: if only a single set of nodes $s$ $(\in S^{SIU}_{S_{P_{[1,2]}}})$ is taken into account, there exist at least $M_{[1]} \cdot N_{s,1} + M_{[2]} \cdot N_{s,2}$ mutually non-overlapping time beats for transmitting the corresponding information blocks.

**Case 1:** $s \in S^{SIU}_{S_{P_{[1,2]}}}$ and $N_{s,1} \cdot N_{s,2} > 0$

In this case, $N_{s,1} \geq 1$ and $N_{s,2} \geq 1$ hold. According to **Definition X.1**, we can respectively obtain the corresponding information block transmission time trace matrices $\boldsymbol{T}_{IB,1}(t,b)$ and $\boldsymbol{T}_{IB,2}(t,b)$ of a primary path $P_{[1]}$ and a primary path $P_{[2]}$ within time interval $D^{int}_{t,b}$ (**Definition IX.1**). Let $S'_i$ be the second-order interference subset formed by all nodes belonging to both the primary path $P_{[i]}$ and the set $s$, i.e.,

$$S'_i \triangleq \left\{n_{i,j_k} \middle| n_{i,j_k} \in s, n_{i,j_k} \in S_{P_{[i]}}, 1 \leq j_k \leq N_{[i]}, 1 \leq k \leq N_{s,i}\right\} \ (i = 1,2)$$

(X.13)

For convenience in recording the sequence numbers of nodes within set $S'_i$, we further define set $S_{I,i}$ as

$$S_{I,i} \triangleq \{j_1, \dots, j_{N_{s,i}}\} \ (i = 1,2)$$

(X.14)

Using the same analytical approach as in **Lemma X.1.1**, we can prove (proof omitted): for the same matrix $\boldsymbol{T}_{IB,i}(t,b)\ (i = 1,2)$, if $k \neq m\ (k,m \in S_{I,i}, 1 \leq l,o \leq M_{[i]})$ or $l \neq o\ (k,m \in S_{I,i}, 1 \leq l,o \leq M_{[i]})$, then $|t_{i_(k,l)} - t_{i_(m,o)}| \geq \Delta t$ holds.

We next prove that for two distinct matrices $\boldsymbol{T}_{IB,1}(t,b)$ and $\boldsymbol{T}_{IB,2}(t,b)$, $|t_{1_(k,l)} - t_{2_(m,o)}| \geq \Delta t$ holds provided $(k \in S_{I,1}) \wedge (m \in S_{I,2}) \wedge (1 \leq l \leq M_{[1]}) \wedge (1 \leq o \leq M_{[2]})$. We use proof by contradiction. The sending node $n_{1,k}$ within $P_{[1]}$ and the sending node $n_{2,m}$ within $P_{[2]}$ both belong to the second-order interference subset $s$, so an interference relationship exists between them. If $|t_{1_(k,l)} - t_{2_(m,o)}| < \Delta t$, this implies that the time intervals during which node $n_{1,k}$ and node $n_{2,m}$ are in the sending state overlap, which contradicts item 5 of **Rule X.1**. Consequently, $|t_{1_(k,l)} - t_{2_(m,o)}| \geq \Delta t$ holds.

Based on the above analysis, the following conclusion can be drawn: from the corresponding entries in matrix $\boldsymbol{T}_{IB,1}(t,b)$ and matrix $\boldsymbol{T}_{IB,2}(t,b)$ (partly from all $M_{[1]} \cdot N_{s,1}$ entries in the $N_{s,1}$ rows of $\boldsymbol{T}_{IB,1}(t,b)$ corresponding to set $S_{I,1}$, i.e., $t_{1_(k,l)}\ (k \in S_{I,1}, 1 \leq l \leq M_{[1]})$; and partly from all $M_{[2]} \cdot N_{s,2}$ entries in the $N_{s,2}$ rows of $\boldsymbol{T}_{IB,2}(t,b)$ corresponding to set $S_{I,2}$, i.e., $t_{2_(m,o)}\ (m \in S_{I,2}, 1 \leq o \leq M_{[2]})$), we can obtain at least $M_{[1]} \cdot N_{s,1} + M_{[2]} \cdot N_{s,2}$ pairwise non-overlapping time intervals for transmitting the corresponding information blocks (namely $[t_{1_(k,l)}, t_{1_(k,l)} + \Delta t)\ (k \in S_{I,1}, 1 \leq l \leq M_{[1]})$ and $[t_{2_(m,o)}, t_{2_(m,o)} + \Delta t)\ (m \in S_{I,2}, 1 \leq o \leq M_{[2]})$).

**Case 2:** $s \in S_{SP_{[1,2]}}^{SIU}$ and $N_{s,1} \cdot N_{s,2} = 0$

Since $N_{s,1} + N_{s,2} \geq 2$, only two cases are possible in this scenario: (1) $N_{s,1} \geq 2$ and $N_{s,2} = 0$; (2) $N_{s,2} \geq 2$ and $N_{s,1} = 0$. Without loss of generality, we only need to analyze the case where $N_{s,1} \geq 2$ and $N_{s,2} = 0$ in the following. Using the same analytical approach as in **Lemma X.1.1**, we can prove (proof omitted) that for the same matrix $\boldsymbol{T}_{IB,1}(t,b)$, $|t_{1_(k,l)} - t_{1_(m,o)}| \geq \Delta t$ holds if $k \neq m$ $(k, m \in S_{I,1}, 1 \leq l, o \leq M_{[1]})$ or $l \neq o$ $(k, m \in S_{I,1}, 1 \leq l, o \leq M_{[1]})$.

Based on the above analysis, we conclude that from merely the corresponding entries in matrix $\boldsymbol{T}_{IB,1}(t,b)$ (all $M_{[1]} \cdot N_{s,1}$ entries in the $N_{s,1}$ rows of $\boldsymbol{T}_{IB,1}(t,b)$ corresponding to set $S_{I,1}$, i.e., $t_{1_(k,l)}$ $(k \in S_{I,1}, 1 \leq l \leq M_{[1]})$ ), we can obtain at least $M_{[1]} \cdot N_{s,1} = M_{[1]} \cdot N_{s,1} + M_{[2]} \cdot N_{s,2}$ $(N_{s,2} = 0)$ pairwise non-overlapping time intervals for transmitting the corresponding information blocks (namely $[t_{1_(k,l)}, t_{1_(k,l)} + \Delta t)$ $(k \in S_{I,1}, 1 \leq l \leq M_{[1]})$).

Combining the above analyses of **Case 1** and **Case 2**, it can be seen that for any second-order interference subset $s$ in $S_{S_{P_{[1,2]}}}^{SIU}$, at least $M_{[1]} \cdot N_{s,1} + M_{[2]} \cdot N_{s,2}$ pairwise non-overlapping time intervals for transmitting the corresponding information blocks can be obtained. That is, for any $s \in S_{S_{P_{[1,2]}}}^{SIU}$, the following relation holds:

$$b \geq M_{[1]} \cdot N_{s,1} + M_{[2]} \cdot N_{s,2} \tag{X.15}$$

It readily follows that

$$b \geq \max_{s \in S_{S_{P_{[1,2]}}}^{SIU}} \left\{ M_{[1]} \cdot N_{s,1} + M_{[2]} \cdot N_{s,2} \right\} \tag{X.16}$$

■

**Lemma X.4.2: Suppose the intrinsic interference intensity of a pair of paths $\boldsymbol{P}_{[1,2]}$ is $\boldsymbol{I}^{*}_{S_{P_{[1,2]}}}$ $(> 1)$, and its set of sending nodes is $\boldsymbol{S}_{P_{[1,2]}}$. Let the total numbers of information blocks successfully received within time interval $\boldsymbol{D}_{t,b}^{int}$ by the respective destination nodes of a primary path $\boldsymbol{P}_{[1]}$ (with $\boldsymbol{N}_{[1]}$ sending nodes) and a primary path $\boldsymbol{P}_{[2]}$ (with $\boldsymbol{N}_{[2]}$ sending nodes) contained in the pair of paths $\boldsymbol{P}_{[1,2]}$ be $\boldsymbol{M}_{[1]}$ $(\in \mathbb{Z}^{+})$ and $\boldsymbol{M}_{[2]}$ $(\in \mathbb{Z}^{+})$, respectively ( $\boldsymbol{b}$ denotes the total number of time beats spanned by $\boldsymbol{D}_{t,b}^{int}$ ). The corresponding information block transmission time trace matrices of these two primary paths are $\boldsymbol{T}_{IB,1}(\boldsymbol{t},\boldsymbol{b})$ and $\boldsymbol{T}_{IB,2}(\boldsymbol{t},\boldsymbol{b})$, respectively. Let $\boldsymbol{N}_{s,1}$ and $\boldsymbol{N}_{s,2}$ denote the numbers of nodes in a set of nodes $\boldsymbol{s}$ belonging to the primary path $\boldsymbol{P}_{[1]}$ and the primary path $\boldsymbol{P}_{[2]}$,** **respectively.** $\boldsymbol{S_{S_{P_{[1,2]}}}^{SSIU}}$ **denotes the universal set of saturated second-order interference subsets of the set of sending nodes** $\boldsymbol{S_{P_{[1,2]}}}$**. Then we have the relation**

$$\boldsymbol{b \geq \max_{s \in S_{S_{P_{[1,2]}}}^{SSIU}} \left\{M_{[1]} \cdot N_{s,1} + M_{[2]} \cdot N_{s,2}\right\}}$$

**(X.17)**

**Proof:**

For convenience, let

$$\alpha \triangleq \max_{s \in S_{S_{P_{[1,2]}}}^{SIU}} \left\{M_{[1]} \cdot N_{s,1} + M_{[2]} \cdot N_{s,2}\right\}$$

(X.18.1)

$$\beta \triangleq \max_{s \in S_{S_{P_{[1,2]}}}^{SSIU}} \left\{M_{[1]} \cdot N_{s,1} + M_{[2]} \cdot N_{s,2}\right\}$$

(X.18.2)

From **Definition V.30**, we have $S_{S_{P_{[1,2]}}}^{SSIU} \subseteq S_{S_{P_{[1,2]}}}^{SIU}$. Therefore, the following relation holds from the above two equations

$$\beta \leq \alpha$$

(X.19)

We now prove that $\beta < \alpha$ does not hold. The proof proceeds by contradiction. If $\beta < \alpha$ holds, there exists at least one second-order interference subset $s \in S_{S_{P_{[1,2]}}}^{SIU}$ and $s \notin S_{S_{P_{[1,2]}}}^{SSIU}$ such that the following equation holds

$$M_{[1]} \cdot N_{s,1} + M_{[2]} \cdot N_{s,2} = \alpha > \beta$$

(X.20)

From **Theorem VI.6**, since $s$ is not a saturated second-order interference subset, there necessarily exists a second-order interference subset $s'$ in the set $S_{P_{[1,2]}}$ of sending nodes of the pair of paths that contains $s$ and is not equal to $s$, i.e., there exist $s' \in S_{P_{[1,2]}}$ and $s' \in S_{S_{P_{[1,2]}}}^{SIU}$ such that $s \subset s'$. It then follows that

$$M_{[1]} \cdot N_{s',1} + M_{[2]} \cdot N_{s',2} > M_{[1]} \cdot N_{s,1} + M_{[2]} \cdot N_{s,2} = \alpha$$

(X.21)

This contradicts Equation X.18.1.

Since $\beta < \alpha$ does not hold, combining **Equation X.19**, $\beta = \alpha$ must hold. Further, together with **Lemma X.4.1**, the desired result follows. ■

**Theorem X.4: Existence theorem for the upper bound of asymptotic joint throughput of a pair of paths.**

**Suppose the intrinsic interference intensity of a pair of paths $P_{[1,2]}$ is $I^*_{S_{P_{[1,2]}}}$, and its set of sending nodes is $S_{P_{[1,2]}}$. Let the asymptotic (joint) throughput of the pair of paths $P_{[1,2]}$, a primary path $P_{[1]}$ and a primary path $P_{[2]}$ contained therein be $R^{\infty}_{S_{P_{[1,2]}}}$, $R^{\infty}_{S_{P_{[1]}}}$ and $R^{\infty}_{S_{P_{[2]}}}$, respectively ($R^{\infty}_{S_{P_{[1]}}} \cdot R^{\infty}_{S_{P_{[2]}}} > 0$). Let $r^{i}_{R^{\infty}_{S_{P_{[1,2]}}}}$ $(i = 1, 2)$ denote the ratio of the asymptotic throughput $R^{\infty}_{S_{P_{[i]}}}$ of $P_{[i]}$ to the asymptotic joint throughput $R^{\infty}_{S_{P_{[1,2]}}}$ of $P_{[1,2]}$. Let $N_{s,1}$ and $N_{s,2}$ denote the numbers of nodes in set of nodes $s$ belonging to the primary path $P_{[1]}$ and the primary path $P_{[2]}$, respectively. $S^{SSIU}_{S_{P_{[1,2]}}}$ denotes the universal set of saturated second-order interference subsets of the set of sending nodes $S_{P_{[1,2]}}$. Then the asymptotic joint throughput of the pair of paths $P_{[1,2]}$ satisfies the following relations**

**Case 1: $I^*_{S_{P_{[1,2]}}} > 1$**

$$R^{\infty}_{S_{P_{[1,2]}}} \leq \frac{1}{\max\limits_{s \in S^{SSIU}_{S_{P_{[1,2]}}}} \left\{ r^{1}_{R^{\infty}_{S_{P_{[1,2]}}}} \cdot N_{s,1} + r^{2}_{R^{\infty}_{S_{P_{[1,2]}}}} \cdot N_{s,2} \right\}} \tag{X.22}$$

**Case 2: $I^*_{S_{P_{[1,2]}}} = 1$**

$$R^{\infty}_{S_{P_{[1,2]}}} \leq 2 \tag{X.23}$$

**Proof:**

**Case 1:** $I^*_{S_{P_{[1,2]}}} > 1$

Let the total numbers of information blocks successfully received within time interval $D^{int}_{t,b}$ by the respective destination nodes of the primary path $P_{[1]}$ and the primary path $P_{[2]}$ contained in the pair of paths $P_{[1,2]}$ be $M_{[1]}$ $(\in \mathbb{Z}^+)$ and $M_{[2]}$ $(\in \mathbb{Z}^+)$, respectively. The following inequality follows from **Lemma X.4.2**

$$
\begin{aligned}
\frac{M_{[1]}+M_{[2]}}{b} &\le \frac{M_{[1]}+M_{[2]}}{\max\limits_{s\in S_{SP_{[1,2]}}^{SSIU}}\left\{M_{[1]}\cdot N_{s,1}+M_{[2]}\cdot N_{s,2}\right\}} \\
&= \frac{1}{\max\limits_{s\in S_{SP_{[1,2]}}^{SSIU}}\left\{\frac{M_{[1]}/b}{(M_{[1]}+M_{[2]})/b}\cdot N_{s,1}+\frac{M_{[2]}/b}{(M_{[1]}+M_{[2]})/b}\cdot N_{s,2}\right\}}
\end{aligned}
$$

(X.24)

As the quantities $M_{[1]}$ and $M_{[2]}$ of information blocks received by the two corresponding destination nodes of the two primary paths keep increasing, the total number of time beats $b$ consumed for transmitting information blocks increases accordingly, while the above inequality always holds. According to the definition of the asymptotic joint throughput of a pair of paths (**Definition IX.5**), to acquire the asymptotic joint throughput of the pair of paths $P_{[1,2]}$, we need to take the limit with respect to the total number of time beats $b$ contained in the time interval $D_{t,b}^{int}$. Based on **Equation X.24** and **Definition IX.5** (i.e., the definition concerning $r^{i}_{R^{\infty}_{SP_{[1,2]}}}$), it readily follows that

$$
\begin{aligned}
R^{\infty}_{SP_{[1,2]}} &\triangleq \lim_{b\to\infty}\left(\frac{M_{[1]}+M_{[2]}}{b}\right) \\
&\le \lim_{b\to\infty}\left(\frac{1}{\max\limits_{s\in S_{SP_{[1,2]}}^{SSIU}}\left\{\frac{M_{[1]}/b}{(M_{[1]}+M_{[2]})/b}\cdot N_{s,1}+\frac{M_{[2]}/b}{(M_{[1]}+M_{[2]})/b}\cdot N_{s,2}\right\}}\right) \\
&= \frac{1}{\max\limits_{s\in S_{SP_{[1,2]}}^{SSIU}}\left\{r^{1}_{R^{\infty}_{SP_{[1,2]}}}\cdot N_{s,1}+r^{2}_{R^{\infty}_{SP_{[1,2]}}}\cdot N_{s,2}\right\}}
\end{aligned}
$$

(X.25)

**Case 2:** $I^{*}_{SP_{[1,2]}}=1$

In this case, there is no interference relationship at all between the two primary paths. Therefore, they can be treated as two mutually independent primary paths each with an intrinsic interference intensity of 1. Considering that the asymptotic joint throughput of a pair of paths equals the sum of the asymptotic throughputs of the two primary paths contained therein, together with **Theorem X.1**, it readily follows that

$$
R^{\infty}_{SP_{[1,2]}} = R^{\infty}_{SP_{[1]}} + R^{\infty}_{SP_{[2]}} \le 2
$$

(X.26)

■

**Note:**

1) It is generally recognized that effectively working pair of paths corresponds to the case of $R^{\infty}_{S_{P_{[1]}}} \cdot R^{\infty}_{S_{P_{[2]}}} > 0$. For the special case of $R^{\infty}_{S_{P_{[1]}}} \cdot R^{\infty}_{S_{P_{[2]}}} = 0$, we can further divide the discussion into three subcases: (1) $R^{\infty}_{S_{P_{[1]}}} = 0$ and $R^{\infty}_{S_{P_{[2]}}} > 0$. Since $R^{\infty}_{S_{P_{[1,2]}}} = R^{\infty}_{S_{P_{[1]}}} + R^{\infty}_{S_{P_{[2]}}}$ and $R^{\infty}_{S_{P_{[2]}}} \le \frac{1}{I^*_{S_{P_{[2]}}}}$, it readily follows that $R^{\infty}_{S_{P_{[1,2]}}} \le \frac{1}{I^*_{S_{P_{[2]}}}}$; (2) $R^{\infty}_{S_{P_{[2]}}} = 0$ and $R^{\infty}_{S_{P_{[1]}}} > 0$. Similarly, we obtain $R^{\infty}_{S_{P_{[1,2]}}} \le \frac{1}{I^*_{S_{P_{[1]}}}}$; (3) $R^{\infty}_{S_{P_{[1]}}} = 0$ and $R^{\infty}_{S_{P_{[2]}}} = 0$. It readily follows that $R^{\infty}_{S_{P_{[1,2]}}} = 0$;

2) It can be seen from the conclusions presented in the above theorem that, in general, the upper bound on the asymptotic joint throughput of a pair of paths depends on the composition of the universal set of saturated second-order interference subsets of the pair of paths. This constitutes the decisive structure governing the end-to-end throughput performance of a pair of paths.

**Corollary X.4.1: Existence theorem for the upper bound of asymptotic joint throughput of a pair of paths (*Continued*).**

**Suppose the intrinsic interference intensity of a pair of paths $P_{[1,2]}$ is $I^*_{S_{P_{[1,2]}}}$, and its set of sending nodes is $S_{P_{[1,2]}}$. Let the asymptotic (joint) throughput of the pair of paths $P_{[1,2]}$, a primary path $P_{[1]}$ and a primary path $P_{[2]}$ contained therein be $R^{\infty}_{S_{P_{[1,2]}}}$, $R^{\infty}_{S_{P_{[1]}}}$ and $R^{\infty}_{S_{P_{[2]}}}$, respectively ($R^{\infty}_{S_{P_{[1]}}} \cdot R^{\infty}_{S_{P_{[2]}}} > 0$). If $r^1_{R^{\infty}_{S_{P_{[1,2]}}}} = r^2_{R^{\infty}_{S_{P_{[1,2]}}}} = 1/2$, then the asymptotic joint throughput of the pair of paths $P_{[1,2]}$ satisfies the following relation**

$$R^{\infty}_{S_{P_{[1,2]}}} \le \frac{2}{I^*_{S_{P_{[1,2]}}}} \tag{X.27}$$

**Proof:**

**Case 1:** $I^*_{S_{P_{[1,2]}}} > 1$

Since $r^{1}_{R^{\infty}_{S_{P_{[1,2]}}}} = r^{2}_{R^{\infty}_{S_{P_{[1,2]}}}} = 1/2$, it follows from **Case 1** of **Theorem X.4** that

$$R^{\infty}_{S_{P_{[1,2]}}} \leq \frac{2}{\max\limits_{s \in S^{SSIU}_{S_{P_{[1,2]}}}} \{N_{s,1} + N_{s,2}\}}$$

(X.28)

From **Theorem VI.6.1**, the maximum second-order interference subset of a pair of paths must also be a saturated second-order interference subset of the pair of paths. Hence, we have

$$\max_{s \in S^{SSIU}_{S_{P_{[1,2]}}}} \{N_{s,1} + N_{s,2}\} = I^{*}_{S_{P_{[1,2]}}}$$

(X.29)

Combining **Equation X.28** and **Equation X.29**, it readily follows that **Equation X.27** holds.

**Case 2:** $I^{*}_{S_{P_{[1,2]}}} = 1$

**Equation X.27** follows directly from **Case 2** of **Theorem X.4**. ■

**Lemma X.5.1: The asymptotic joint throughput of a pair of paths given its reachable joint period and activation times.**

**Given the reachable joint period of a pair of paths $\boldsymbol{P_{[1,2]}}$ is $\boldsymbol{T_{S_{P_{[1,2]}}}}$, and in each joint period (i.e. $\boldsymbol{T_{S_{P_{[1,2]}}}}$ continuous time beats), the subsets of equally spaced nodes belonging to the primary paths $\boldsymbol{P_{[1]}}$ and $\boldsymbol{P_{[2]}}$ are activated $\boldsymbol{L_1}$ $(\in \mathbb{Z}^{+})$ and $\boldsymbol{L_2}$ $(\in \mathbb{Z}^{+})$ times, respectively. The asymptotic joint throughput of $\boldsymbol{P_{[1,2]}}$ is**

$$\boldsymbol{R^{\infty}_{S_{P_{[1,2]}}} = \frac{L_1 + L_2}{T_{S_{P_{[1,2]}}}}}$$

**(X.30)**

**Proof:**

For a primary path, in steady state, every time all the subsets of equally spaced nodes contained in it are activated once (that is, one traversal of all the subsets of equally spaced nodes contained in the primary path is completed), its corresponding destination node will receive an information block. Therefore, in a reachable joint period (i.e. $T_{S_{P_{[1,2]}}}$ continuous time beats), if the subsets of equally spaced nodes belonging to the primary paths $P_{[1]}$ and $P_{[2]}$ are activated $L_1$ and $L_2$ times respectively, it indicates that the corresponding destination nodes in $P_{[1]}$ and $P_{[2]}$ have respectively received $L_1 + L_2$ information blocks in total. Therefore, the average joint throughput of a pair of paths in a reachable joint period is $R_{SP_{[1,2]}}^{av} = \frac{L_1+L_2}{T_{SP_{[1,2]}}}$. Considering that the pair of paths $P_{[1,2]}$ transmits information blocks strictly according to the period of $T_{SP_{[1,2]}}$, let $n\ (\in \mathbb{Z}^+)$ represent the number of periods experienced, and according to the definition of the asymptotic joint throughput of a pair of paths (**Definition IX.5**), it can be obtained that

$$\begin{aligned} R_{SP_{[1,2]}}^{\infty} &= \lim_{n\to\infty} \frac{n \cdot (L_1 + L_2)}{n \cdot T_{SP_{[1,2]}}} \\ &= \frac{L_1 + L_2}{T_{SP_{[1,2]}}} \end{aligned}$$

(X.31)

■

**Algorithm X.2: The algorithm for the cooperative volatility transmission of information blocks in a pair of paths based on the set of maximum supporting elements in the case of equal opportunity traversals.**

Let the intrinsic interference intensity of a pair of paths $P_{[1,2]}$ be $I_{SP_{[1,2]}}^*$, and assume that the reachable periods of the primary paths $P_{[1]}$ and $P_{[2]}$ are $T_{SP_{[1]}}$ and $T_{SP_{[2]}}$, respectively. Let $\boldsymbol{C}$ be the matrix of joint concurrency relationship between subsets of equally spaced nodes of a pair of paths $P_{[1,2]}$. And the maximum reachable supporting number of $\boldsymbol{C}$ is $U_{\boldsymbol{C}_{sp}}^*$. Determine one set $S_{\boldsymbol{C}_{sp}}^*$ as the set of maximum supporting elements of $\boldsymbol{C}$, and assign incremental sequence numbers to all of its $U_{\boldsymbol{C}_{sp}}^*$ elements (i.e. element No. 1, ..., and element No. $U_{\boldsymbol{C}_{sp}}^*$). Allocate $T_{SP_{[1,2]}}$ consecutive time beats as follows to equally activate all the subsets of equally spaced nodes from both $P_{[1]}$ and $P_{[2]}$ for $L\ (\in \mathbb{Z}^+)$ times:

1: BEGIN

2: Make the time beat counter $m$ indicate the sequence number of the current time beat;

3: FOR (the counter of traversal times = 1 to $L\ (\in \mathbb{Z}^+)$) DO 4: As for the No. 1 element in $S^*_{\boldsymbol{C}_{sp}}$, i.e. $(i_1, j_1)$ $(\in S^*_{\boldsymbol{C}_{sp}})$, assume that its corresponding non-zero element in $\boldsymbol{C}$ is $\boldsymbol{C}[i_1, j_1]$ $(= 1)$, and arrange the subset $\Phi'_{S_{P_{[1]}}}(i_1, T_{S_{P_{[1]}}})$ of equally spaced nodes in the primary path $P_{[1]}$ and the subset $\Phi'_{S_{P_{[2]}}}(j_1, T_{S_{P_{[2]}}})$ of equally spaced nodes in the primary path $P_{[2]}$ to be activated concurrently within the time beat No. $m + 1$. Likewise, as for the No. 2 element in $S^*_{\boldsymbol{C}_{sp}}$, i.e. $(i_2, j_2)$ $(\in S^*_{\boldsymbol{C}_{sp}})$, assume that its corresponding non-zero element in $\boldsymbol{C}$ is $\boldsymbol{C}[i_2, j_2]$ $(= 1)$, and arrange the subset $\Phi'_{S_{P_{[1]}}}(i_2, T_{S_{P_{[1]}}})$ of $P_{[1]}$ and the subset $\Phi'_{S_{P_{[2]}}}(j_2, T_{S_{P_{[2]}}})$ of $P_{[2]}$ to be activated concurrently within the time beat No. $m + 2$. By analogy, we can allocate $U^*_{\boldsymbol{C}_{sp}}$ continuous time beats to realize the concurrent activation of the subsets of equally spaced nodes respectively from $P_{[1]}$ and $P_{[2]}$ (i.e. $U^*_{\boldsymbol{C}_{sp}}$ subsets are from $P_{[1]}$, and $U^*_{\boldsymbol{C}_{sp}}$ subsets are from $P_{[2]}$). And mark all these $U^*_{\boldsymbol{C}_{sp}}$ time beats as time beats of "Category 1";

5: Through the above processing for arranging time beats, there are still $T_{S_{P_{[1]}}} - U^*_{\boldsymbol{C}_{sp}}$ subsets of equally spaced nodes in $P_{[1]}$, having not been assigned to their corresponding time beats. In order to avoid mutual interference, these remaining $T_{S_{P_{[1]}}} - U^*_{\boldsymbol{C}_{sp}}$ subsets of equally spaced nodes in $P_{[1]}$ are arranged to be activated in the subsequent $T_{S_{P_{[1]}}} - U^*_{\boldsymbol{C}_{sp}}$ time beats without overlapping to each other and without any omission (that is, only one subset of equally spaced nodes is activated in each time beat). And mark all these $T_{S_{P_{[1]}}} - U^*_{\boldsymbol{C}_{sp}}$ time beats as time beats of "Category 2";

6: Through the above processing for arranging time beats, there are still $T_{S_{P_{[2]}}} - U^*_{\boldsymbol{C}_{sp}}$ subsets of equally spaced nodes in $P_{[2]}$, having not been assigned to their corresponding time beats. In order to avoid mutual interference, these remaining $T_{S_{P_{[2]}}} - U^*_{\boldsymbol{C}_{sp}}$ subsets of equally spaced nodes in $P_{[2]}$ are arranged to be activated in the subsequent $T_{S_{P_{[2]}}} - U^*_{\boldsymbol{C}_{sp}}$ time beats without overlapping to each other and without any omission. And mark all these $T_{S_{P_{[2]}}} - U^*_{\boldsymbol{C}_{sp}}$ time beats as time beats of "Category 3";

7: Increase the value of the time beat counter $m$ by $T_{S_{P_{[1]}}} + T_{S_{P_{[2]}}} - U^*_{C_{sp}}$;

//Note: after the above three steps, the total allocated time beats are $U^*_{C_{sp}} + (T_{S_{P_{[1]}}} - U^*_{C_{sp}}) + (T_{S_{P_{[2]}}} - U^*_{C_{sp}})$, i.e. $T_{S_{P_{[1]}}} + T_{S_{P_{[2]}}} - U^*_{C_{sp}}$.

8: END FOR;

9: So far, the allocation of $T_{S_{P_{[1,2]}}} = L(T_{S_{P_{[1]}}} + T_{S_{P_{[2]}}} - U^*_{C_{sp}})$ time beats contained in a reachable joint period has been completed. The next transmissions will be executed in periods of $T_{S_{P_{[1,2]}}}$;

10: END

**Note:**

1) This algorithm serves as the specific implementation procedure of the construction method presented in **Lemma VIII.5.1**, and furthermore it is extended to support traversals with equal opportunities;
2) According to the concurrent transmission theorem (**Theorem VI.4**), to ensure the effectiveness of concurrent transmission of the activated subset of sending nodes, the remaining sending nodes on the pair of paths are kept silent during the executions of the algorithm.

**Lemma X.5.2: Given that the reachable periods of primary paths $P_{[i]}$ $(i = 1, 2)$ contained in a pair of paths $P_{[1,2]}$ are $T_{S_{P_{[1]}}}$ and $T_{S_{P_{[2]}}}$, respectively. Suppose that, subject to the definition of the reachable joint period of a pair of paths (Definition VII.3), subsets of equally spaced nodes are allocated within one reachable joint period, i.e., within $T$ consecutive time beats, such that all subsets of equally spaced nodes contained in primary paths $P_{[1]}$ and $P_{[2]}$ are activated $L$ $(\in \mathbb{Z}^+)$ times. Then the $T$ time beats encompassed by this reachable joint period can always be partitioned into $L$ mutually disjoint groups. Within each group of time beats, all subsets of equally spaced nodes belonging to primary paths $P_{[1]}$ and $P_{[2]}$ can be activated once (i.e., traversed completely once).**

**Proof:**

For ease of analysis, we take the initial phase of a subset of equally spaced nodes on a primary path as the sequence number of this equally spaced subset. For instance, the subset of equally spaced nodes on primary path $P_{[1]}$ with initial phase $\theta_1$ $(1 \le \theta_1 \le T_{S_{P_{[2]}}})$ is referred to as subset $\theta_1$ of equally spaced nodes in $P_{[1]}$. Furthermore, we use sequence number $t_l$ $(t_l \in \mathbb{Z}^+, 1 \le l \le T)$ to denote each time beat. And, each time beat $t_l$ among the $T$ time beats is associated with an ordered pair, namely $(l, j_l, k_l)$ $(0 \le j_l \le T_{S_{P_{[1]}}}, 0 \le k_l \le T_{S_{P_{[2]}}})$, according to the subset(s) of equally spaced nodes to which it is assigned (hereinafter, we also state that a time beat contains the corresponding subset of equally spaced nodes). Specifically, the ordered pair $(l, j_l, k_l)$ signifies that the subset of equally spaced nodes $j_l$ on $P_{[1]}$ and the subset of equally spaced nodes $k_l$ on $P_{[2]}$ are assigned to time beat $t_l$. In particular, if $j_l = 0$ or $k_l = 0$, the corresponding subset of equally spaced nodes is not allocated to time beat $t_l$. Furthermore, by **Definition VII.3**, each time beat $t_l$ $(1 \le l \le T)$ must be assigned at least one subset of equally spaced nodes.

Next, to prove this proposition, we present an algorithm that partitions the $T$ time beats assigned with subsets of equally spaced nodes into $L$ mutually disjoint group sets. These group sets are indexed in ascending order and referred to as group set $s_1$, …, group set $s_L$. At the initialization of the algorithm, let set $S_a$ contain all the $T$ time beats ($S_a$ and each group set store the sequence numbers of the corresponding time beats), and all $L$ group sets are empty, i.e., $s_i = \emptyset$ $(1 \le i \le L)$. Overall, the algorithm proceeds in three rounds, which are elaborated sequentially below.

We begin with the first round of the algorithm, which comprises $T_{S_{P_{[1]}}}$ steps. In **Step 1**, each of the $L$ time beats allocated to subset 1 of equally spaced nodes in primary path $P_{[1]}$ is placed into the $L$ group sets, with exactly one time beat per group set. Subsequently, these $L$ time beats are eliminated from set $S_a$. This operation can certainly be completed, as subset 1 of equally spaced nodes in $P_{[1]}$ shall be activated $L$ times. Similarly, in **Step $\boldsymbol{\theta_1}$** $(1 < \theta_1 \le T_{S_{P_{[1]}}})$, each of the $L$ time beats assigned to subset $\theta_1$ of equally spaced nodes in $P_{[1]}$ is placed into the $L$ group sets, with exactly one time beat per group set. These $L$ time beats are subsequently removed from set $S_a$. Such an operation can also be accomplished, as subset $\theta_1$ of equally spaced nodes in $P_{[1]}$ is activated $L$ times… After this round terminates, we guarantee that in each group set $s_i$ $(1 \le i \le L)$, all $T_{S_{P_{[1]}}}$ subsets of equally spaced nodes belonging to primary path $P_{[1]}$ are each activated exactly once.

Next comes the second round of operations of the algorithm. We examine all group sets $s_i$ $(1 \le i \le L)$ to check whether any group set contains a time beat $t_{l_1}$ $(t_{l_1} \in s_i, 1 \le i \le L)$ such that the subset of equally spaced nodes from primary path $P_{[2]}$ assigned to this time beat is identical to the subset of equally spaced nodes from primary path $P_{[2]}$ assigned to at least one other time beat $t_{l_2}(t_{l_2} \in s_i, 1 \le i \le L, t_{l_1} \ne t_{l_2})$ within the same group set $s_i$. That is, in the ordered pair $(l_1, j_{l_1}, k_{l_1})$ corresponding to time beat $t_{l_1}$ and the ordered pair $(l_2, j_{l_2}, k_{l_2})$ corresponding to time beat $t_{l_2}$, the case $(j_{l_1} \ne j_{l_2}) \wedge (k_{l_1} = k_{l_2}) \wedge (0 < k_{l_1} \le T_{S_{P_{[2]}}})$ occurs. In such a case, subset $k_{l_1}$ of equally spaced nodes of $P_{[2]}$ is activated more than once within $s_i$, which violates the grouping requirement that each equally spaced subset of nodes shall be activated exactly once in a single group set. To resolve this issue, we search for another group set $s_m$ $(1 \le m \le L, m \ne i)$ such that none of its contained time beats are assigned subset $k_{l_1}$ of equally spaced nodes from primary path $P_{[2]}$. Such a group set is guaranteed to exist. The reasoning is as follows: if no such group set can be found, then subset $k_{l_1}$ of equally spaced nodes of primary path $P_{[2]}$ is activated at least $L + 1$ times across all the assigned time beats in all the group sets (at least twice within group set $s_i$, and at least once in each of the remaining $L - 1$ group sets). This contradicts the hypothesis. Next, within the identified group set $s_m$, locate the time beat assigned subset $j_{l_1}$ of equally spaced nodes from primary path $P_{[1]}$. From the operations in the first round, it follows that such a time beat (denoted as time beat $t_{l_3}$) is guaranteed to exist within group set $s_m$. We then swap the group sets to which time beat $t_{l_1}$ and time beat $t_{l_3}$ belong; that is, assign time beat $t_{l_1}$ to group set $s_m$ and time beat $t_{l_3}$ to group set $s_i$. Repeat the above inspection and the necessary swapping operations for the group sets containing the relevant time beats until none of the aforementioned issues persist across all group sets $s_i$ $(1 \le i \le L)$. (The number of problematic time beats is finite, and each swapping operation resolves one such issue. Hence, the above procedure is guaranteed to terminate.) … Upon completion of this round, we guarantee that within each group set $s_i$ $(1 \le i \le L)$, every subset of equally spaced nodes from both primary path $P_{[1]}$ and primary path $P_{[2]}$ assigned to the contained time beats is activated exactly once. If $S_a = \emptyset$, the algorithm terminates. At this point, the algorithm completes the task of allocating the $T$ time beats into $L$ mutually disjoint group sets. This concludes the proof of the proposition. If $S_a \ne \emptyset$, we proceed to the third round of the algorithm.

After the above two rounds are completed, it is easy to see that among the $T$ time beats to be grouped, the remaining time beats that have not yet been placed into the $L$ group sets (these remaining time beats still belong to set $S_a$) are each assigned only one subset of equally spaced nodes from primary path $P_{[2]}$. In the third round, we place all remaining time beats within $S_a$ into the $L$ group sets in $L$ steps. In **Step 1**, from all $T_{S_{P_{[2]}}}$ subsets of equally spaced nodes of primary path $P_{[2]}$, we filter out those subsets of equally spaced nodes that are not contained in any time beat already affiliated to group set $s_1$ (these equally spaced subsets of nodes are referred to as "to-be-matched subsets of equally spaced nodes"). If the number of filtered-out to-be-matched subsets of equally spaced nodes satisfies $n_T = 0$, we directly proceed to the next step of this round. Otherwise (i.e., condition $n_T > 0$ holds), for these to-be-matched subsets of equally spaced nodes, we extract from set $S_a$ the time beats allocated to these $n_T$ to-be-matched subsets of equally spaced nodes (there are correspondingly $n_T$ such time beats) and place them into group set $s_1$. This extraction implements the matching between the to-be-matched subsets of equally spaced nodes and specific time beats in set $S_a$. Thereafter, these time beats that have been placed into set $s_1$ are removed from $S_a$. This step can always be completed successfully. Otherwise, for some to-be-matched subset of equally spaced nodes (among $n_T$ to-be-matched subsets of equally spaced nodes), no corresponding time beat within $S_a$ can be matched. This implies that this equally spaced subset of nodes cannot be activated $L$ times, which contradicts the hypothesis. Similarly, we repeat the above operations in **Step $\boldsymbol{i}$** $(1 \le i \le L)$...Upon completion of this round, the $T$ time beats awaiting grouping will be exactly allocated into the $L$ group sets. Moreover, the time beats contained within each group set achieve a one-time activation of all subsets of equally spaced nodes contained in both primary paths $P_{[1]}$ and $P_{[2]}$. Furthermore, the scenario corresponding to set $S_a \neq \emptyset$ no longer exists after this round. The reason is that the occurrence of such a scenario would contradict the hypothesis. ■

**Lemma X.5.3: Algorithm X.2 can achieve the minimal reachable joint period of a pair of paths.**

**Let the reachable periods of the primary paths $\boldsymbol{P_{[1]}}$ and $\boldsymbol{P_{[2]}}$, which are contained in**

**a pair of paths $P_{[1,2]}$, be $T_{S_{P_{[1]}}}$ and $T_{S_{P_{[2]}}}$, respectively. Let $C$ be the matrix of joint concurrency relationship between subsets of equally spaced nodes of a pair of paths $P_{[1,2]}$. And the maximum reachable supporting number of $C$ is $U^*_{C_{sp}}$. If time beats are allocated so that all the subsets of equally spaced nodes in $P_{[1]}$ and $P_{[2]}$ are activated $L\ (\in \mathbb{Z}^+)$ times. Then under equal opportunity traversals, Algorithm X.2 can achieve the minimal reachable joint period of the pair of paths, i.e.**

$$\boldsymbol{T_{S_{P_{[1,2]}}} = L\left(T_{S_{P_{[1]}}} + T_{S_{P_{[2]}}} - U^*_{C_{sp}}\right)}$$

**(X.32)**

**Proof:**

According to the definition of reachable joint period of a pair of paths (**Definition VII.3**), it is known that the time beats contained in a joint period must belong to one of the following three categories, that is, in time beats of category 1, one subset of equally spaced nodes from the primary path $P_{[1]}$ and one subset of equally spaced nodes from the primary path $P_{[2]}$ are activated at the same time, and the two subsets of nodes together constitute a second-order concurrency subset of nodes; In time beats of category 2, only one subset of equally spaced nodes from $P_{[1]}$ is activated; In time beats of category 3, only one subset of equally spaced nodes from $P_{[2]}$ is activated. If **Algorithm X.2** is executed, it can be seen that the number of time beats of category 1 within a joint period is $L \cdot U^*_{C_{sp}}$; The number of time beats of category 2 is $L(T_{S_{P_{[1]}}} - U^*_{C_{sp}})$, and the number of time beats of category 3 is $L(T_{S_{P_{[2]}}} - U^*_{C_{sp}})$. Therefore, by executing **Algorithm X.2**, the total number of time beats within a joint period is

$$T_{S_{P_{[1,2]}}} = L\left(T_{S_{P_{[1]}}} + T_{S_{P_{[2]}}} - U^*_{C_{sp}}\right)$$

(X.33)

It is hereinafter proved that the above joint period $T_{S_{P_{[1,2]}}} = L(T_{S_{P_{[1]}}} + T_{S_{P_{[2]}}} - U^*_{C_{sp}})$ is the achievable minimum. Since all equally spaced second-order concurrency subsets of nodes contained in primary path $P_{[1]}$ and $P_{[2]}$ are activated through $L$ traversals, according to **Lemma X.5.2**, the $T_{S_{P_{[1,2]}}}$ time beats allocated by **Algorithm X.2** can be divided into non-overlapping $L$ groups, with each group corresponding to one complete traversal and activation of $P_{[1]}$ and $P_{[2]}$. Let the quantities of category 1, category 2 and category 3 time beats contained in the $i\ (1 \le i \le L)$ -th group be $m_{i,1}$, $m_{i,2}$ and $m_{i,3}$, respectively. Then we have

$$T_{S_{P_{[1]}}} = m_{i,1} + m_{i,2}$$

(X.34.1)

and

$$T_{S_{P_{[2]}}} = m_{i,1} + m_{i,3}$$

(X.34.2)

Furthermore, the total number of time beats contained in the $i\ (1 \le i \le L)$ -th group is derived as

$$\sum_{j=1}^{3} m_{i,j} = T_{S_{P_{[1]}}} + T_{S_{P_{[2]}}} - m_{i,1}$$

(X.35)

According to the definition of the maximum reachable supporting number $U^*_{C_{sp}}$ of matrix $\boldsymbol{C}$ (**Definition VII.11**), we have

$$T_{S_{P_{[1]}}} + T_{S_{P_{[2]}}} - m_{i,1} \ge T_{S_{P_{[1]}}} + T_{S_{P_{[2]}}} - U^*_{C_{sp}}$$

(X.36)

Therefore, the total number of time beats required for $L$ traversal activations of all equally spaced second-order concurrency subsets of nodes in primary path $P_{[1]}$ and $P_{[2]}$ satisfies the relation

$$\sum_{i=1}^{L} (T_{S_{P_{[1]}}} + T_{S_{P_{[2]}}} - m_{i,1}) \ge L\left(T_{S_{P_{[1]}}} + T_{S_{P_{[2]}}} - U^*_{C_{sp}}\right) = T_{S_{P_{[1,2]}}}$$

(X.37)

Hence, the joint period $T_{S_{P_{[1,2]}}} = L(T_{S_{P_{[1]}}} + T_{S_{P_{[2]}}} - U^*_{C_{sp}})$ achieved by **Algorithm X.2** is the minimum achievable value. ■

**Theorem X.5: Reachability theorem for the upper bound of the asymptotic joint throughput of a pair of paths (in the case of equal opportunity traversals).**

**Let the intrinsic interference intensity of a pair of paths $P_{[1,2]}$ be $I^*_{S_{P_{[1,2]}}}$, and assume that the reachable periods of the primary paths $P_{[1]}$ and $P_{[2]}$, which are contained in**

**$P_{[1,2]}$, are $T_{S_{P_{[1]}}}$ and $T_{S_{P_{[2]}}}$, respectively. Let $C$ be the matrix of joint concurrency relationship between subsets of equally spaced nodes of $P_{[1,2]}$. And the maximum reachable supporting number of $C$ is $U^*_{C_{sp}}$. If time beats are allocated by using Algorithm X.2, so that the joint period of $P_{[1,2]}$ is $T_{S_{P_{[1,2]}}}$, and within which all the subsets of equally spaced nodes in $P_{[1]}$ and $P_{[2]}$ are activated $L\ (\in \mathbb{Z}^+)$ times. It can be concluded that the upper bound of the asymptotic joint throughput $R^\infty_{S_{P_{[1,2]}}}$ of the pair of paths $P_{[1,2]}$ is independent of the activation times $L$, and**

$$R^\infty_{S_{P_{[1,2]}}} \le \frac{2}{I^*_{S_{P_{[1,2]}}}} \tag{X.38}$$

**When $U^*_{C_{sp}} = T_{S_{P_{[1]}}} + T_{S_{P_{[2]}}} - I^*_{S_{P_{[1,2]}}}$, the equality relation holds (namely, the upper bound of the asymptotic joint throughput of a pair of paths can be reached).**

**Proof:**

According to **Lemma X.5.1** and **Lemma X.5.3**, we know that the asymptotic joint throughput of the pair of paths $P_{[1,2]}$ under **Algorithm X.2** is

$$R^\infty_{S_{P_{[1,2]}}} = \frac{2L}{T_{S_{P_{[1,2]}}}} = \frac{2}{T_{S_{P_{[1]}}} + T_{S_{P_{[2]}}} - U^*_{C_{sp}}} \tag{X.39}$$

According to **Lemma VIII.5.4**, it can be found that $U^*_{C_{sp}} \le T_{S_{P_{[1]}}} + T_{S_{P_{[2]}}} - I^*_{S_{P_{[1,2]}}}$. Therefore, by combing with the above formula, it is obtained that

$$R^\infty_{S_{P_{[1,2]}}} = \frac{2}{T_{S_{P_{[1]}}} + T_{S_{P_{[2]}}} - U^*_{C_{sp}}} \le \frac{2}{I^*_{S_{P_{[1,2]}}}} \tag{X.40}$$

Furthermore, the equality relation holds when $U^*_{C_{sp}} = T_{S_{P_{[1]}}} + T_{S_{P_{[2]}}} - I^*_{S_{P_{[1,2]}}}$. ■

**Note:**

1) **Corollary X.4.1** indicates that when $r^1_{R^\infty_{S_{P_{[1,2]}}}} = r^2_{R^\infty_{S_{P_{[1,2]}}}} = 1/2$ (i.e., corresponding to the "equal opportunity traversals" scenario considered in this theorem), the upper bound on the asymptotic joint throughput of pair of paths $P_{[1,2]}$ is $R^{\infty}_{S_{P_{[1,2]}}} \leq \frac{2}{I^{*}_{S_{P_{[1,2]}}}}$. Hence, this theorem shows that such an upper bound on asymptotic joint throughput is achievable by using **Algorithm X.2**;

2) From the perspective of allocating time-beat resources within a reachable joint period of a pair of paths, this theorem establishes the upper-bound performance of the asymptotic joint throughput of a pair of paths under equal opportunity traversals. **Corollary X.4.1**, by contrast, analyzes the upper-bound performance of the asymptotic joint throughput for a pair of paths based on the notion of the information block transmission time trace matrix of a primary path. Hence, this theorem and **Corollary X.4.1** employ distinct methods and yield fully consistent results. That is to say, this theorem and **Corollary X.4.1** mutually corroborate their validity!

**Algorithm X.3: The algorithm for the cooperative volatility transmission of information blocks in a pair of paths based on the set of maximum supporting elements in the case of unequal opportunity traversals.**

Let the reachable periods of the primary paths $P_{[1]}$ and $P_{[2]}$, which are contained in a pair of paths $P_{[1,2]}$, be $T_{S_{P_{[1]}}}$ and $T_{S_{P_{[2]}}}$, respectively. Let $\boldsymbol{C}$ be the matrix of joint concurrency relationship between subsets of equally spaced nodes of a pair of paths $P_{[1,2]}$. Allocate $T_{S_{P_{[1,2]}}}$ consecutive time beats as follows, so as to activate the subsets of equally spaced nodes from $P_{[1]}$ for $L_1\ (\in \mathbb{Z}^+)$ times, and to activate the subsets of equally spaced nodes from $P_{[2]}$ for $L_2\ (\in \mathbb{Z}^+)$ times.

1: BEGIN

2: Generate the $L_1 \times L_2$-th continuation matrix $\boldsymbol{C}^{L_1 \times L_2}$ of the matrix $\boldsymbol{C}$;

//Note: According to the definitions of the matrix $\boldsymbol{C}$ of joint concurrency relationship between subsets of equally spaced nodes and its $L_1 \times L_2$-th continuation matrix $\boldsymbol{C}^{L_1 \times L_2}$ (**Definition IX.7**), any element $\boldsymbol{C}^{L_1 \times L_2}[i,j]\ (1 \leq i \leq L_1 \cdot T_{S_{P_{[1]}}}, 1 \leq j \leq L_2 \cdot T_{S_{P_{[2]}}})$ in $\boldsymbol{C}^{L_1 \times L_2}$, represents the joint concurrency coefficient (**Definition VII.6**) between the subset $\Phi'_{S_{P_{[1]}}}(((i-1) \bmod T_{S_{P_{[1]}}}) + 1, T_{S_{P_{[1]}}})$ of equally spaced nodes in the primary path $P_{[1]}$ and the subset $\Phi'_{S_{P_{[2]}}}(((j-1) \bmod T_{S_{P_{[2]}}})+1, T_{S_{P_{[2]}}})$ of equally spaced nodes in the primary path $P_{[2]}$.

3: Determine one set $S^*_{\boldsymbol{C}_{sp}^{L_1 \times L_2}}$ as the set of maximum supporting elements of $\boldsymbol{C}^{L_1 \times L_2}$. Let $U^*_{\boldsymbol{C}_{sp}^{L_1 \times L_2}}$ denote the maximum reachable supporting number of $\boldsymbol{C}^{L_1 \times L_2}$, and assign incremental sequence numbers to all of the $U^*_{\boldsymbol{C}_{sp}^{L_1 \times L_2}}$ elements of $S^*_{\boldsymbol{C}_{sp}^{L_1 \times L_2}}$ (i.e. element No. 1, ..., and element No. $U^*_{\boldsymbol{C}_{sp}^{L_1 \times L_2}}$);

4: As for the No. 1 element in $S^*_{\boldsymbol{C}_{sp}^{L_1 \times L_2}}$, i.e. $(i_1, j_1)$ ($\in S^*_{\boldsymbol{C}_{sp}^{L_1 \times L_2}}$), assume that its corresponding non-zero element in $\boldsymbol{C}^{L_1 \times L_2}$ is $\boldsymbol{C}^{L_1 \times L_2}[i_1, j_1]$ $(=1)$, and arrange the subset $\Phi'_{S_{P_{[1]}}}(((i_1 - 1) \bmod T_{S_{P_{[1]}}})+1, T_{S_{P_{[1]}}})$ of equally spaced nodes in $P_{[1]}$ and the subset $\Phi'_{S_{P_{[2]}}}(((j_1 - 1) \bmod T_{S_{P_{[2]}}})+1, T_{S_{P_{[2]}}})$ of equally spaced nodes in $P_{[2]}$ to be activated concurrently within the time beat No. 1. Likewise, as for the No. 2 element in $S^*_{\boldsymbol{C}_{sp}^{L_1 \times L_2}}$, i.e. $(i_2, j_2)$ ($\in S^*_{\boldsymbol{C}_{sp}^{L_1 \times L_2}}$), assume that its corresponding non-zero element in $\boldsymbol{C}^{L_1 \times L_2}$ is $\boldsymbol{C}^{L_1 \times L_2}[i_2, j_2]$ $(=1)$, and arrange the subset $\Phi'_{S_{P_{[1]}}}(((i_2 - 1) \bmod T_{S_{P_{[1]}}})+1, T_{S_{P_{[1]}}})$ in $P_{[1]}$ and the subset $\Phi'_{S_{P_{[2]}}}(((j_2 - 1) \bmod T_{S_{P_{[2]}}})+1, T_{S_{P_{[2]}}})$ in $P_{[2]}$ to be activated concurrently within the time beat No. 2. By analogy, we can allocate $U^*_{\boldsymbol{C}_{sp}^{L_1 \times L_2}}$ continuous time beats to realize the concurrent activation of the subsets of equally spaced nodes respectively from $P_{[1]}$ and $P_{[2]}$. And mark all these $U^*_{\boldsymbol{C}_{sp}^{L_1 \times L_2}}$ time beats as time beats of "Category 1";

5: Through the above processing for arranging time beats, there are still $L_1 \cdot T_{S_{P_{[1]}}} - U^*_{\boldsymbol{C}_{sp}^{L_1 \times L_2}}$ subsets of equally spaced nodes in $P_{[1]}$, having not been assigned to their corresponding time beats. In order to avoid mutual interference, these remaining $L_1 \cdot T_{S_{P_{[1]}}} - U^*_{\boldsymbol{C}_{sp}^{L_1 \times L_2}}$ subsets of equally spaced nodes in $P_{[1]}$ are arranged to be activated in the subsequent $L_1 \cdot T_{S_{P_{[1]}}} - U^*_{\boldsymbol{C}_{sp}^{L_1 \times L_2}}$ time beats without overlapping to each other and without any omission (that is, only one subset of equally spaced nodes is activated in each time beat). And mark all these $L_1 \cdot T_{S_{P_{[1]}}} - U^*_{C_{sp}^{L_1 \times L_2}}$ time beats as time beats of "Category 2";

6: Through the above processing for arranging time beats, there are still $L_2 \cdot T_{S_{P_{[2]}}} - U^*_{C_{sp}^{L_1 \times L_2}}$ subsets of equally spaced nodes in $P_{[2]}$, having not been assigned to their corresponding time beats. In order to avoid mutual interference, these remaining $L_2 \cdot T_{S_{P_{[2]}}} - U^*_{C_{sp}^{L_1 \times L_2}}$ subsets of equally spaced nodes in $P_{[2]}$ are arranged to be activated in the subsequent $L_2 \cdot T_{S_{P_{[2]}}} - U^*_{C_{sp}^{L_1 \times L_2}}$ time beats without overlapping to each other and without any omission (that is, only one subset of equally spaced nodes is activated in each time beat). And mark all these $L_2 \cdot T_{S_{P_{[2]}}} - U^*_{C_{sp}^{L_1 \times L_2}}$ time beats as time beats of "Category 3";

7: So far, the allocation of $T_{S_{P_{[1,2]}}} = L_1 \cdot T_{S_{P_{[1]}}} + L_2 \cdot T_{S_{P_{[2]}}} - U^*_{C_{sp}^{L_1 \times L_2}}$ time beats contained in a reachable joint period has been completed. The next transmissions will be executed in periods of $T_{S_{P_{[1,2]}}}$;

//Note: From the above step 4 to step 6, the total allocated time beats are $U^*_{C_{sp}^{L_1 \times L_2}} + (L_1 \cdot T_{S_{P_{[1]}}} - U^*_{C_{sp}^{L_1 \times L_2}}) + (L_2 \cdot T_{S_{P_{[2]}}} - U^*_{C_{sp}^{L_1 \times L_2}})$, i.e. $L_1 \cdot T_{S_{P_{[1]}}} + L_2 \cdot T_{S_{P_{[2]}}} - U^*_{C_{sp}^{L_1 \times L_2}}$.

8: END

**Note:**

1) This algorithm is an extension of **Algorithm X.2**.

**Theorem X.6: Algorithm X.3 can achieve the minimal reachable joint period of a pair of paths.**

**Let the reachable periods of the primary paths $P_{[1]}$ and $P_{[2]}$, which are contained in a pair of paths $P_{[1,2]}$, be $T_{S_{P_{[1]}}}$ and $T_{S_{P_{[2]}}}$, respectively. Let $C$ be the matrix of joint concurrency relationship between subsets of equally spaced nodes of a pair of paths $P_{[1,2]}$. $C^{L_1 \times L_2}$ is the $L_1 \times L_2$-th continuation matrix of $C$, and its maximum reachable supporting number is $U^*_{C_{sp}^{L_1 \times L_2}}$. If time beats are allocated by using Algorithm X.3, so that the joint** **period of $P_{[1,2]}$ is $T_{S_{P_{[1,2]}}}$, and within which subsets of equally spaced nodes in $P_{[1]}$ and $P_{[2]}$ are activated $L_1\ (\in \mathbb{Z}^+)$ and $L_2\ (\in \mathbb{Z}^+)$ times ($L_1$ and $L_2$ need not be equal), respectively. Then in the case of unequal opportunity traversals, Algorithm X.3 can achieve the minimal reachable joint period of the pair of paths, i.e.**

$$\boldsymbol{T_{S_{P_{[1,2]}}} = L_1 \cdot T_{S_{P_{[1]}}} + L_2 \cdot T_{S_{P_{[2]}}} - U^{*}_{C_{sp}^{L_1 \times L_2}}}$$

**(X.41)**

**Proof:**

According to the definition of reachable joint period of a pair of paths (**Definition VII.3**), it is known that the time beats contained in a joint period must belong to one of the following three categories, that is, in time beats of category 1, one subset of equally spaced nodes from the primary path $P_{[1]}$ and one subset of equally spaced nodes from the primary path $P_{[2]}$ are activated at the same time, and the two subsets of nodes together constitute a second-order concurrency subset of nodes; In time beats of category 2, only one subset of equally spaced nodes from $P_{[1]}$ is activated; In time beats of category 3, only one subset of equally spaced nodes from $P_{[2]}$ is activated. If **Algorithm X.3** is executed, it can be seen that the number of time beats of category 1 within a joint period is $U^{*}_{C_{sp}^{L_1 \times L_2}}$; The number of time beats of category 2 is $L_1 \cdot T_{S_{P_{[1]}}} - U^{*}_{C_{sp}^{L_1 \times L_2}}$, and the number of time beats of category 3 is $L_2 \cdot T_{S_{P_{[2]}}} - U^{*}_{C_{sp}^{L_1 \times L_2}}$. Therefore, by executing **Algorithm X.3**, the total number of time beats within a joint period is

$$T_{S_{P_{[1,2]}}} = L_1 \cdot T_{S_{P_{[1]}}} + L_2 \cdot T_{S_{P_{[2]}}} - U^{*}_{C_{sp}^{L_1 \times L_2}}$$

(X.42)

It is hereinafter proved that the above joint period $T_{S_{P_{[1,2]}}} = L_1 \cdot T_{S_{P_{[1]}}} + L_2 \cdot T_{S_{P_{[2]}}} - U^{*}_{C_{sp}^{L_1 \times L_2}}$ is the achievable minimum. Let the quantities of category 1, category 2 and category 3 time beats contained in any joint period $T$ be $m_1$, $m_2$ and $m_3$, respectively. Then, we have

$$L_1 \cdot T_{S_{P_{[1]}}} = m_1 + m_2$$

(X.43.1)

and

$$L_2 \cdot T_{S_{P_{[2]}}} = m_1 + m_3$$

(X.43.2)

The identical term $m_1$ is present in both foregoing equations. The reason is that, from the perspective of either the primary path $P_{[1]}$ or the primary path $P_{[2]}$, the count of subsets of equally spaced nodes paired with subsets of equally spaced nodes in the counterpart primary path is exactly the same. Furthermore, it can be obtained that

$$T = \sum_{i=1}^{3} m_i = L_1 \cdot T_{S_{P_{[1]}}} + L_2 \cdot T_{S_{P_{[2]}}} - m_1$$

(X.44)

According to the definition of the maximum reachable supporting number $U^*_{C_{sp}^{L_1 \times L_2}}$ of matrix $C^{L_1 \times L_2}$ (**Definition IX.7**), we have

$$\begin{aligned} T &= L_1 \cdot T_{S_{P_{[1]}}} + L_2 \cdot T_{S_{P_{[2]}}} - m_1 \\ &\geq L_1 \cdot T_{S_{P_{[1]}}} + L_2 \cdot T_{S_{P_{[2]}}} - U^*_{C_{sp}^{L_1 \times L_2}} \\ &= T_{S_{P_{[1,2]}}} \end{aligned}$$

(X.45)

Hence, the joint period $T_{S_{P_{[1,2]}}} = L_1 \cdot T_{S_{P_{[1]}}} + L_2 \cdot T_{S_{P_{[2]}}} - U^*_{C_{sp}^{L_1 \times L_2}}$ achieved by **Algorithm X.3** is the minimum achievable value. ■

**Corollary X.6.1: The asymptotic joint throughput of a pair of paths achieved by Algorithm X.3.**

**Let the reachable periods of the primary paths $P_{[1]}$ and $P_{[2]}$, which are contained in a pair of paths $P_{[1,2]}$, be $T_{S_{P_{[1]}}}$ and $T_{S_{P_{[2]}}}$, respectively. Let $C$ be the matrix of joint concurrency relationship between subsets of equally spaced nodes of a pair of paths $P_{[1,2]}$. $C^{L_1 \times L_2}$ is the $L_1 \times L_2$-th continuation matrix of $C$, and its maximum reachable supporting number is $U^*_{C_{sp}^{L_1 \times L_2}}$. If time beats are allocated by using Algorithm X.3, so that the joint period of $P_{[1,2]}$ is $T_{S_{P_{[1,2]}}}$, and within which subsets of equally spaced nodes in $P_{[1]}$ and $P_{[2]}$ are activated $L_1\ (\in \mathbb{Z}^+)$ and $L_2\ (\in \mathbb{Z}^+)$ times ($L_1$ and $L_2$ need not be equal), respectively. It can be concluded that the asymptotic joint throughput of $P_{[1,2]}$ is**

$$R^{\infty}_{S_{P_{[1,2]}}} = \frac{L_1 + L_2}{L_1 \cdot T_{S_{P_{[1]}}} + L_2 \cdot T_{S_{P_{[2]}}} - U^{*}_{C_{sp}^{L_1 \times L_2}}}$$

**(X.46)**

**Proof:**

According to the above descriptions of **Algorithm X.3**, the reachable joint period achieved by **Algorithm X.3** is $T_{S_{P_{[1,2]}}} = L_1 \cdot T_{S_{P_{[1]}}} + L_2 \cdot T_{S_{P_{[2]}}} - U^{*}_{C_{sp}^{L_1 \times L_2}}$. Furthermore, by combining with **Lemma X.5.1**, **Equation X.46** can be obtained. ■

**Corollary X.6.2: The upper bound of the asymptotic joint throughput of a pair of paths achieved by Algorithm X.3.**

**Let the reachable periods of the primary paths $P_{[1]}$ and $P_{[2]}$, which are contained in a pair of paths $P_{[1,2]}$, be $T_{S_{P_{[1]}}}$ and $T_{S_{P_{[2]}}}$, respectively. Let $C$ be the matrix of joint concurrency relationship between subsets of equally spaced nodes of a pair of paths $P_{[1,2]}$. $C^{L_1 \times L_2}$ is the $L_1 \times L_2$-th continuation matrix of $C$, and its maximum reachable supporting number is $U^{*}_{C_{sp}^{L_1 \times L_2}}$. If time beats are allocated by using Algorithm X.3, so that the joint period of $P_{[1,2]}$ is $T_{S_{P_{[1,2]}}}$, and within which subsets of equally spaced nodes in $P_{[1]}$ and $P_{[2]}$ are activated $L_1 \, (\in \mathbb{Z}^{+})$ and $L_2 \, (\in \mathbb{Z}^{+})$ times ($L_1$ and $L_2$ need not be equal), respectively. The asymptotic joint$T_{S_{P_{[1,2]}}}$ throughput of $P_{[1,2]}$ is**

$$R^{\infty}_{S_{P_{[1,2]}}} \leq \frac{1}{T_{S_{P_{[1]}}}} + \frac{1}{T_{S_{P_{[2]}}}}$$

**(X.47)**

**When $U^{*}_{C_{sp}^{L_1 \times L_2}} = L_1 \cdot T_{S_{P_{[1]}}} = L_2 \cdot T_{S_{P_{[2]}}}$, the equality relation holds.**

**Proof:**

**Case 1:** $L_1 \cdot T_{S_{P_{[1]}}} \geq L_2 \cdot T_{S_{P_{[2]}}}$

In this case, $U^{*}_{C_{sp}^{L_1 \times L_2}} \leq L_2 \cdot T_{S_{P_{[2]}}}$ and $L_2 \leq \frac{L_1 \cdot T_{S_{P_{[1]}}}}{T_{S_{P_{[2]}}}}$. Therefore, according to **Corollary X.6.1**, we have

$$
\begin{aligned}
R^{\infty}_{SP_{[1,2]}} &= \frac{L_1 + L_2}{L_1 \cdot T_{SP_{[1]}} + L_2 \cdot T_{SP_{[2]}} - U^*_{C_{sp}^{L_1 \times L_2}}} \\
&\le \frac{L_1 + L_2}{L_1 \cdot T_{SP_{[1]}}} \le \frac{L_1\left(1 + \frac{T_{SP_{[1]}}}{T_{SP_{[2]}}}\right)}{L_1 \cdot T_{SP_{[1]}}} \\
&= \frac{1}{T_{SP_{[1]}}} + \frac{1}{T_{SP_{[2]}}}
\end{aligned}
\tag{X.48}
$$

It is evident that when $U^*_{C_{sp}^{L_1 \times L_2}} = L_1 \cdot T_{SP_{[1]}} = L_2 \cdot T_{SP_{[2]}}$, the equality relation holds.

**Case 2:** $L_1 \cdot T_{SP_{[1]}} < L_2 \cdot T_{SP_{[2]}}$

In this case, $U^*_{C_{sp}^{L_1 \times L_2}} \le L_1 \cdot T_{SP_{[1]}}$ and $L_1 < \frac{L_2 \cdot T_{SP_{[2]}}}{T_{SP_{[1]}}}$. Therefore, according to **Corollary X.6.1**, we have

$$
\begin{aligned}
R^{\infty}_{SP_{[1,2]}} &= \frac{L_1 + L_2}{L_1 \cdot T_{SP_{[1]}} + L_2 \cdot T_{SP_{[2]}} - U^*_{C_{sp}^{L_1 \times L_2}}} \\
&\le \frac{L_1 + L_2}{L_2 \cdot T_{SP_{[2]}}} < \frac{L_2\left(\frac{T_{SP_{[2]}}}{T_{SP_{[1]}}} + 1\right)}{L_2 \cdot T_{SP_{[2]}}} \\
&= \frac{1}{T_{SP_{[1]}}} + \frac{1}{T_{SP_{[2]}}}
\end{aligned}
\tag{X.49}
$$

The proposition is proved by combining the above two cases. ■

**Corollary X.6.3: Relationship between $U^*_{C_{sp}}$ and $U^*_{C_{sp}^{L \times L}}$.**

**Let the reachable periods of the primary paths $P_{[1]}$ and $P_{[2]}$, which are contained in a pair of paths $P_{[1,2]}$, be $T_{SP_{[1]}}$ and $T_{SP_{[2]}}$, respectively. Let $C$ be the matrix of joint concurrency relationship between subsets of equally spaced nodes of a pair of paths $P_{[1,2]}$. $C^{L \times L}$ is the $L \times L$ -th continuation matrix of $C$. The maximum reachable supporting numbers of $C$ and $C^{L \times L}$ are $U^*_{C_{sp}}$ and $U^*_{C_{sp}^{L \times L}}$, respectively. There is**

$$
U^*_{C_{sp}} = \frac{U^*_{C_{sp}^{L \times L}}}{L}
$$

| (X.50) |
|---:|

**Proof:**

Consider the case $L_1 = L_2 = L$, from **Corollary X.6.1**, the asymptotic joint throughput of the pair of paths $P_{[1,2]}$ attained under **Algorithm X.3** is

$$\begin{aligned} R^{\infty}_{S_{P_{[1,2]}},a3} &= \frac{2L}{L\left(T_{S_{P_{[1]}}} + T_{S_{P_{[2]}}} - \frac{U^{*}_{C^{L\times L}_{sp}}}{L}\right)} \\ &= \frac{2}{T_{S_{P_{[1]}}} + T_{S_{P_{[2]}}} - \frac{U^{*}_{C^{L\times L}_{sp}}}{L}} \end{aligned}$$

(X.51)

Moreover, by **Lemma X.5.1** and **Lemma X.5.3**, the asymptotic joint throughput of $P_{[1,2]}$ attained under **Algorithm X.2** is

$$R^{\infty}_{S_{P_{[1,2]}},a2} = \frac{2}{T_{S_{P_{[1]}}} + T_{S_{P_{[2]}}} - U^{*}_{C_{sp}}}$$

(X.52)

By **Lemma X.5.3** and **Theorem X.6**, **Algorithm X.2** and **Algorithm X.3** both attain the maximum asymptotic joint throughput of $P_{[1,2]}$ under the case of equal opportunity traversals. Hence,

$$R^{\infty}_{S_{P_{[1,2]}},a2} = R^{\infty}_{S_{P_{[1,2]}},a3}$$

(X.53)

Combining **Equation X.51**, **Equation X.52** and **Equation X.53**, it can be concluded that $U^{*}_{C_{sp}} = \frac{U^{*}_{C^{L\times L}_{sp}}}{L}$. ■

**Note:**

1) From **Lemma X.5.3** and **Theorem X.6**, **Algorithm X.2** and **Algorithm X.3** are equivalent under equal opportunity traversals. Therefore, **Algorithm X.3** may be viewed as a generalized extension of **Algorithm X.2**.

**Algorithm X.4: The algorithm flow of maximizing end-to-end asymptotic joint throughput of a pair of paths.**

Given the distribution of the geographical locations of network nodes, and given the pairs of source nodes and destination nodes belonging to the two primary paths respectively, the following algorithm flow is executed so as to maximize end-to-end asymptotic joint throughput of a pair of paths:

BEGIN

WHILE($R_{SP_{[1,2]}}^{\infty}$ is not yet optimal and can be further improved by adjusting routing selections for the pair of paths) DO

For two pairs of source nodes and destination nodes, two multi-hop routes, which are used to support the transmissions of information blocks of two primary paths, are selected to improve the asymptotic joint throughput of the pair of paths;

Based on the generated multi-hop routes, the transmission coverage relationships and the interference relationships between nodes of the pair of paths are formed;

Based on the obtained transmission coverage relationships and interference relationships between nodes, the intrinsic interference intensities $I_{SP_{[1]}}^{*}$, $I_{SP_{[2]}}^{*}$ and $I_{SP_{[1,2]}}^{*}$ corresponding to $P_{[1]}$, $P_{[2]}$ and $P_{[1,2]}$ are calculated, respectively;

WHILE($R_{SP_{[1,2]}}^{\infty}$ is not yet optimal and can be further improved by adjusting the reachable periods of the two primary paths) DO

Based on the obtained intrinsic interference intensities $I_{SP_{[1]}}^{*}$ and $I_{SP_{[2]}}^{*}$, the reachable periods $T_{SP_{[1]}}$ ($\geq I_{SP_{[1]}}^{*}$) and $T_{SP_{[2]}}$ ($\geq I_{SP_{[2]}}^{*}$) adopted by $P_{[1]}$ and $P_{[2]}$ respectively, which can further improve the asymptotic joint throughput of the pair of paths, are obtained;

Based on the obtained reachable periods $T_{SP_{[1]}}$ and $T_{SP_{[2]}}$, the subsets $\Phi'_{SP_{[1]}}(i, T_{SP_{[1]}})$ $(1 \leq i \leq T_{SP_{[1]}})$ and $\Phi'_{SP_{[2]}}(j, T_{SP_{[2]}})$ $(1 \leq j \leq T_{SP_{[2]}})$ of equally spaced nodes in the primary paths $P_{[1]}$ and $P_{[2]}$ are generated respectively;

Based on the obtained subsets of equally spaced nodes and the interference relationships among them, the matrix $\boldsymbol{C}$ of joint concurrency relationship between subsets of equally spaced nodes of $P_{[1,2]}$ is obtained;

WHILE($R_{SP_{[1,2]}}^{\infty}$ is not yet optimal and can be further improved by adjusting the activation times of the two primary paths) DO

11: The activation times $L_1$ and $L_2$ of the subsets of equally spaced nodes in $P_{[1]}$ and $P_{[2]}$, which can improve the asymptotic joint throughput of a pair of paths within a reachable joint period, is obtained respectively;

12: Based on the obtained matrix $\boldsymbol{C}$ and the selected activation times $L_1$ and $L_2$, generated the corresponding $L_1 \times L_2$-th continuation matrix $\boldsymbol{C}^{L_1 \times L_2}$, and calculated out its maximum reachable supporting numbers $U^*_{\boldsymbol{C}_{sp}^{L_1 \times L_2}}$;

13: Based on the above determined parameters (i.e. end-to-end routings, $T_{S_{P_{[1]}}}$, $T_{S_{P_{[2]}}}$, $L_1$, $L_2$, and $U^*_{\boldsymbol{C}_{sp}^{L_1 \times L_2}}$), calculated the corresponding asymptotic joint throughput of a pair of paths as: $R^{\infty}_{S_{P_{[1,2]}}} = \frac{L_1 + L_2}{L_1 \cdot T_{S_{P_{[1]}}} + L_2 \cdot T_{S_{P_{[2]}}} - U^*_{\boldsymbol{C}_{sp}^{L_1 \times L_2}}}$;

14: END WHILE

15: END WHILE

16: END WHILE

17: After the relevant parameters of maximizing the asymptotic joint throughput are determined, execute the time beat allocation algorithm (such as the one given in **Algorithm X.3**), so as to implement the optimal allocation of $T_{S_{P_{[1,2]}}} = L_1 \cdot T_{S_{P_{[1]}}} + L_2 \cdot T_{S_{P_{[2]}}} - U^*_{\boldsymbol{C}_{sp}^{L_1 \times L_2}}$ time beats within a joint period. And based on the optimized allocation of time beats, take $T_{S_{P_{[1,2]}}}$ time beats as a cycle to activate the corresponding subsets of equally spaced nodes, so as to realize the volatility transmission of information blocks on the pair of paths.

18: END

**Note:**

1) The above algorithm flow can be realized in a centralized way or in a more distributed way (i.e. not a purely distributed implementation). For example, in a possible centralized implementation, a centralized control node can be selected to collect necessary information, complete relevant calculations, and issue time beat allocation instructions and so on. In a more distributed implementation, one possible way is to select one appropriate node from each of the two primary paths to serve as the coordination nodes between the two primary paths. The selected coordination nodes are responsible for: collecting the relevant information of nodes in the corresponding primary path, interacting with the coordination node of another primary path, calculating the time beat allocation of the corresponding primary path according to the same algorithm, and finally issuing the instructions of the time beat allocation in the corresponding primary path.

## XI. A general method for solving the intrinsic period of a primary path

In the previous part of the paper, **Theorem VIII.1** gives an important conclusion that the intrinsic period of a primary path is equal to its intrinsic interference intensity. However, this conclusion is based on the assumption that a primary path should follow the interference decreasing rule between nodes (i.e. **Rule IV.1**). Considering the fact that in the real world not all the primary paths follow this assumption, in this section we introduce the concept of the Distribution Spectrum of Interference-Spacing of nodes (i.e. DSIS) of a primary path to find out a more general method for solving the intrinsic period of a primary path. In this section, we first define the concept of the DSIS of a primary path and its related basic concepts (such as ***Definition XI.1***). Next, we analyze several basic properties related to the DSIS of a primary path, and in ***Corollary XI.3.1*** the sufficient and necessary condition for a positive integer $T$ to be the intrinsic period of a primary path is proved. Finally, based on the obtained fundamental properties of the DSIS of a primary path, we propose a simple and general algorithm for solving the intrinsic period of a primary path (i.e. **Algorithm XI.1**).

### $A$. The definition of the DSIS of a primary path

Consider that in a primary path (**Definition III.1**) consisting of $N+1$ nodes, information blocks are relayed from the node 1 (i.e. the source node) through the node 2, the node 3,..., the node $i$,..., and the node $N\ (\geq 1)$ to the final node $N+1$ (i.e. the destination node). If two different sending nodes, like $k\ (1 \leq k \leq N)$ and $l\ (1 \leq l \leq N)$, in $P$ are activated within a same time beat, which brings about the interference with the correct receptions of information blocks of the corresponding receiving nodes, these two nodes cannot be concurrent within the same time beat, i.e. there is an interference relationship between the two sending nodes (**Definition IV.3**). According to **Definition III.4**, the spacing (i.e. "distance") between these two mutually interfering nodes in the primary path is $|k-l|$.

**Definition XI.1：The Distribution Spectrum of Interference-Spacing of nodes (i.e. DSIS) of a primary path**

For a primary path $P$, let us construct a function mapping relationship $F_{S_P}^{I_d}(\eta)$, where the independent variable $\eta\ (1 \leq \eta \leq N-1, \eta \in \mathbb{Z})$ represents the spacing between mutually interfering sending node pairs on $P$. The value $F_{S_P}^{I_d}(\eta)\ (\geq 0)$ represents the total number of pairs of mutually interfering sending nodes with a spacing of $\eta$ in the set $S_P$ of all the sending nodes of $P$. The function mapping relationship $F_{S_P}^{I_d}(\eta)$ is referred to as the Distribution Spectrum of Interference-Spacing of nodes (i.e. DSIS) of a primary path $P$. $\eta$ is the sequence number of a spectral line, and the corresponding value $F_{S_P}^{I_d}(\eta)$ is defined as the intensity of a spectral line with the sequence number of $\eta$. For the spectrum $F_{S_P}^{I_d}(\eta)$, if all the intensities of its spectral lines are greater than 0 (i.e. $F_{S_P}^{I_d}(\eta) > 0\ (1 \leq \eta \leq N-1)$), it is called as a pure non-zero distribution spectrum of interference-spacing of nodes (referred to as "pure non-zero spectrum" for short).

**Note:**

1) Considering that the interference relationship given in **Definition IV.3** is mutual, when counting the number of pairs of mutually interfering nodes, the interfering node pair $(k, l)$ and the interfering node pair $(l, k)$ will be treated equally and only counted as one pair of interfering nodes.

**Definition XI.2：A zero point of the DSIS of a primary path**

If the intensity $F_{S_P}^{I_d}(\eta)$ of a spectral line with sequence number $\eta\ (1 \leq \eta \leq N-1)$ is 0, the spectral line with sequence number $\eta$ is called as a zero point of the spectrum $F_{S_P}^{I_d}(\eta)$.

**Definition XI.3：An equally spaced cumulative zero point of the DSIS of a primary path with a spacing of $T$**

For a spectral line with sequence number $\eta$ in a spectrum $F_{S_P}^{I_d}(\eta)$, if $\sum_{i=0}^{\left\lfloor \frac{(N-1)-\eta}{T} \right\rfloor} F_{S_P}^{I_d}(\eta + i \cdot T) = 0\ (1 \leq \eta \leq N-1, i \in \mathbb{Z})$ holds (where $T \geq 1, T \in \mathbb{Z}$), then the spectral line $\eta$ is said to be an equally spaced cumulative zero point of the spectrum $F_{S_P}^{I_d}(\eta)$ with a spacing of $T$.

**Definition XI.4：A periodic equally spaced cumulative zero point of the DSIS of a primary**

**path**

If the spectral line $\eta$ is an equally spaced cumulative zero point of the spectrum $F_{S_P}^{I_d}(\eta)$ with a spacing of $T = \eta$, then it is said to be a periodic equally spaced cumulative zero point of the spectrum $F_{S_P}^{I_d}(\eta)$ with a period of $\eta$.

**Definition XI.5：The minimum periodic equally spaced cumulative zero point of the DSIS of a primary path**

Among all the periodic equally spaced cumulative zero points of a spectrum $F_{S_P}^{I_d}(\eta)$, the spectral line with the minimum sequence number is called as the minimum periodic equally spaced cumulative zero point of the spectrum $F_{S_P}^{I_d}(\eta)$, and its corresponding sequence number is denoted as $\eta_{Az}^{min}$.

**Definition XI.6：A point of returning to zero of the DSIS of a primary path**

For a spectrum $F_{S_P}^{I_d}(\eta)$, if the spectral line $\eta\ (1 \leq \eta \leq N-1)$ satisfies the condition of $F_{S_P}^{I_d}(m) = 0\ (\eta \leq m \leq N-1, m \in \mathbb{Z})$, then the spectral line $\eta$ is said to be a point of returning to zero of the spectrum $F_{S_P}^{I_d}(\eta)$.

**Definition XI.7：The minimum point of returning to zero of the DSIS of a primary path**

For a spectrum $F_{S_P}^{I_d}(\eta)$ with some points of returning to zero, the spectral line with the minimum sequence number among all its points of returning to zero is said to be the minimum point of returning to zero of the spectrum $F_{S_P}^{I_d}(\eta)$, and its corresponding sequence number is denoted as $\eta_{Bz}^{min}$.

**Definition XI.8：A zeroing primary path and a non-zeroing primary path**

If the number of points of returning to zero in a spectrum $F_{S_P}^{I_d}(\eta)$ is greater than 0, then the spectrum $F_{S_P}^{I_d}(\eta)$ is called as a zeroing spectrum, and its corresponding primary path $P$ is called as a zeroing primary path. Otherwise, the spectrum $F_{S_P}^{I_d}(\eta)$ is called as a non-zeroing spectrum, and its corresponding primary path $P$ is called as a non-zeroing primary path.

**Definition XI.9: An interference continuous primary path and an interference non-continuous primary path**

As for any two sending nodes, like $k$ and $l$ $(k < l)$, in a primary path $P$ having interference relationships between them, if the subset of nodes composed of all the sending nodes (including $k$ and $l$) with their sequence numbers between $k$ and $l$ is a continuous second-order interference subset (**Definition V.23**), $P$ is defined as an interference continuous primary path, otherwise $P$ is called as an interference non-continuous primary path.

**Definition XI.10: A compact primary path and a non-compact primary path**

If the intrinsic period (**Definition VII.4**) of a primary path $P$ is equal to its intrinsic interference intensity (**Definition V.14**), $P$ is called as a compact primary path, otherwise $P$ is called as a non-compact primary path.

### *B*. Several basic properties of the DSIS of a primary path

**Theorem XI.1: A basic property of an equally spaced cumulative zero point of a spectrum $F_{S_P}^{I_d}(\eta)$ with a spacing of $T$.**

**Assume that the DSIS of a primary path $P$ (including $N$ sending nodes) is $F_{S_P}^{I_d}(\eta)$. If the spectral line with sequence number $\eta$ is an equally spaced cumulative zero point of the spectrum $F_{S_P}^{I_d}(\eta)$ with a spacing of $T$, and $1 \le \eta + i \cdot T \le N - 1$ ($i \in \mathbb{Z}^+$), then the spectral line with sequence number $\eta + i \cdot T$ is also an equally spaced cumulative zero point of the spectrum $F_{S_P}^{I_d}(\eta)$ with a spacing of $T$.**

**Proof:**

Due to the fact that the spectral line with the sequence number $\eta$ is an equally spaced cumulative zero point of the spectrum $F_{S_P}^{I_d}(\eta)$ with a spacing of $T$, according to **Definition XI.3**, the following equation holds

$$\sum_{j=0}^{\left\lfloor \frac{(N-1)-\eta}{T} \right\rfloor} F_{S_P}^{I_d}(\eta + j \cdot T) = 0 \tag{XI.1}$$

According to **Definition XI.1**, the intensities of all the spectral lines are greater than or equal to 0, therefore there is

$$\sum_{j=0}^{i-1} F_{S_P}^{I_d}(\eta + j \cdot T) = 0 \tag{XI.2}$$

where $i - 1 \le \left\lfloor \frac{(N-1)-\eta}{T} \right\rfloor$. By combining **Equation XI.1** and **Equation XI.2**, we can further obtain

$$\begin{aligned} \sum_{j=0}^{\left\lfloor \frac{(N-1)-\eta}{T} \right\rfloor} F_{S_P}^{I_d}(\eta + j \cdot T) &= \sum_{j=0}^{i-1} F_{S_P}^{I_d}(\eta + j \cdot T) + \sum_{j=i}^{\left\lfloor \frac{(N-1)-\eta}{T} \right\rfloor} F_{S_P}^{I_d}(\eta + j \cdot T) \\ &= \sum_{j=i}^{\left\lfloor \frac{(N-1)-\eta}{T} \right\rfloor} F_{S_P}^{I_d}(\eta + j \cdot T) \end{aligned}$$

$$
\begin{aligned}
&= \sum_{j=i}^{\left\lfloor\frac{(N-1)-\eta}{T}\right\rfloor} F_{S_P}^{I_d}((\eta + i \cdot T) + (j-i) \cdot T) \\
&= 0
\end{aligned}
$$

(XI.3)

Let $k = j - i$, and by substituting $k$ into the above equation, it can be obtained that

$$
\begin{aligned}
\sum_{j=0}^{\left\lfloor\frac{(N-1)-\eta}{T}\right\rfloor} F_{S_P}^{I_d}(\eta + j \cdot T) &= \sum_{k=0}^{\left\lfloor\frac{(N-1)-\eta}{T}\right\rfloor - i} F_{S_P}^{I_d}((\eta + i \cdot T) + k \cdot T) \\
&= \sum_{k=0}^{\left\lfloor\frac{(N-1)-(\eta+i\cdot T)}{T}\right\rfloor} F_{S_P}^{I_d}((\eta + i \cdot T) + k \cdot T) \\
&= 0
\end{aligned}
$$

(XI.4)

According to **Definition XI.3**, the spectral line $\eta + i \cdot T$ is also an equally spaced cumulative zero point of the spectrum $F_{S_P}^{I_d}(\eta)$ with a spacing of $T$. ■

> **Theorem XI.2: A basic property of a periodic equally spaced cumulative zero point of a spectrum $F_{S_P}^{I_d}(\eta)$.**
>
> **Assume that the DSIS of a primary path $P$ (including $N$ sending nodes) is $F_{S_P}^{I_d}(\eta)$. If the spectral line with sequence number $\eta$ is a periodic equally spaced cumulative zero point of the spectrum $F_{S_P}^{I_d}(\eta)$ with a period of $\eta$, and $1 \le i \cdot \eta \le N-1$ $(i \in \mathbb{Z}^+, i > 1)$, then the spectral line with sequence number $i \cdot \eta$ is also a periodic equally spaced cumulative zero point of the spectrum $F_{S_P}^{I_d}(\eta)$ with a period of $i \cdot \eta$.**

**Proof:**

Proof by contradiction is adopted. Assuming that spectral line $i \cdot \eta$ is not a periodic equally spaced cumulative zero point of the spectrum $F_{S_P}^{I_d}(\eta)$ with a period of $i \cdot \eta$, according to **Definition XI.4**, it can be inferred that there is at least one spectral line $\eta' = i \cdot \eta + j' \cdot (i \cdot \eta)$ $(i \cdot \eta \le \eta' \le N-1, 0 \le j' \le \left\lfloor\frac{(N-1)-i\cdot\eta}{i\cdot\eta}\right\rfloor)$ among spectral lines $i \cdot \eta + j \cdot (i \cdot \eta)$ $(0 \le j \le \left\lfloor\frac{(N-1)-i\cdot\eta}{i\cdot\eta}\right\rfloor)$, with its corresponding intensity of the spectral line satisfying $F_{S_P}^{I_d}(\eta') > 0$. Therefore, it can be obtained that

$$
\begin{aligned}
F_{S_P}^{I_d}(\eta') &= F_{S_P}^{I_d}\left(i \cdot \eta + j' \cdot (i \cdot \eta)\right) \\
&= F_{S_P}^{I_d}(\eta + (i-1) \cdot \eta + j' \cdot i \cdot \eta) \\
&= F_{S_P}^{I_d}\left(\eta + \left((j'+1)i - 1\right) \cdot \eta\right) \\
&= F_{S_P}^{I_d}(\eta + i'' \cdot \eta) \\
&> 0
\end{aligned}
$$

(XI.5)

where $i'' \triangleq (j'+1)i - 1$. Since $j' \geq 0$ and $i > 1$, we have $i'' \geq 1$. On the other hand, according to **Definition XI.4**, since spectral line $\eta$ is a periodic equally spaced cumulative zero point of the spectrum $F_{S_P}^{I_d}(\eta)$ with a period of $\eta$, the intensities of all the spectral lines, corresponding to the sequence numbers which can be represented in the form of $\eta + i \cdot \eta\ (i \geq 0)$ in the spectrum $F_{S_P}^{I_d}(\eta)$, must be equal to 0. Therefore, it can be found that the above assumption is contradictory.

■

**Theorem XI.3: The sufficient and necessary condition for a positive integer $T$ to be a reachable period (Definition VII.2) of a primary path $P$.**

**Assume that the DSIS of a primary path $P$ (including $N$ sending nodes) is $F_{S_P}^{I_d}(\eta)$. The sufficient and necessary condition for a positive integer $T\ (1 \leq T \leq N-1)$ to be a reachable period of a primary path $P$ is that the spectral line with sequence number $\eta = T$ is a periodic equally spaced cumulative zero point of the spectrum $F_{S_P}^{I_d}(\eta)$ with a period of $T$.**

**Proof:**

Firstly, let us prove that if $T$ is a reachable period of $P$, then the spectral line with sequence number $\eta = T$ is a periodic equally spaced cumulative zero point of the spectrum $F_{S_P}^{I_d}(\eta)$ with a period of $T$. Proof by contradiction is adopted. Assuming that the spectral line $\eta = T$ is not a periodic equally spaced cumulative zero point of the spectrum $F_{S_P}^{I_d}(\eta)$, according to **Definition XI.4**, there is at least one spectral line $\eta' = T + j' \cdot T\ (T \leq \eta' \leq N-1, 0 \leq j' \leq \left\lfloor \frac{(N-1)-T}{T} \right\rfloor)$ among the spectral lines $T + j \cdot T\ (0 \leq j \leq \left\lfloor \frac{(N-1)-T}{T} \right\rfloor, j \in \mathbb{Z})$, with the intensity of its spectral line satisfying $F_{S_P}^{I_d}(\eta') > 0$. That is to say, in the set $S_P$ of all the sending nodes of $P$, there is at least one pair of interfering nodes $(k,l)$ with a spacing between them being $d_{S_P}^{k,l} = |k-l| = \eta' = (1+j') \cdot T$ (**Definition III.4**). Because the spacing between the two interfering sending nodes is an integer multiple of the reachable period $T$ of $P$, these two interfering sending nodes must share a common subset of equally spaced nodes with the spacing between adjacent nodes being $T$ in the primary path $P$, which contradicts to the premise that all the subsets of equally spaced nodes with the spacing between adjacent nodes being $T$ in $P$ are second-order concurrency subsets.

Secondly, let us prove that if the spectral line with sequence number $\eta = T$ is a periodic equally spaced cumulative zero point of the spectrum $F_{S_P}^{I_d}(\eta)$ with a period of $T$, then $T$ is a reachable period of $P$. Again, proof by contradiction is adopted. Assuming that $T$ is not a reachable period of $P$, there is at least one subset $\Phi'_{S_P}$ of equally spaced nodes with the spacing between adjacent nodes being $T$ in the primary path $P$, which is not a second-order concurrency subset. That is to say, in the set $\Phi'_{S_P}$, there is at least one pair of interfering nodes $(k', l')$ with a spacing $d_{S_P}^{k',l'} = |k'-l'| = T + j' \cdot T \ (j' \geq 0, j' \in \mathbb{Z})$ between them. According to the definition of a subset of equally spaced nodes with the spacing between adjacent nodes being $T$ (**Definition VII.1**), it is known that $T \leq d_{S_P}^{k',l'} \leq N-1$. Therefore, it can be obtained that the intensity $F_{S_P}^{I_d}(d_{S_P}^{k',l'})$ of the spectral line $\eta = d_{S_P}^{k',l'}$ is larger than 0, i.e. $F_{S_P}^{I_d}(d_{S_P}^{k',l'}) = F_{S_P}^{I_d}(T + j' \cdot T) > 0$, which contradicts to the premise that the spectral line with sequence number $\eta = T$ is a periodic equally spaced cumulative zero point of the spectrum $F_{S_P}^{I_d}(\eta)$ with a period of $T$.■

**Corollary XI.3.1: The sufficient and necessary condition for a positive integer $T$ to be the intrinsic period (Definition VII.4) of a primary path $P$.**

**Assume that the DSIS of a primary path $P$ (including $N$ sending nodes) is $F_{S_P}^{I_d}(\eta)$. The sufficient and necessary condition for a positive integer $T$ $(1 \leq T \leq N-1)$ to be the intrinsic period of a primary path $P$ is that the spectral line with sequence number $\eta = T$ is the minimum periodic equally spaced cumulative zero point of the spectrum $F_{S_P}^{I_d}(\eta)$.**

**Proof:**

Firstly, let us prove that if $T$ is the intrinsic period of $P$, then the spectral line with sequence number $\eta = T$ is the minimum periodic equally spaced cumulative zero point of the spectrum $F_{S_P}^{I_d}(\eta)$. Proof by contradiction is adopted. Assuming that the spectral line with sequence number $\eta = T$ is not the minimum periodic equally spaced cumulative zero point of the spectrum $F_{S_P}^{I_d}(\eta)$, there must exist a spectral line $\eta' < T$ $(\eta' \in \mathbb{Z}^+)$, which is a periodic equally spaced cumulative zero point of the spectrum $F_{S_P}^{I_d}(\eta)$ with a period of $\eta'$. According to **Theorem XI.3**, the positive integer $\eta'$ must be a reachable period of $P$, which contradicts to the premise that $T$ is the intrinsic period of $P$.

Secondly, let us prove that if the spectral line with sequence number $\eta = T$ is the minimum periodic equally spaced cumulative zero point of the spectrum $F_{S_P}^{I_d}(\eta)$, then $T$ is the intrinsic period of $P$. Proof by contradiction is adopted. According to **Theorem XI.3**, since the spectral line with sequence number $\eta = T$ is the minimum periodic equally spaced cumulative zero point of the spectrum $F_{S_P}^{I_d}(\eta)$ (i.e. the spectral line $\eta = T$ is also a periodic equally spaced cumulative zero point of $F_{S_P}^{I_d}(\eta)$), $T$ is at least a reachable period of $P$. Assuming that $T$ is not the intrinsic period of $P$, there must exist a positive integer $T' < T$ is another reachable period of $P$. According to **Theorem XI.3**, the spectral line with sequence number $\eta' = T'$ is a periodic equally spaced cumulative zero point of the spectrum $F_{S_P}^{I_d}(\eta)$ with a period of $T'$, which contradicts to the premise that the spectral line with sequence number $\eta = T$ is the minimum periodic equally spaced cumulative zero point of $F_{S_P}^{I_d}(\eta)$.■

**Corollary XI.3.2: If a positive integer $T$ is a reachable period of a primary path $P$, then the positive integer $i \cdot T$ is also a reachable period of the primary path $P$.**

**If a positive integer $T$ $(1 \leq T \leq N-1)$ is a reachable period of a primary path $P$ (including $N$ sending nodes), and $1 \leq i \cdot T \leq N-1$, then the positive integer $i \cdot T$ is also a reachable period of $P$.**

**Proof:**

According to **Theorem XI.3**, if $T$ is a reachable period of $P$, then the spectral line with sequence number $\eta = T$ is a periodic equally spaced cumulative zero point of the spectrum $F_{S_P}^{I_d}(\eta)$ with a period of $T$. Furthermore, according to **Theorem XI.2**, the spectral line with sequence number $\eta = i \cdot T\ (1 \leq i \cdot T \leq N-1)$ is also a periodic equally spaced cumulative zero point of the spectrum $F_{S_P}^{I_d}(\eta)$ with a period of $i \cdot T$. Finally, again based on **Theorem XI.3**, it can be obtained that the positive integer $i \cdot T$ is also a reachable period of $P$.■

**Corollary XI.3.3: The intrinsic period of a primary path with a pure non-zero spectrum is $N$.**

**If the DSIS of a primary path $P$ (including $N$ sending nodes) is a pure non-zero spectrum (Definition XI.1), the intrinsic period of $P$ is $N$.**

**Proof:**

According to the definition of a reachable period of a primary path (**Definition VII.2**), the positive integer $N$ is a reachable period of a primary path $P$ (i.e. it is evident that each sending node in $P$ can individually serve as a second-order concurrency subset of nodes). Since the spectrum $F_{S_P}^{I_d}(\eta)$ is a pure non-zero spectrum, for a positive integer $T\ (1 \leq T \leq N-1)$, the intensity of the corresponding spectral line $\eta = T$ is greater than 0. Therefore, the spectral line with sequence number $\eta = T$ is not a periodic equally spaced cumulative zero point of the spectrum $F_{S_P}^{I_d}(\eta)$. Furthermore, according to **Theorem XI.3**, positive integer $T\ (1 \leq T \leq N-1)$ is not a reachable period of $P$. Therefore, the positive integer $N$ is the minimum reachable period of $P$, i.e. the intrinsic period of $P$.■

**Lemma XI.4.1: A point of returning to zero of the DSIS of a primary path is a periodic equally spaced cumulative zero point of it.**

**Assume that the DSIS of a primary path $P$ (including $N$ sending nodes) is $F_{S_P}^{I_d}(\eta)$. If a spectral line with sequence number $\eta = T$ is a point of returning to zero of the spectrum $F_{S_P}^{I_d}(\eta)$, it must be a periodic equally spaced cumulative zero point of $F_{S_P}^{I_d}(\eta)$ with a period of $T$.**

**Proof:**

According to the definition of a point of returning to zero of the DSIS (**Definition XI.6**), we have

$$\sum_{i=0}^{\left\lfloor\frac{(N-1)-T}{T}\right\rfloor} F_{S_P}^{I_d}(T+i\cdot T)=0$$

(XI.6)

Based on the definition of a periodic equally spaced cumulative zero point of the DSIS (**Definition XI.4**), the spectral line with sequence number $\eta = T$ is a periodic equally spaced cumulative zero point of $F_{S_P}^{I_d}(\eta)$ with a period of $T$. ■

**Lemma XI.4.2: The sequence number of a spectral line of a point of returning to zero of the DSIS of a primary path is a reachable period of the primary path.**

**Assume that the DSIS of a primary path $P$ (including $N$ sending nodes) is $F_{S_P}^{I_d}(\eta)$. If the spectral line with sequence number $\eta = T$ is a point of returning to zero of the spectrum $F_{S_P}^{I_d}(\eta)$, then $T$ is a reachable period of the primary path $P$.**

**Proof:**

According to **Lemma XI.4.1**, the spectral line with sequence number $\eta = T$ is a periodic equally spaced cumulative zero point of $F_{S_P}^{I_d}(\eta)$ with a period of $T$. Furthermore, according to **Theorem XI.3**, $T$ is a reachable period of $P$.■

**Lemma XI.4.3: The intrinsic period of a primary path is not larger than the sequence number of a spectral line of the minimum point of returning to zero of the DSIS of the primary path.**

**Assume that the DSIS of a primary path $P$ (including $N$ sending nodes) is $F_{S_P}^{I_d}(\eta)$. The intrinsic period of $P$ is $T_{S_P}^{*}$. If the spectrum $F_{S_P}^{I_d}(\eta)$ is a zeroing spectrum (Definition XI.8), and the corresponding sequence number of its minimum point of returning to zero is $\eta_{Bz}^{min}$, then $T_{S_P}^{*} \leq \eta_{Bz}^{min}$ holds.**

**Proof:**

According to **Lemma XI.4.2**, $\eta_{Bz}^{min}$ is a reachable period of $P$. Furthermore, based on the definition of the intrinsic period of a primary path (**Definition VII.4**), it can be concluded that $T_{S_P}^* \leq \eta_{Bz}^{min}$ holds. ■

**Theorem XI.4: The intrinsic interference intensity of a primary path is not larger than its intrinsic period, and the intrinsic period of the primary path is not larger than the sequence number of a spectral line of the minimum point of returning to zero of the DSIS.**

**Assume that the DSIS of a primary path $P$ (including $N$ sending nodes) is $F_{S_P}^{I_d}(\eta)$. The intrinsic interference intensity and the intrinsic period of $P$ are $I_{S_P}^*$ and $T_{S_P}^*$, respectively. If the spectrum $F_{S_P}^{I_d}(\eta)$ is a zeroing spectrum, and the corresponding sequence number of its minimum point of returning to zero is $\eta_{Bz}^{min}$, then $I_{S_P}^* \leq T_{S_P}^* \leq \eta_{Bz}^{min}$ holds.**

**Proof:**

By combing **Lemma VIII.1.1** and **Lemma XI.4.3**, it can be readily concluded that $I_{S_P}^* \leq T_{S_P}^* \leq \eta_{Bz}^{min}$ holds. ■

**Lemma XI.5.1: An interference continuous primary path must satisfy the interference decreasing rule between nodes.**

**If a primary path $P$ (including $N$ sending nodes) is an interference continuous primary path, it must satisfy the interference decreasing rule between nodes (i.e. Rule IV.1).**

**Proof:**

Firstly, we prove that if two different nodes $k$ and $l\,(k < l)$ in $P$ have concurrency relationship, and node $l + 1$ is another sending node of $P$, then it can be deduced that node $k$ and node $l + 1$ also have concurrency relationship. Proof by contradiction is adopted. If node $k$ and node $l + 1$ interfere with each other, it can be inferred from the fact of $P$ being an interference continuous primary path that node $k$ and node $l$ must interfere with each other, which contradicts the fact that node $k$ and node $l$ have concurrency relationship.

Secondly, we prove that if two different nodes $k$ and $l\,(k < l)$ in $P$ have concurrency relationship, and node $k - 1$ is another sending node of $P$, then it can be deduced that node $k -$ $1$ and node $l$ also have concurrency relationship. Proof by contradiction is adopted. If node $k-1$ and node $l$ interfere with each other, it can be inferred from the fact of $P$ being an interference continuous primary path that node $k$ and node $l$ must interfere with each other, which contradicts the fact that nodes $k$ and $l$ have concurrency relationship. ■

**Theorem XI.5: The necessary and sufficient condition for a primary path to be an interference continuous primary path is that it satisfies the interference decreasing rule between nodes.**

**The necessary and sufficient condition for a primary path $\boldsymbol{P}$ (including $\boldsymbol{N}$ sending nodes) to be an interference continuous primary path is that it satisfies the interference decreasing rule between nodes (i.e. Rule IV.1).**

**Proof:**

It can be readily proved by combing **Theorem VI.2** and **Lemma XI.5.1**. ■

**Corollary XI.5.1: An interference continuous primary path must also be a compact primary path.**

**If a primary path $\boldsymbol{P}$ (including $\boldsymbol{N}$ sending nodes) is an interference continuous primary path, then it must be a compact primary path.**

**Proof:**

Since $P$ is an interference continuous primary path, it must follow the interference decreasing rule between nodes. Furthermore, in this case, according to **Theorem VIII.1**, the intrinsic interference intensity of $P$ equals to its intrinsic period, which indicates that $P$ is a compact primary path (**Definition XI.10**). ■

**Theorem XI.6: The non-zero characteristics for intensities of partial spectral lines of the DSIS of an interference continuous primary path.**

**Assume that a primary path $\boldsymbol{P}$ (including $\boldsymbol{N}$ sending nodes) is an interference continuous primary path, and its DSIS is $\boldsymbol{F_{S_P}^{I_d}(\eta)}$. The spectrum $\boldsymbol{F_{S_P}^{I_d}(\eta)}$ is a zeroing spectrum, and the corresponding sequence number of its minimum point of returning to**

**zero is $\eta_{Bz}^{min}$ (> 1), then $F_{S_P}^{I_d}(\eta) > 0$ $(1 \le \eta < \eta_{Bz}^{min})$ holds.**

**Proof:**

According to the definition of the minimum point of returning to zero of the spectrum $F_{S_P}^{I_d}(\eta)$ (**Definition XI.7**), the intensity $F_{S_P}^{I_d}(\eta_{Bz}^{min} - 1)$ of the spectral line with sequence number $\eta = \eta_{Bz}^{min} - 1$ is larger than 0, which suggests that there is at least one pair of mutually interfering sending nodes $k$ and $l$ $(k < l)$ in $P$ with a spacing between them being equal to $d_{S_P}^{k,l} = |k - l| = \eta_{Bz}^{min} - 1$. Since $P$ is an interference continuous primary path, the subset of nodes composed of all the sending nodes (including $k$ and $l$) with their sequence numbers between $k$ and $l$ is a continuous second-order interference subset $S'_P \triangleq \{i | i \in S_P; k \le i \le l\}$ ($S_P$ is the set of nodes composed of all the sending nodes in $P$). Therefore, at least one sending node $m$ $(k < m \le l)$ that interferes with node $k$ can be found in $S'_P$, so that $d_{S_P}^{k,m} = |k - m| = \eta$ $(1 \le \eta \le \eta_{Bz}^{min} - 1)$, which means that $F_{S_P}^{I_d}(\eta) > 0$ $(1 \le \eta < \eta_{Bz}^{min})$ holds. ■

**Theorem XI.7: The strict monotonic decreasing property for intensities of partial spectral lines of the DSIS of an interference continuous primary path.**

**Assume that a primary path $P$ (including $N$ sending nodes) is an interference continuous primary path, and its DSIS is $F_{S_P}^{I_d}(\eta)$. The spectrum $F_{S_P}^{I_d}(\eta)$ is a zeroing spectrum, and the corresponding sequence number of its minimum point of returning to zero is $\eta_{Bz}^{min}$ (> 1), then $F_{S_P}^{I_d}(\eta) > F_{S_P}^{I_d}(\eta + 1)$ $(1 \le \eta < \eta_{Bz}^{min})$ holds.**

**Proof:**

Firstly, let us prove the case of $\eta = \eta_{Bz}^{min} - 1$. According to the definition of the minimum point of returning to zero of the DSIS of a primary path, in this case, $F_{S_P}^{I_d}(\eta) > F_{S_P}^{I_d}(\eta + 1) = F_{S_P}^{I_d}(\eta_{Bz}^{min}) = 0$ must hold.

Secondly, let us prove the case of $1 \le \eta < \eta_{Bz}^{min} - 1$. Assuming that $F_{S_P}^{I_d}(\eta + 1) = m$ $(m > 0)$, in the set of sending nodes in $P$, there are $m$ pairs of mutually interfering node pairs with spacing between them being $\eta + 1$. Without loss of generality, the $m$ pairs of mutually interfering node pairs can be respectively expressed as: $(k_1, k_1+\eta+1)$, $(k_2, k_2+\eta+1)$, ..., and $(k_m, k_m+\eta+1)$ $(k_1 < k_2 <\ldots< k_m)$. Based on these listed node pairs, and combing with the fact that the primary path $P$ is an interference continuous primary path, therefore, at least $m+1$ pairs of mutually interfering node pairs with spacing between them being $\eta$ can be found. This is because that $m+1$ pairs of mutually interfering node pairs can be easily constructed and represented as: $(k_1, k_1+\eta)$, $(k_2, k_2+\eta)$, ..., $(k_m, k_m+\eta)$, and $(k_m+1, k_m+\eta+1)$ $(k_1 < k_2 <\ldots< k_m)$. Therefore, $F_{S_P}^{I_d}(\eta) > m$ $(m > 0)$ holds, i.e. $F_{S_P}^{I_d}(\eta) > F_{S_P}^{I_d}(\eta+1)$.■

**Corollary XI.7.1: The intrinsic interference intensity, the intrinsic period, and the sequence number of a spectral line of the minimum point of returning to zero of the DSIS of an interference continuous primary path are equal to each other.**

**Assume that a primary path $P$ (including $N$ sending nodes) is an interference continuous primary path, and its intrinsic interference intensity, intrinsic period, and DSIS is $I_{S_P}^*$, $T_{S_P}^*$, and $F_{S_P}^{I_d}(\eta)$, respectively. The spectrum $F_{S_P}^{I_d}(\eta)$ is a zeroing spectrum, and the corresponding sequence number of its minimum point of returning to zero is $\eta_{Bz}^{min}$, then $I_{S_P}^* = T_{S_P}^* = \eta_{Bz}^{min}$ holds.**

**Proof:**

According to **Theorem VIII.1**, it can be readily obtained that $I_{S_P}^* = T_{S_P}^*$.

Below, we just need to prove that $T_{S_P}^* = \eta_{Bz}^{min}$. According to **Lemma XI.4.3**, we know that $T_{S_P}^* \le \eta_{Bz}^{min}$.

As for the case of $\eta_{Bz}^{min} = 1$, since an intrinsic period should not less than 1, and $\eta_{Bz}^{min}$ is a reachable period of $P$ (according to **Lemma XI.4.2**), therefore, we conclude that $T_{S_P}^* = \eta_{Bz}^{min}$, i.e. $I_{S_P}^* = T_{S_P}^* = \eta_{Bz}^{min}$ holds.

As for the case of $\eta_{Bz}^{min} > 1$, it can be obtained from **Theorem XI.6** that $F_{S_P}^{I_d}(\eta) > 0$ $(1 \le \eta < \eta_{Bz}^{min})$. Therefore, for a sequence number $\eta$ $(1 \le \eta < \eta_{Bz}^{min})$, its corresponding spectral line should not be a periodic equally spaced cumulative zero point of the spectrum $F_{S_P}^{I_d}(\eta)$ with a period of $\eta$. Furthermore, according to **Theorem XI.3**, it can be seen that the positive integer $\eta$ $(1 \le \eta < \eta_{Bz}^{min})$ must not be a reachable period of $P$, and on the other hand $\eta_{Bz}^{min}$ is a reachable period of $P$ (according to **Lemma XI.4.2**), therefore, the intrinsic period $T_{S_P}^*$ of $P$ must be equal to $\eta_{Bz}^{min}$, i.e. $I_{S_P}^* = T_{S_P}^* = \eta_{Bz}^{min}$ holds. ■

### *C*. A general algorithm for solving the intrinsic period of a primary path

In this section, based on **Corollary XI.3.1**, a general algorithm for solving the intrinsic period of a given primary path $P$ is given. That is to say, in the proposed algorithm, we no longer restrict a primary path $P$ to follow the interference decreasing rule between nodes (i.e. **Rule IV.1**). According to **Corollary XI.3.1**, the intrinsic period of a given primary path $P$ is equal to the sequence number of the minimum periodic equally spaced cumulative zero point of its DSIS (i.e. the spectrum $F_{S_P}^{I_d}(\eta)$). Therefore, the problem of solving the intrinsic period is transformed into the problem of searching the minimum periodic equally spaced cumulative zero point of the spectrum $F_{S_P}^{I_d}(\eta)$. The detailed algorithm flow is shown in **Algorithm XI.1** as follows.

**Algorithm XI.1: A general algorithm for solving the intrinsic period of a primary path.**

BEGIN

  Initialize the intrinsic period as $T_{S_P}^{*} = N$;

  // $N$ denotes the number of sending nodes in $P$

  // If a periodic equally spaced cumulative zero point cannot be found within the range of

  // 1 to $N-1$, the intrinsic period must be equal to $N$

  Initialize the sequence number of the spectral line being considered as $\eta = 1$;

  WHILE ($\eta \leq N-1$) DO

    $i = 0$;

    $FLAG = 1$;

    // If the indicator $FLAG$ remains to be 1, it shows that the spectral line being

    // considered is a periodic equally spaced cumulative zero point. Otherwise, it will be

    // set to be 0

    WHILE ($i \leq \left\lfloor \frac{(N-1)-\eta}{\eta} \right\rfloor$) DO

      IF ($F_{S_P}^{I_d}(\eta + i \cdot \eta) > 0$) THEN

        $FLAG = 0$;

        BREAK;

// It shows that the spectral line $\eta$ being considered is not a periodic equally
// spaced cumulative zero point. Therefore, the current WHILE loop should be
// early ended

END IF

$i = i + 1$;

END WHILE // The WHILE loop for $i$

IF $(FLAG = 1)$ THEN

$T_{S_P}^{*} = \eta$;

BREAK;

// It shows that the minimum periodic equally spaced cumulative zero point is
// found, and the intrinsic period $T_{S_P}^{*}$ is set to be the sequence number $\eta$ of
// the spectral line being considered. Then, the current WHILE loop should be
// early ended, and the algorithm flow is ended

END IF

$\eta = \eta + 1$;

END WHILE // The WHILE loop for $\eta$

END

## XII. Conclusions

Aiming at the disorder problem (i.e. uncertainty problem) of the utilization of network resources commonly existing in multi-hop transmission networks, the paper proposes the idea and the corresponding supporting theory, i.e. theory of network wave, by constructing volatility information transmission mechanism between the sending nodes and their corresponding receiving nodes of a pair of paths (composed of two primary paths), so as to improve the orderliness of the utilization of network resources. The research results of the paper lay an ideological and theoretical foundation for further exploring more general methods that can improve the orderly utilization of network resources.

Some important conclusions obtained in the paper are summarized as follows:

1) The concurrent transmission theorem (**Theorem VI.4**) is proved.
2) It is proved that the intrinsic period of a primary path is equal to its intrinsic interference intensity (**Theorem VIII.1**).
3) In **Theorem X.1**, the existence theorem for the upper bound of the asymptotic throughput of a primary path is proved. Moreover, in **Theorem X.2**, it is pointed out that the upper bound of the asymptotic throughput of a primary path can be reached.
4) In **Theorem X.3**, theorem of unique universal optimal partition is proved, i.e. the partition of equally spaced second-order concurrency subsets of nodes is the unique universal partitioning method to achieve the maximum asymptotic throughput of a primary path. Therefore, it is revealed by this theorem that the adoption of network oscillation is not one of optional schemes for multi-hop transmission but the unique universally optimal alternative. This conclusion is consistent with various oscillatory phenomena ubiquitous in nature!
5) We introduce the concept of the Distribution Spectrum of Interference-Spacing of nodes (i.e. DSIS) of a primary path, and several basic properties related to the DSIS of a primary path are analyzed. In **Corollary XI.3.1**, the sufficient and necessary condition for a positive integer $T$ to be the intrinsic period of a primary path is given. Finally, based on the obtained fundamental properties of the DSIS of a primary path, we propose a simple and general algorithm flow for solving the intrinsic period of a primary path (i.e. **Algorithm XI.1**).
6) Necessary and sufficient condition under which the joint period of a pair of paths can reach its intrinsic interference intensity is proved in **Theorem VIII.5**.

7) For a pair of paths, we establish the existence theorem for the upper bound on its asymptotic joint throughput (**Theorem X.4**). We further prove the reachability of this upper bound under equal opportunity traversals (**Theorem X.5**).
8) As for the more general cases of traversals with unequal opportunities, an algorithm for the cooperative volatility transmission of information blocks in a pair of paths based on the set of maximum supporting elements is also proposed (**Algorithm X.3**). The upper bound of the asymptotic joint throughput of a pair of paths achieved by using this algorithm is obtained (**Corollary X.6.1** and **Corollary X.6.2**).

Based on the conclusions obtained in this paper, we can further summarize the advantages of volatility transmission method as: **it can maximize the end-to-end asymptotic throughput of a primary path/a pair of paths while maintaining the orderly utilization of network resources as much as possible.** In the future work, based on the work proposed in this paper, we will further study how to extend the theory of network wave to more general cases, such as the cases including any multiple primary paths (i.e. group of paths).

**Appendix A：Symbol List**

| | |
|---|---|
| $P_{[i]}$ | A primary path with its index number being $i\,(\in \mathbb{Z}^{+})$ (refer to the Definition III.1) |
| $\Delta t$ | Duration of one time beat (refer to the Definition III.1) |
| $S_{P_{[i]}}^{U}$ | The universal node set of a primary path (refer to the Definition III.2) |
| $n_{i,j}$ | A specific node on a primary path $P_{[i]}$ (refer to the Definition III.3) |
| $n_{i,j}^{-}$ | The predecessor node of node $n_{i,j}$ in $P_{[i]}$ (refer to the Definition III.3) |
| $n_{i,j}^{+}$ | The successor node of node $n_{i,j}$ in $P_{[i]}$ (refer to the Definition III.3) |
| $d_{P_{[i]}}^{k,l}$ | The spacing between two nodes on a primary path $P_{[i]}$ (refer to the Definition III.4) |
| $S_{P_{[i]}}$ | The set of sending nodes of a primary path $P_{[i]}$ (refer to the Definition III.5) |
| $S'_{P_{[i]}}$ | A subset of sending nodes of a primary path $P_{[i]}$ (refer to the Definition III.6) |
| $P_{[1,2]}$ | A pair of paths consisting of primary paths $P_{[1]}$ and $P_{[2]}$ (refer to the Definition III.7) |
| $S_{P_{[1,2]}}^{U}$ | The universal set of nodes for a pair of paths $P_{[1,2]}$ (refer to the Definition III.8) |
| $S_{P_{[1,2]}}$ | The set of sending nodes of a pair of paths $P_{[1,2]}$ (refer to the Definition III.9) |
| $S'_{P_{[1,2]}}$ | A subset of sending nodes of a pair of paths $P_{[1,2]}$ (refer to the Definition III.10) |
| $r_{n_{i,j}}^{TC}$ | Transmission coverage radius of node $n_{i,j}$ (refer to the Definition IV.1) |
| $n_{i_1,j_1} \| n_{i_2,j_2}$ | There is a concurrency relationship between two nodes $n_{i_1,j_1}$ and $n_{i_2,j_2}$ (refer to the Definition IV.3) |
| $n_{i_1,j_1} >< n_{i_2,j_2}$ | There is an interference relationship between two nodes $n_{i_1,j_1}$ and $n_{i_2,j_2}$ (refer to the Definition IV.3) |
| $S_{n_{i,j}}^{CU}$ | The universal set of concurrency nodes of node $n_{i,j}$ (refer to the Definition IV.4) |
| $S_{n_{i,j}}^{IU}$ | The universal set of interfering nodes of node $n_{i,j}$ (refer to the Definition IV.5) |
| $C_{S'_{P_{[1,2]}}}^{*}$ | The intrinsic concurrency intensity of a nonempty subset $S'_{P_{[1,2]}}$ of sending nodes (refer to the Definition V.3) |
| $C_{S_{P_{[i]}}}^{*}$ | The intrinsic concurrency intensity of a primary path $P_{[i]}$ (refer to the Definition V.4) |
| $C_{S_{P_{[1,2]}}}^{*}$ | The intrinsic concurrency intensity of a pair of paths $P_{[1,2]}$ (refer to the Definition V.5) |
| $D_{S'_{P_{[1,2]}}\vert\vert n_{i,j}}$ | The concurrency connection degree of a node $n_{i,j}$ in a subset $S'_{P_{[1,2]}}$ of sending nodes (refer to the Definition V.6) |
| $D_{S'_{P_{[1,2]}}\vert\vert}^{*}$ | The intrinsic concurrency connection degree of nodes in a subset $S'_{P_{[1,2]}}$ of sending nodes (refer to the Definition V.7) |
| $D_{S_{P_{[i]}}\vert\vert}^{*}$ | The intrinsic concurrency connection degree of nodes in a primary path $P_{[i]}$ (refer to the Definition V.8) |
| $D_{S_{P_{[1,2]}}\vert\vert}^{*}$ | The intrinsic concurrency connection degree of nodes in a pair of paths $P_{[1,2]}$ (refer to the Definition V.9) |
| $I_{S'_{P_{[1,2]}}}^{*}$ | The intrinsic interference intensity of a nonempty subset $S'_{P_{[1,2]}}$ of sending nodes (refer to the Definition V.13) |
| $I_{S_{P_{[i]}}}^{*}$ | The intrinsic interference intensity of a primary path $P_{[i]}$ (refer to the Definition V.14) |
| $I_{S_{P_{[1,2]}}}^{*}$ | The intrinsic interference intensity of a pair of paths $P_{[1,2]}$ (refer to the Definition V.15) |
| $D_{S'_{P_{[1,2]}}><n_{i,j}}$ | The interference connection degree of a node $n_{i,j}$ in a subset $S'_{P_{[1,2]}}$ of nodes (refer to the Definition V.16) |
| $D_{S'_{P_{[1,2]}}><}^{*}$ | The intrinsic interference connection degree of nodes in a nonempty subset $S'_{P_{[1,2]}}$ of sending nodes (refer to the Definition V.17) |
| $D_{S_{P_{[i]}}><}^{*}$ | The intrinsic interference connection degree of nodes in a primary path $P_{[i]}$ (refer to the Definition V.18) |
| $D_{S_{P_{[1,2]}}><}^{*}$ | The intrinsic interference connection degree of nodes in a pair of paths |
| | $P_{[1,2]}$ (refer to the Definition V.19) |
| $S_{S'}^{CU}$ | The universal set of concurrency nodes for a subset $S'$ of sending nodes on a pair of paths $P_{[1,2]}$ (refer to the Definition V.24) |
| $S_{S'}^{IU}$ | The universal set of interfering nodes for a subset $S'$ of sending nodes on a pair of paths $P_{[1,2]}$ (refer to the Definition V.25) |
| $S_{P_{[1,2]}}^{\backslash S'}$ | The complement of a subset $S'$ of sending nodes on a pair of paths $P_{[1,2]}$ (refer to the Definition V.26) |
| $S_{S'}^{SCU}$ | The universal set of second-order concurrency subsets of a set of sending nodes $S'$ (refer to the Definition V.29) |
| $S_{S'}^{SSCU}$ | The universal set of saturated second-order concurrency subsets of a set of sending nodes $S'$ (refer to the Definition V.29) |
| $S_{S'}^{SIU}$ | The universal set of second-order interference subsets of a set of sending nodes $S'$ (refer to the Definition V.30) |
| $S_{S'}^{SSIU}$ | The universal set of saturated second-order interference subsets of a set of sending nodes $S'$ (refer to the Definition V.30) |
| $N_{S',i}$ | The number of nodes belonging to a certain primary path $P_{[i]}$ $(i = 1,2)$ within a set of nodes $S'$ (refer to the Definition V.31) |
| $\Phi'_{S_{P_{[i]}}}(\theta_{S_{P_{[i]}}}, T_{S_{P_{[i]}}})$ | The subset of equally spaced nodes with its initial phase being $\theta_{S_{P_{[i]}}}$ and the spacing between adjacent nodes being $T_{S_{P_{[i]}}}$ in a primary path $P_{[i]}$ (refer to the Definition VII.1) |
| $T_{S_{P_{[i]}}}$ | The period of a primary path $P_{[i]}$ (refer to the Definition VII.2) |
| $T_{S_{P_{[1,2]}}}$ | The joint period of a pair of paths $P_{[1,2]}$ (refer to the Definition VII.3) |
| $T^*_{S_{P_{[i]}}}$ | The intrinsic period of a primary path $P_{[i]}$ (refer to the Definition VII.4) |
| $T^*_{S_{P_{[1,2]}}}$ | The intrinsic period of a pair of paths $P_{[1,2]}$ (refer to the Definition VII.5) |
| $c(\theta_{S_{P_{[1]}}}, \theta_{S_{P_{[2]}}})$ | The joint concurrency coefficient between subsets of equally spaced nodes of two primary paths $P_{[1]}$ and $P_{[2]}$ (refer to the Definition VII.6) |
| $\boldsymbol{C}$ | The matrix of joint concurrency relationship between subsets of equally |
| | spaced nodes of a pair of paths (refer to the Definition VII.7) |
| $\boldsymbol{C}^*$ | The matrix of intrinsic joint concurrency relationship between subsets of equally spaced nodes of a pair of paths (refer to the Definition VII.8) |
| $S_{\boldsymbol{B}_{sp}}$ | The set of supporting elements of a binary matrix $\boldsymbol{B}$ (refer to the Definition VII.9) |
| $S^*_{\boldsymbol{B}_{sp}}$ | The set of maximum supporting elements of a binary matrix $\boldsymbol{B}$ (refer to the Definition VII.10) |
| $U_{\boldsymbol{B}_{sp}}$ | The reachable supporting number of a binary matrix $\boldsymbol{B}$ (refer to the Definition VII.11) |
| $U^*_{\boldsymbol{B}_{sp}}$ | The maximum reachable supporting number of a binary matrix $\boldsymbol{B}$ (refer to the Definition VII.11) |
| $D^{int}_{t,b}$ | The time interval starting from time instant $t$ and lasting for $b$ ($b \in \mathbb{Z}, b \geq 0$) time beats (refer to the Definition IX.1) |
| $R^{av}_{S_{P_{[i]}}}$ | The average throughput of a primary path $P_{[i]}$ in a specific time interval (refer to the Definition IX.2) |
| $R^{av}_{S_{P_{[1,2]}}}$ | The average joint throughput of a pair of paths $P_{[1,2]}$ in a specific time interval (refer to the Definition IX.3) |
| $R^{\infty}_{S_{P_{[i]}}}$ | The asymptotic throughput of a primary path $P_{[i]}$ (refer to the Definition IX.4) |
| $R^{\infty}_{S_{P_{[1,2]}}}$ | The asymptotic joint throughput of a pair of paths $P_{[1,2]}$ (refer to the Definition IX.5) |
| $r^{i}_{R^{\infty}_{S_{P_{[1,2]}}}}$ | The ratio between the asymptotic throughput of a primary path $P_{[i]}$ and the asymptotic joint throughput of pair of paths $P_{[1,2]}$ (refer to the Definition IX.5) |
| $\boldsymbol{M}^{L_1 \times L_2}$ | The $L_1 \times L_2$-th continuation matrix of a matrix $\boldsymbol{M}$ (refer to the Definition IX.7) |
| $S^*_{\boldsymbol{M}^{L_1 \times L_2}_{sp}}$ | The set of maximum supporting elements of $\boldsymbol{M}^{L_1 \times L_2}$ (refer to the Definition IX.7) |
| $U^*_{\boldsymbol{M}^{L_1 \times L_2}_{sp}}$ | The maximum reachable supporting number of $\boldsymbol{M}^{L_1 \times L_2}$ (refer to the |
| | Definition IX.7) |
| $D^{ed}_{S_{P_{[i]}}}$ | One time end-to-end delay of an information block on a primary path $P_{[i]}$ (refer to the Definition IX.8) |
| $\boldsymbol{T}_{IB,i}(t,b)$ | The information block transmission time trace matrix of a primary path $P_{[i]}$ within time interval $D^{int}_{t,b}$ (refer to the Definition X.1) |
| $F^{I_d}_{S_P}(\eta)$ | The Distribution Spectrum of Interference-Spacing of nodes (i.e. DSIS) of a primary path$P$ (refer to the Definition XI.1) |
| $\eta$ | The sequence number of a spectral line of the spectrum $F^{I_d}_{S_P}(\eta)$ (refer to the Definition XI.1) |
| $\eta^{min}_{Az}$ | The sequence number of the minimum periodic equally spaced cumulative zero point of the spectrum $F^{I_d}_{S_P}(\eta)$ (refer to the Definition XI.5) |
| $\eta^{min}_{Bz}$ | The sequence number of the minimum point of returning to zero of the spectrum $F^{I_d}_{S_P}(\eta)$ (refer to the Definition XI.6) |

**Appendix B：List of Abbreviations**

| | |
|---|---|
| DSIS | the Distribution Spectrum of Interference-Spacing of nodes of a primary path (refer to the Section VIII) |
| | |